\documentclass[aps,prb,twocolumn,superscriptaddress, longbibliography]{revtex4-2}
\usepackage{graphicx}
\usepackage{latexsym}
\usepackage{amssymb}
\usepackage{amsmath}
\usepackage{amsfonts}
\usepackage[vcentermath]{youngtab}
\usepackage{upgreek}
\usepackage{bm,dsfont}
\usepackage{multirow}
\usepackage{enumitem}
\usepackage[usenames, dvipsnames]{color}
\usepackage[svgnames]{xcolor}
\usepackage{hyperref}
\hypersetup{
colorlinks = true,
linkcolor = [rgb]{0.70,0.13,0.13}, 
citecolor = [rgb]{0.13,0.55,0.13},
urlcolor = [rgb]{0.25, 0.41, 0.88}}
\usepackage{makecell}

\newcommand{\bra}[1]{\langle #1|}
\newcommand{\ket}[1]{|#1 \rangle}

\newcommand{\dd}{\mathrm{d}}

\newcommand{\ii}{\mathrm{i}}
\newcommand{\e}{\mathrm{e}}

\newcommand{\E}{\mathop{\mathbb{E}}}
\newcommand{\LU}{\mathrm{LU}}
\newcommand{\U}{\mathrm{U}}
\newcommand{\SU}{\mathrm{SU}}

\newcommand{\dsZ}{\mathbb{Z}}

\newcommand{\scP}{\mathcal{P}}

\renewcommand{\Re}{\operatorname{Re}}
\renewcommand{\Im}{\operatorname{Im}}
\newcommand{\vect}[1]{{\bm{#1}}}

\newcommand{\eq}[1]{\begin{equation}#1\end{equation}}
\newcommand{\eqs}[1]{\begin{equation}\begin{split}#1\end{split}\end{equation}}
\newcommand{\eqnref}[1]{Eq.\,\eqref{#1}}
\newcommand{\figref}[1]{Fig.\,\ref{#1}}

\newcommand{\secref}[1]{Sec.\,\ref{#1}}

\newcommand{\refcite}[1]{Ref.\,\onlinecite{#1}}

\newcommand{\bea}{\begin{eqnarray}}
\newcommand{\eea}{\end{eqnarray}}
\def\be{\begin{equation}}
\def\ee{\end{equation}}
\newcommand{\beq}{\begin{equation}}
\newcommand{\eeq}{\end{equation}}
\newcommand{\beqn}{\begin{eqnarray}}
\newcommand{\eeqn}{\end{eqnarray}}

\newcommand{\bk}{{\vect{k}}}
\newcommand{\br}{{\vect{r}}}
\newcommand{\bR}{{\vect{R}}}
\newcommand{\bq}{\vect{q}}
\newcommand{\bQ}{\vect{Q}}

\newcommand{\ssb}{\text{SSB}}

\newcommand{\smg}{\text{SMG}}

\usepackage{ulem}
\begin{document}

\title{Green's Function Zeros in Fermi Surface Symmetric Mass Generation}

\author{Da-Chuan Lu}
\affiliation{Department of Physics, University of California, San Diego, CA 92093, USA}
\author{Meng Zeng}
\affiliation{Department of Physics, University of California, San Diego, CA 92093, USA}
\author{Yi-Zhuang You}
\affiliation{Department of Physics, University of California, San Diego, CA 92093, USA}


\begin{abstract}
The Fermi surface symmetric mass generation (SMG) is an intrinsically interaction-driven mechanism that opens an excitation gap on the Fermi surface without invoking symmetry-breaking or topological order. We explore this phenomenon within a bilayer square lattice model of spin-1/2 fermions, where the system can be tuned from a metallic Fermi liquid phase to a strongly-interacting SMG insulator phase by an inter-layer spin-spin interaction. The SMG insulator preserves all symmetries and has no mean-field interpretation at the single-particle level. It is characterized by zeros in the fermion Green's function, which encapsulate the same Fermi volume in momentum space as the original Fermi surface, a feature mandated by the Luttinger theorem. Utilizing both numerical and field-theoretical methods, we provide compelling evidence for these Green's function zeros across both strong and weak coupling regimes of the SMG phase. Our findings highlight the robustness of the zero Fermi surface, which offers promising avenues for experimental identification of SMG insulators through spectroscopy experiments despite potential spectral broadening from noise or dissipation. 
\end{abstract}
\maketitle


\section{Introduction}

Symmetric mass generation (SMG) \cite{Fid2010ISPT, Fid2011ISPT, Wang1307.7480, Slagle1409.7401, Ayyar1410.6474, Catterall1510.04153, Tong2022SMG, YZY2022SMGreview} is an interaction-driven mechanism that creates many-body excitation gaps in anomaly-free fermion systems \emph{without} condensing any fermion bilinear operator or developing topological orders. It has emerged as a alternative symmetry-preserving approach for mass generation in relativistic fermion systems, which is distinct from the traditional symmetry-breaking Higgs mechanism \cite{Nambu1960higgs,Nambu1961higgs,Goldstone1962higgs,Anderson1963higgs,Englert1964higgs,Higgs1964higgs}. The prospect of SMG offering a potential solution to the long-standing fermion doubling problem \cite{Nielsen1981b,Nielsen1981a,Nielsen1981NoGo,Swift1984,Eichten1986Chiral,Smit1986,KaplanRLB1992} has sparked significant interest in the lattice gauge theory community \cite{Wen1305.1045,You1402.4151,You1412.4784,BenTov1412.0154,Ayyar1511.09071,Ayyar1606.06312,Ayyar1611.00280,DeMarco1706.04648,Ayyar1709.06048,Schaich1710.08137,Butt1811.01015,Butt1810.06117,Kikukawa1710.11618,Kikukawa1710.11101,Wang1807.05998,Wang1809.11171,Catterall2002.00034,Razamat2009.05037,Catterall2021SMG,Butt2101.01026,Butt2111.01001,ZM2022SMG,Catterall2201.00750,Catterall2023KDfermion,Guo2023S2306.17420}. In condensed matter physics, SMG was initially explored within the framework of the interaction-reduced classification of fermionic symmetry protected topological (SPT) states \cite{Fid2010ISPT,Fid2011ISPT,Turner1008.4346,Ryu1202.4484,Qi1202.3983,Yao1202.5805,Gu1304.4569,Wang1401.1142,Metlitski1406.3032,Kapustin2015IFSPT,YZY2014,Cheng1501.01313,Yoshida1505.06598,Gu1512.04919,Song1609.07469,Queiroz1601.01596,Witten1605.02391,Wang1703.10937,Kapustin1701.08264,Wang2018Tunneling1801.05416,Wang1811.00536,Guo1812.11959,Aasen2109.10911,Barkeshli2109.11039}, and has been recently extended to systems with Fermi surfaces \cite{Zou2004.14391,yahui2020-1,Zou2002.02972,yahui2020-2,yahui2021,YZY2022FSSMG}, given the growing understanding that Fermi liquids can be perceived as fermionic SPT states within the phase space \cite{Bulmash1410.4202,yzy2023FSanomaly}.

One important feature of the SMG gapped state lies in the zeros of fermion Green's function \cite{Gurarie1011.2273, yzy2014greenfunctionzero, Catterall1609.08541, YZY2018SMGDQCP-1, Catterall1708.06715, Xu2103.15865} at low-energy. Investigations reveal that the poles of the fermion Green's function in the pristine gapless fermion state will be replaced by zeros in the gapped SMG state as the fermion system goes across the SMG transition upon increasing the interaction strength. This pole-to-zero transition was postulated \cite{yzy2014greenfunctionzero} as a direct indicator of the SMG transition \cite{YZY2018SMGDQCP-1, YZY2018SMGDQCP-2} that can be probed by spectroscopy experiments. However, the presence of similar zeros in the Green's function within Fermi surface SMG states has not been investigated yet, and it is the focus of our present research.

Fermi surface SMG \cite{YZY2022FSSMG} refers to the occurrence of SMG phenomena on Fermi surfaces with non-zero Fermi volumes. It describes scenarios where the fermion interaction transforms a gapless Fermi liquid state (metal) into a non-degenerate, gapped, direct product state (trivial insulator), without breaking any symmetry (for example, without invoking Cooper pairing or density wave orders). Such a metal-insulator transition is viable when Fermi surfaces collaboratively cancel the Fermi surface anomaly \cite{Chong2021FSanomaly, XiaoGang2021FSanomaly, YZY2022FSSMG}. This anomaly can be perceived as a mixed anomaly between the translation symmetry and the charge conservation $\U(1)$ symmetry on the lattice \cite{Meng2016Topo, Shinsei2017FSLSM, Bultinck1808.00324, Else2018FSLSMcrystalline, XiaoGang2021FSanomaly, Chong2021FSanomaly, Seiberg2022emanant}, or as an anomaly of an emergent loop $\LU(1)$ symmetry \cite{Senthil2007.07896,Else2010.10523,Darius-Shi2204.07585} in the infrared theory.

In this work, we present evidence of robust Green's function zeros in Fermi surface SMG states. Let $t$ be the energy scale of band dispersion and $J$ be the energy scale of SMG gapping interaction, we investigate the problem from two parameter regimes:
\begin{itemize}
\item Deep in the SMG phase ($J/t\gg 1$), we start with an exact-solvable SMG product state in a lattice model and calculate the fermion Green's function by treating the fermion hopping as perturbation \cite{CPT2000}. We find that the Green's function $G_{\smg}(\omega,\bk)$ deep in the SMG phase takes the following form 
\begin{equation}\label{eq: GSMG strong}
    G_{\smg}(\omega,\bk) = \frac{\omega+\alpha \epsilon_{\bk}/J^2}{(\omega-\epsilon_\bk/2)^2-J^2},
\end{equation}
where $(\omega,\bk)$ labels the frequency-momentum of the the fermion. $\epsilon_\bk$ is the energy dispersion of the original band structure in the free-fermion limit,  and $\alpha$ is an order-one number depending on other details of the system. One salient feature of $G_{\smg}$ is that it has a series of zeros at $\omega=-\alpha \epsilon_\bk/J^2$ in the frequency-momentum space. At $\omega=0$, the Green's function zeros form a zero Fermi surface that replaces the original Fermi surface.

\item If the SMG phase is adjacent to a spontaneous symmetry breaking (SSB) phase, we use perturbative field theory to argue that the Green's function in the SMG phase near the symmetry-breaking transition ($J/t\gtrsim 1$) should take the form of
\eq{\label{eq: GSMG weak}
G'_\smg(\omega,\bk)=\frac{\omega+\epsilon_\bk}{\omega^2-\epsilon_\bk^2-\Delta_0^2}}
where we assume that the SSB order parameter retains a finite amplitude $\Delta_0$ in the SMG phase, but its phase is randomly fluctuating \cite{Wang2023P2212.05737}. Again, $G'_\smg$ features a series of zeros at $\omega=-\epsilon_\bk$, with the same zero Fermi surface.

\end{itemize}

Many previous works \cite{Altshuler1998Lcond-mat/9703120, Oshikawacond-mat/0002392, 2003LuttingerSurface, Gnezdilov2022S2111.09906} suggest that the Luttinger theorem \cite{LuttingerRP1960} will not be violated in the presence of the interaction that preserves the translation and charge conservation symmetry. However, quasi-particles (poles of Green's function) may not exist in the strongly correlated systems, the Fermi surface is instead defined by the surface of Green's function zeros at zero frequency, i.e.,~$G(0,\bk)=0$, and the Green's function changes sign on the two sides of the \emph{zero Fermi surface}, or the so-called Luttinger surface \cite{Stanescu2007Tcond-mat/0602280, 2003LuttingerSurface, Senthil2007.07896, Sachdev1711.09925, xiang2022d}. This can be regarded as the remnant of the conventional Fermi surface in the strongly interacting gapped phase. Our analysis shows that the volume enclosed by the zeros of the Green's function in the SMG phase is the same as the Fermi volume in the Fermi liquid phase, which agrees with the Luttinger theorem.

The paper will be structured as follows. We start by introducing a concrete lattice model for Fermi surface SMG in \secref{sec: model} and briefly discussing its phase diagram. We give theoretical arguments for Green's function zeros in the SMG phase from the Luttinger theorem in \secref{sec: Luttinger} (general), and the particle-hole symmetry in \secref{sec: symmetry} (specific). We provide numerical and field theoretical evidence of Green's function zeros from both the strong coupling \secref{sec: strong} and the weak coupling \secref{sec: weak} perspectives. We comment on the robustness of probing the zero structure in spectroscopy experiments in \secref{sec: probe}. We conclude in \secref{sec: summary} with a discussion of the relevance of our model to the nickelate superconductor La$_3$Ni$_2$O$_7$.


\section{Argument For Green's Function Zeros}

\subsection{Lattice Model and Phase Diagram}\label{sec: model}

\begin{figure}[htbp]
\begin{center}
\includegraphics[scale=0.65]{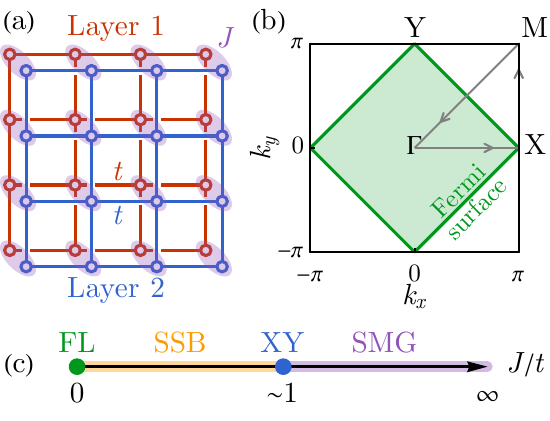}
\caption{(a) Bilayer square lattice model with intra-layer hopping and inter-layer spin interaction. (b) Fermi sea and Fermi surface at $J=0$ in the Brillouin zone. A high-symmetry path is traced out in gray. (c) A conjectured phase diagram consist of a Fermi liquid (FL) fixed point, a spontaneous symmetry breaking (SSB) phase, a XY transition, and a SMG insulating phase.}
\label{fig: lattice}
\end{center}
\end{figure}

As a specific example of Fermi surface SMG, we consider a bilayer square lattice \cite{Zhai2009A0905.1711, Ruger2014T1311.6504, Rhim2019C1808.05926} model of spin-1/2 fermions, as illustrated in \figref{fig: lattice}(a). Let $c_{il\sigma}$ be the fermion annihilation operator on site-$i$ layer-$l$ ($l=1,2$) and spin-$\sigma$ ($\sigma=\uparrow,\downarrow$). The model is described by the following Hamiltonian
\begin{equation}\label{eq: H}
H = -t\sum_{\langle ij \rangle,l,\sigma} (c_{il\sigma}^\dagger c_{jl\sigma}+ \text{h.c.})+ J \sum_{i} \vect{S}_{i1}\cdot \vect{S}_{i2},
\end{equation}
where $\vect{S}_{il}:=\tfrac{1}{2}c_{il\sigma}^\dagger\vect{\sigma}_{\sigma\sigma'}c_{il\sigma'}$ denotes the spin operator with $\vect{\sigma}:=(\sigma^1,\sigma^2,\sigma^3)$ being the Pauli matrices. The Hamiltonian $H$ contains a nearest-neighbor hopping $t$ of the fermions within each layer and an inter-layer Heisenberg spin-spin interaction with antiferromagnetic coupling $J>0$. The Heisenberg interaction should be understood as a four-fermion interaction, that there is no explicitly formed local moment degrees of freedom. Unlike the standard $t$-$J$ model \cite{Chao1978tJ}, we do \emph{not} impose any on-site single-occupancy constraint \cite{Gutzwiller1964E} here. We assume that the fermions are half-filled in each layer. 

In the non-interacting limit ($J/t\to 0$), the ground state of the tight-binding Hamiltonian in \eqnref{eq: H} is a Fermi liquid with a four-fold degenerated (two layers and two spins) square-shaped Fermi surface in the Brillouin zone, as shown in \figref{fig: lattice}(b). The fermion system is gapless in this limit. However, given that the fermion carries one unit charge under the $\U(1)$ symmetry, the Fermi surface anomaly vanishes due to \cite{Shinsei2017FSLSM,yzy2023FSanomaly} 
\begin{equation}
    \sum_{a=1}^{4} q_a \nu_a = 4\times 1 \times \frac{1}{2}=0 \mod 1,
\end{equation}
where $a$ indexes the four-fold degenerated Fermi surface with $q_a=1$ being the $\U(1)$ charge carried by the fermion and $\nu_a=1/2$ being the filling fraction. This implies there must be a way to gap out the Fermi surface into a trivial insulator while preserving both the translation and the $\U(1)$ charge conservation symmetries. Nevertheless, these symmetry requirements are restrictive enough to rule out all possible fermion bilinear gapping mechanisms, leaving Fermi surface SMG the only available option.

One possible SMG gapping interaction is the interlayer Heisenberg spin-spin interaction $J$ in \eqnref{eq: H}. In the strong interaction limit ($J/t\to\infty$), the system has a unique ground state, given by
\eq{\label{eq: SMG GS}\ket{0}=\bigotimes_i (c_{i1\uparrow}^\dagger c_{i2\downarrow}^\dagger - c_{i1\downarrow}^\dagger c_{i2\uparrow}^\dagger)\ket{\text{vac}},}
which is a direct product of the inter-layer spin-singlet state on every site. $\ket{\text{vac}}$ stands for the vacuum state of fermions (i.e.~$c_{il\sigma}\ket{\text{vac}}=0$). The SMG ground state $\ket{0}$ does not break any symmetry and does not have topological order. All excitations are gapped by an energy of the order $J$ from the ground state. Any local perturbation far below the energy scale $J$ can not close this excitation gap, so the SMG phase is expected to be stable in a large parameter regime as long as $J\gg t$.

Given the distinct ground states in the two limits of $J/t$, we anticipate at least one quantum phase transition separating the Fermi liquid and the SMG insulator. However, due to the perfect nesting of the Fermi surface, the Fermi liquid state is unstable towards spontaneous symmetry breaking (SSB) upon infinitesimal interaction, so a more plausible phase diagram should look like \figref{fig: lattice}(c), where an intermediate SSB phase sets in. A mean-field analysis based on the Fermi liquid fixed point shows that there are two degenerated leading instabilities: (i) the inter-layer exciton condensation (EC) and (ii) the inter-layer superconductivity (SC). They are respectively described by the following order parameters
\eq{\label{eq: def phi}
\phi_\text{EC}=\sum_{i,\sigma}(-)^{i} c_{i1\sigma}^\dagger c_{i2\sigma},\quad \phi_\text{SC}=\sum_{i,\sigma}(-)^{\sigma} c_{i1\sigma}^\dagger c_{i2\bar{\sigma}}^\dagger.}
Here, $(-)^i$ denotes the stagger sign on the square lattice of lattice momentum $(\pi,\pi)$. $(-)^\sigma=+1$ for $\sigma=\uparrow$ and $-1$ for $\sigma=\downarrow$. $\bar{\sigma}$ stands for the opposite spin of $\sigma$. 

The energetic degeneracy of these two SSB orders can be explained by the fact that their order parameters $\phi_\text{EC}$ and $\phi_\text{SC}$ are related by a particle-hole transformation $c_{i2\sigma}\to(-)^{i}(-)^{\sigma}c_{i2\bar{\sigma}}^\dagger$ in the second layer only, which is a symmetry of the model Hamiltonian in \eqnref{eq: H}. The EC $\langle\phi_\text{EC}\rangle\neq 0$ spontaneously breaks the translation and interlayer $\U(1)$ symmetry, and the SC $\langle\phi_\text{SC}\rangle\neq 0$ spontaneously breaks the total $\U(1)$ symmetry. Both of them gap out the Fermi surfaces fully, leading to an SSB insulator (or superconductor). The SSB and SMG phases are likely separated by an XY transition, at which the symmetry gets restored. We will leave the numerical verification of the proposed phase diagram \figref{fig: lattice}(c) for future study, as the main focus of this research is to investigate the structure of fermion Green's function in the SMG insulating phase.

We note that the model \eqnref{eq: H} was also introduced as the ``coupled ancilla qubit'' model to describe the pseudo-gap physics in the recent literature \cite{yahui2020-1,yahui2020-2,yahui2021}. Its honeycomb lattice version has been investigated in recent numerical simulations \cite{YZY2022nuSMG}, where a direct quantum phase transition between semimetal and insulator phases was observed.

\subsection{Luttinger Theorem and Green's Function Zeros}
\label{sec: Luttinger}

The Luttinger theorem \cite{Luttinger1960G, LuttingerRP1960} asserts that in a fermion many-body system with lattice translation and charge $\U(1)$ symmetries, the ground state charge density $\langle N\rangle/V$ (i.e.,~the $\U(1)$ charge per unit cell) is tied to the momentum space volume in which the real part of the zero-frequency fermion Green's function is positive $\Re G(0,\bk)>0$. This can be formally expressed as
\eq{\label{eq: Luttinger}
\frac{\langle N\rangle}{V}=N_f\int_{\Re G(0,\bk)>0}\frac{\dd^2\bk}{(2\pi)^2}.}
Here, the $\U(1)$ symmetry generator $N=\sum_{i,l,\sigma}c_{il\sigma}^\dagger c_{il\sigma}$ measures the total charge, and the volume $V=\sum_{i}1$ is defined as the number of unit cells in the lattice system. $N_f=4$ counts the fermion flavor number (or the Fermi surface degeneracy), including two layers and two spins. The Green's function $G(\omega,\bk)$ in \eqnref{eq: Luttinger} is defined by the fermion two-point correlation as 
\eq{\label{eq: def G}
\langle c_{l\sigma}(\omega,\bk)c_{l'\sigma'}(\omega,\bk)^\dagger\rangle=G(\omega,\bk)\delta_{ll'}\delta_{\sigma\sigma'}.}
The correlation function is proportional to an identity matrix in the flavor (layer-spin) space because of the layer $\U(1):c_{l\sigma}\to \e^{(-)^l\ii\theta}c_{l\sigma}$, the layer interchange $\dsZ_2:c_{1\sigma}\leftrightarrow c_{2\sigma}$, and the spin $\SU(2):c_{l\sigma}\to (\e^{\ii\vect{\theta}\cdot\vect{\sigma}/2})_{\sigma\sigma'}c_{l\sigma'}$ symmetries.

The Luttinger theorem applies to the Fermi liquid and SMG states in the bilayer square lattice model \eqnref{eq: H}, as both states preserve the translation and charge $\U(1)$ symmetries. Given that the fermions are half-filled ($\nu=1/2$) in the system, the Fermi volume should be
\eq{\int_{\Re G(0,\bk)>0}\frac{\dd^2\bk}{(2\pi)^2}=\frac{\langle N \rangle}{V N_f}=\nu=\frac{1}{2}.} 
The Fermi volume is enclosed by the Fermi surface, across which $\Re G(0,\bk)$ changes sign. The sign change can be achieved either by poles or zeros in the Green's function.

In the Fermi liquid state, the required Fermi volume is satisfied via Green's function poles along the Fermi surface, as pictured in \figref{fig: lattice}(b). However, the SMG insulator is a fully gapped state of fermions that has no low-energy quasi-particles (below the energy scale $J$). Consequently, the Green's function $G(\omega,\bk)$ cannot develop poles at $\omega=0$, meaning the required Fermi volume can only be satisfied by Green's function zeros. Therefore, the Lutinger theorem implies that there must be robust Green's function zeros at low energy in the SMG phase, and the zero Fermi surface must enclose half of the Brillouin zone volume in place of the original pole Fermi surface.

It is known that the Luttinger theorem can be violated in the presence of topological order \cite{Senthil2003Topo, Senthilcond-mat/0305193, Paramekanticond-mat/0406619, Senthil2006Topo, Senthil2012Topo, Meng2016Topo, Sachdev1711.09925, Sachdev2019T1801.01125, Bultinck1808.00324, Skolimowski2022L2202.00426}. However, this concern does not affect our discussion in the SMG phase, because the SMG insulator is a trivial insulator without topological order.

\subsection{Particle-Hole Symmetry and Zero Fermi Surface}
\label{sec: symmetry}

The Luttinger theorem only constrains the Fermi volume but does not impose requirements on the shape of the Fermi surface. However, in this particular example of the bilayer square lattice model \eqnref{eq: H}, the system has sufficient symmetries to determine even the shape of the Fermi surface. 

The key symmetry here is a particle-hole symmetry $\dsZ_2^C$, which acts as
\eq{\label{eq: Z2C}
c_{il\sigma}\to(-)^i(-)^\sigma c_{il\bar{\sigma}}^\dagger.}
The Hamiltonian $H$ in \eqnref{eq: H} is invariant under this transformation. Since the Green's function is an identity matrix in the flavor space \eqnref{eq: def G} which is invariant under any flavor basis transformation, we can omit the flavor indices and focus on the frequency-momentum dependence of the Green's function, written as
\eq{G(\omega,\vect{k})=\sum_{t,\vect{x},t',\vect{x}'}\langle c(t,\vect{x})c(t',\vect{x}')^\dagger\rangle \e^{\ii(\omega (t-t')-\vect{k}\cdot(\vect{x}-\vect{x}'))}.}
Given \eqnref{eq: Z2C}, the fermion field $c(t,\vect{x})$  transforms under the $\dsZ_2^C$ symmetry as 
\eq{c(t,\vect{x})\to c(t,\vect{x})^\dagger \e^{\ii \vect{Q}\cdot\vect{x}},\quad c(t,\vect{x})^\dagger \to c(t,\vect{x}) \e^{-\ii \vect{Q}\cdot\vect{x}},} 
where $\vect{Q}=(\pi,\pi)$ is the momentum associated with the stagger sign factor $(-)^i$ on the square lattice. As a consequence, the Green's function transforms as
\eq{G(\omega,\vect{k})\to -G(-\omega,\vect{Q}-\vect{k}).
}
Furthermore, there are also two diagonal reflection symmetries on the square lattice, which maps $\bk=(k_x,k_y)$ to $(k_y,k_x)$ or $(-k_y,-k_x)$ in the momentum space.

Both the Fermi liquid and the SMG states preserve the particle-hole symmetry $\dsZ_2^C$ and the lattice reflection symmetry, which requires the Green's function to be invariant under the combined symmetry transformations. So the zero-frequency Green's function must satisfy
\eq{G(0,k_x,k_y)=-G(0,\pi\pm k_y,\pi\pm k_x),}
meaning that the sign change of $G(0,\bk)$ should happen along $k_x\pm k_y=\pi\mod 2\pi$, which precisely describes the shape of the Fermi surface. The Fermi surface is pole-like in the Fermi liquid state and becomes zero-like in the SMG state, but its shape and volume remain the same.

However, it should be noted that the precise overlap of the zero Fermi surface in the SMG insulator and the pole Fermi surface in the Fermi liquid is a fine-tuned feature of the bilayer square lattice model \eqnref{eq: H}. In more general cases, such as including further neighbor hopping in the model, the particle-hole symmetry would cease to exist, thus the invariance in the shape of the Fermi surface is no longer guaranteed. Nevertheless, the Luttinger theorem can still ensure the invariance in the Fermi volume, thereby providing the SMG insulator with robust Green's function zeros.

To verify this proposition, we will analyze the behavior of the Green's function in the SMG phase from both strong and weak coupling perspectives in \secref{sec: evidence}. Our calculations suggest that, for this specific model, the SMG state indeed possesses a Fermi surface (of Green's function zeros) that is identical in shape to that in the Fermi liquid state.



\section{Evidence of Green's Function Zeros}\label{sec: evidence}

\subsection{Strong Coupling Analysis}\label{sec: strong}

We will first focus on the strong interaction limit $(J/t\to\infty)$, where the system is deep in the SMG phase and the exact ground state is known (see \eqnref{eq: SMG GS}). We start from this limit and turn on the hopping term as a perturbation. We employ exact diagonalization and cluster perturbation theory (CPT) \cite{CPT2000,senechal2002cluster} to compute the Green's function in the SMG phase. The details of our method are described in Appendix~\ref{append:cpt}. It is valid to use a small cluster to reconstruct the Green's function in the SMG phase since the ground state is close to a product state that does not have long-range correlation or long-range quantum entanglement. This is quite different from the Hubbard model, where the Fermi surface anomaly is non-vanishing, and the infrared phase must be either SSB order or topological order \cite{Senthil2003Topo, Senthil2006Topo, Senthil2012Topo, Meng2016Topo}. In either case, the ground state wave functions cannot be reconstructed from the small clusters due to the long-range correlation/entanglement. This argument has been noted in the original paper on the CPT method \cite{CPT2000}.

\begin{figure}[htbp]
\begin{center}
\includegraphics[scale=0.65]{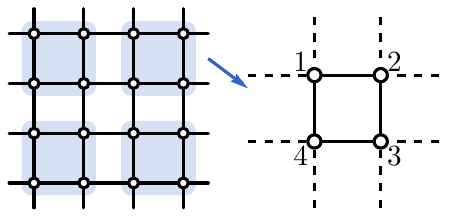}
\caption{Partition the square lattice into $2\times 2$ clusters. The many-body Hamiltonian is exactly diagonalized within each cluster. The effect of inter-cluster hopping is included in an RPA-like approach.}
\label{fig: cluster}
\end{center}
\end{figure}

To be specific, we first partition the square lattice (including both layers) into $2\times 2$ square clusters as shown in \figref{fig: cluster}. Let us first ignore the inter-cluster hopping. Within each cluster, we represent the Hamiltonian in the many-body Hilbert space and use the Lanczos method to obtain the lowest $\sim 2000$ eigenvalues and eigenvectors. The Green's function in the cluster can then be obtained by the K\"all\'en–Lehmann representation
\begin{equation}
    G_0(\omega)_{ij} = \sum_{m>0} \frac{\bra{0}c_{i}\ket{m}\bra{m} c_{j}^\dagger\ket{0}}{\omega -(E_m-E_0)}+  \frac{\bra{m}c_{i}\ket{0}\bra{0} c_{j}^\dagger\ket{m}}{\omega +(E_m-E_0)},
\end{equation}
where $\ket{m}$ is the $m$th excited state with energy $E_m$, and $\ket{0}$ is the ground state with energy $E_0$, whose wave function was previously given in \eqnref{eq: SMG GS}. Since the four fermion flavors (two spins and two layers) are identical under the internal flavor symmetry, we can drop the flavor index in the Green's function and only focus on one particular flavor with the site indices $i,j$, where $i,j=1,2,3,4$ as indicated in \figref{fig: cluster}. 
The convergence of the Green's function can be verified by including more eigenstates from the Lanczos method. We checked that increasing the number of eigenpairs to $\sim 8000$ will not change the result significantly, indicating that the result with $\sim 2000$ eigenpairs has already converged.


Now we restore the inter-cluster hoping to extend the Green's function from small clusters to the infinite lattice. The Green's function of super-lattice momentum $\bk$ can be obtained from the random phase approximation (RPA) approach \cite{CPT2000},
\eq{G(\omega,\bk)_{ij} = \left(\frac{G_0(\omega)}{1-T(\bk)G_0(\omega)}\right)_{ij},}
where the $T(\bk)$ matrix
\begin{equation}
\label{eq-V}
T(\bk) =-t\begin{pmatrix}
0 &  \e^{-\ii 2 k_x} & 0 & \e^{\ii 2 k_y}\\
\e^{\ii 2 k_x} &0 & \e^{\ii 2 k_y} & 0\\
0 &  \e^{-\ii 2 k_y} & 0&  \e^{\ii 2 k_x}\\
 \e^{-\ii 2 k_y} & 0 &  \e^{-\ii 2 k_x} & 0
\end{pmatrix}
\end{equation}
describes the inter-cluster fermion hopping. The resulting Green's function $G(\omega,\bk)_{ij}$ is defined in the folded Brillouin zone $\bk\in(-\pi/2,\pi/2]^{\times 2}$ with sub-lattice indices $i,j$. To unfold the Green's function to the original Brillouin zone $\bk\in(-\pi,\pi]^{\times 2}$, we perform the following (partial) Fourier transform
\begin{equation}\label{eq: GSMG numerics}
    G(\omega,\bk) = \frac{1}{L}\sum_{i,j}\e^{-\ii \bk\cdot (\br_i-\br_j)}G(\omega,\bk)_{ij}.
\end{equation}


\begin{figure}[htbp]
\begin{center}
\includegraphics[scale=0.65]{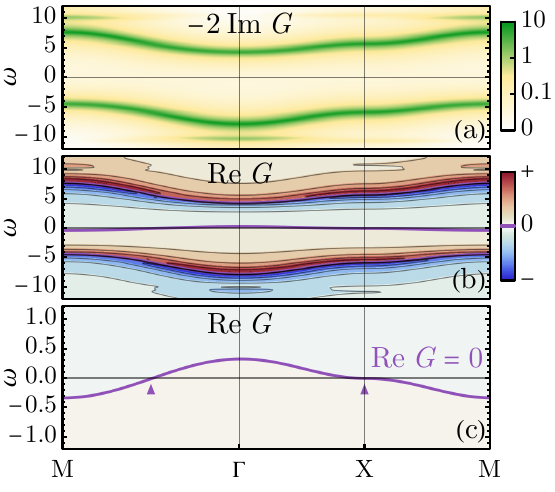}
\caption{Fermion Green's function \eqnref{eq: GSMG numerics} deep in the SMG insulator phase, at $J=8t$. (a) The imaginary part (spectral function) $-2\Im G(\omega+\ii0_+,\vect{k})$ shows the pole (spectral peak) structure. (b) The real part $\Re G(\omega,\vect{k})$ shows the pole (divergence) and zero (purple contour) structures. (c) Same as (b) but zoomed in near $\omega=0$ to show the dispersion of Green's function zeros.}
\label{fig: Gdeep}
\end{center}
\end{figure}

\begin{figure}[htbp]
\begin{center}
\includegraphics[scale=0.65]{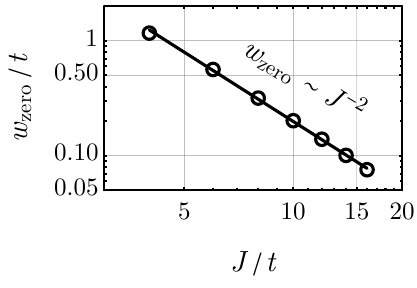}
\caption{Scaling of the Green's function zero ``bandwidth'' $w_\text{zero}$ with the interaction strength $J$. Circles represent the numerically calculated $w_\text{text}$ at different $J$, and the line is a fit to the data.}
\label{fig: wzero}
\end{center}
\end{figure}

We numerically calculated the unfolded Green's function $G(\omega,\bk)$ using the above-mentioned cluster perturbation method. We take a large interaction strength $J/t=8$ deep in the SMG phase and present the resulting Green's function in \figref{fig: Gdeep}. From \figref{fig: Gdeep}(a), the poles of the Green's function form two dispersing bands around $\omega=\pm J$, which resembles the upper and lower Hubbard bands in the Hubbard model. This indicates the quasi-particles are fully gapped in the SMG phase. Meanwhile, from \figref{fig: Gdeep}(b,c), the zeros of the Green's function appear around $\omega= -\alpha \epsilon_\bk/J^2$ with some non-universal but positive coefficient $\alpha>0$. We find that the ``dispersion'' of zeros is reversed compared to the original band dispersion $\epsilon_\bk$. In \figref{fig: wzero}, we also numerically confirmed that the ``bandwidth'' $w_\text{zero}$ of zeros is suppressed by the interaction $J$ as $w_\text{zero}\sim J^{-2}$ as $J\to\infty$.

Building upon the above observation of the poles and zeros of the Green's function, we put forth the following empirical formula:
\begin{equation}\label{eq: GSMG deep}
G_{\smg}(\omega,\bk) = \frac{\omega+\alpha \epsilon_{\bk}/J^2}{(\omega-\epsilon_\bk/2)^2-J^2},
\end{equation}
as an approximate description of our numerical result \eqnref{eq: GSMG numerics}. An important aspect of this formula is the positioning of the Green's function zeros precisely around the initial Fermi surface (where $\epsilon_\bk=0$) at $\omega=0$. This is indicated by the small arrows in \figref{fig: Gdeep}(c). 

Assuming $\Re G_\smg(0,\bk)=0$ as the definition of the zero Fermi surface in the SMG phase, it would encompass the same Fermi volume as the pole Fermi surface in the Fermi liquid phase. As both translation and charge conservation symmetries remain unbroken in the SMG phase, the Luttinger theorem mandates the preservation of the Fermi volume. Given that the SMG state is a fully gapped trivial insulator, there is no pole (no quasi-particle) at low energy, thus the Green's function can only rely on zeros to fulfill the Fermi volume required by the Luttinger theorem, which is explicitly demonstrated by \eqnref{eq: GSMG deep}.


\subsection{Weak Coupling Analysis}\label{sec: weak}

Nevertheless, SMG is not the sole mechanism for gapping out the Fermi surface. SSB might also open a full gap on the Fermi surface, which corresponds to the Higgs mechanism for fermion mass generation. Specifically, in the bilayer square lattice model \eqnref{eq: H}, due to the perfect nesting of the Fermi surface, the Fermi liquid exhibits strong instability toward SSB orders. Without loss of generality, we will focus on the inter-layer exciton condensation in the weak coupling limit. The corresponding order parameter $\phi_\text{EC}$ was introduced in \eqnref{eq: def phi}, which carries momentum $\vect{Q}=(\pi,\pi)$. The exciton condensation leads to an SSB insulating phase, as noted in the phase diagram \figref{fig: lattice}(c). However, there are significant differences between the SMG insulator and the SSB insulator, especially in terms of the structure of Green's function zeros.

\begin{figure}[htbp]
\begin{center}
\includegraphics[scale=0.65]{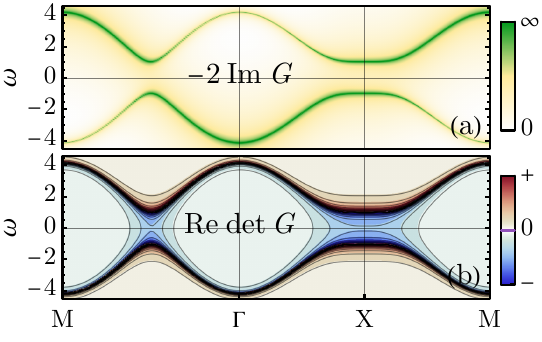}
\caption{Fermion Green's function  \eqnref{eq: GSSB} $G_\ssb$  in the SSB insulator phase, assuming a gap size of $|\Delta|=t$. (a) The imaginary part $-2\Im G(\omega+\ii 0_+,\bk)_{11}$ in the $\langle c_\bk^\dagger c_\bk \rangle$ channel, showing the pole (quasi-particle peak) along gapped bands. (b) The real part of the determinant $\Re\det G(\omega,\bk)$. No zero within the gap. In both plots, the frequency is shifted by a small imaginary part $\omega\to\omega+0.01\ii t$ for better visualization of spectral features.}
\label{fig: Gssb}
\end{center}
\end{figure}

In the SSB insulator phase, the Brillouin zone folds by the nesting vector $\vect{Q}=(\pi,\pi)$. The fermion Green's function can be written in the $(c_{\bk},c_{\bk+\vect{Q}})^\intercal$ basis (omitting layers and spins freedom) as
\eq{\label{eq: GSSB}
G_\ssb(\omega, \bk)=\frac{\omega \sigma^0+\epsilon_\bk \sigma^3+ \Re\Delta\sigma^1+\Im\Delta\sigma^2}{\omega^2-\epsilon_\bk^2-|\Delta|^2},}
where $\Delta= J \langle \phi_\text{EC}\rangle$ denotes the exciton gap induced by the exciton condensation $\langle \phi_\text{EC} \rangle\neq 0$. The properties of $G_\ssb$ are illustrated in \figref{fig: Gssb}. The spectral function in \figref{fig: Gssb}(a) depicts the quasi-particle peak along the band dispersion, reflecting a gapped (insulating) band structure. 

Since $G_\ssb$ is a matrix, its zero structure should be defined by its determinant being zero, i.e.,~$\det G_\ssb(\omega,\bk)=0$, which is the only way to define the zero structure in a basis independent manner. \figref{fig: Gssb}(b) indicates the determinant of $G_\ssb$ remains the same sign within the band gap induced by the exciton condensation. Since $G_\ssb$ does not preserve the translation symmetry (as $\Delta\to-\Delta$ is translation-odd), and $\Delta$ is non-zero, $\det G_\ssb$ does not have zeros crossing $\omega=0$ at the original Fermi surface. These two observations are linked: the absence of translation symmetry makes the Luttinger theorem ineffective, hence there is no expectation for the zero Fermi surface in the SSB insulator.

As the interaction $J$ intensifies, the SSB insulator ultimately transitions into the SMG insulator, as depicted in the phase diagram \figref{fig: lattice}(c). During this transition, the broken symmetry is restored, yet the fermion excitation gap remains intact, similar to the pseudo-gap phenomenon seen in correlated materials \cite{Lee2004Dcond-mat/0410445, Keimer2014H1409.4673}. In the context of modeling fermion spectral functions, the pseudo-gap phenomenon can be interpreted as a consequence of the phase (or orientation) fluctuations of fermion bilinear order parameters \cite{Franz1998Pcond-mat/9805401, Kwon1999Econd-mat/9809225, Kwon2001Ocond-mat/0006290, Franz2001Acond-mat/0012445, Curty2003Tcond-mat/0401124, Li2018P1805.05530, Li2018W1803.08226, Ye2019H1905.11412}. In this picture, the order parameter $\Delta=\Delta_0\e^{\ii\theta}$ maintains a finite amplitude $\Delta_0$ as we enter the SMG phase from the adjacent SSB phase, but its phase $\theta$ is disordered by long-wavelength random fluctuations. Consequently, on the large scale, $\Delta$ cannot condense to form long-range order; but on a smaller scale, $\Delta_0$ still provides a local excitation gap everywhere for fermions.

Based on this picture of the SMG state, the simplest treatment is to focus on the long wavelength fluctuation of $\Delta$ and estimate its self-energy correction for the fermion by
\eq{
\Sigma(\omega,\bk)=
\raisebox{-3pt}{\includegraphics[scale=0.65]{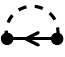}}=\E_\Delta\hat{\Delta}^\dagger G_0(\omega,\bk)\hat{\Delta}=\frac{\Delta_0^2}{\omega\sigma^0+\epsilon_\bk\sigma^3},}
where the vertex operator is $\hat{\Delta}:=\Re \Delta\sigma^1+\Im\Delta\sigma^2$ and the bare Green's function is $G_0(\omega,\bk)=(\omega\sigma^0-\epsilon_\bk\sigma^3)^{-1}$. Here we have assumed that the correlation length $\xi$ of the bosonic field $\Delta$ is long enough that its momentum is negligible for fermions. This assumption is valid near the transition to the SSB phase, as the correlation length diverges ($\xi\to\infty$) at the transition. 

Using this self-energy to correct the bare Green's function, we obtain
\eqs{G(\omega,\bk)&=(G_0(\omega,\bk)^{-1}-\Sigma(\omega,\bk))^{-1}\\
&=\frac{\omega\sigma^0+\epsilon_\bk\sigma^3}{\omega^2-\epsilon_\bk^2-\Delta_0^2}.}
Since the translation symmetry has been restored in the SMG phase, we can unfold the Green's function back to the original Brillouin zone (by taking the $G(\omega,\bk)_{11}$ component), which leads to a weak coupling description of the Green's function in the shallow SMG phase near the transition to the SSB phase
\eq{\label{eq: GSMG shallow}
G'_\smg(\omega,\bk)=\frac{\omega+\epsilon_\bk}{\omega^2-\epsilon_\bk^2-\Delta_0^2}.}
A more rigorous treatment of a similar problem can be found in \refcite{Wang2023P2212.05737}, which includes finite momentum fluctuations of $\Delta$. The major effect of these fluctuations is to introduce a spectral broadening for the fermion Green's function as if replacing $\omega\to\omega+\ii\delta$ in \eqnref{eq: GSMG shallow}. It was also found that the broadening $\delta\sim\xi^{-1}$  scales inversely with the correlation length $\xi$ of the order parameter, which justifies our simple treatment in the large-$\xi$ regime. Similar Green's functions as \eqnref{eq: GSMG shallow} was previously constructed to describe non-Fermi liquid  \cite{2003LuttingerSurface} statisfying the Luttinger theorem. However, its physical meaning is now clarified as Green's function in the SMG phase.

\begin{figure}[htbp]
\begin{center}
\includegraphics[scale=0.65]{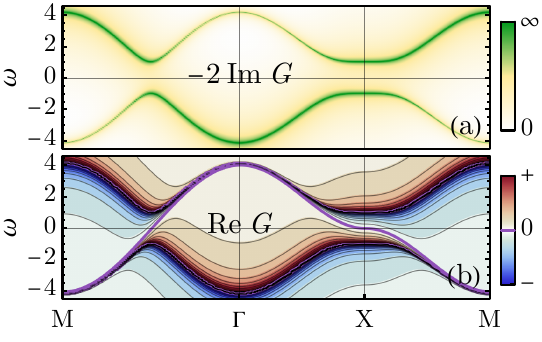}
\caption{Fermion Green's function \eqnref{eq: GSMG shallow} $G'_\smg$ in the SMG insulator phase near the phase transition to an adjacent SSB phase, assuming a local gap size of $\Delta_0=t$. (a) The imaginary part (spectral function) $-2\Im G(\omega+\ii 0_+,\bk)$ shows the pole (quasi-particle peak) along gapped bands. (b) The real part $\Re G(\omega,\bk)$ exhibits the zero (purple contour) crossing $\omega=0$ at the original Fermi surface. In both plots, the frequency is shifted by a small imaginary part $\omega\to\omega+0.01\ii t$ for better visualization of spectral features.}
\label{fig: Gshallow}
\end{center}
\end{figure}

The features of $G'_\smg$ in \eqnref{eq: GSMG shallow} are presented in \figref{fig: Gshallow}. When comparing \figref{fig: Gshallow}(a) and \figref{fig: Gssb}(a), we can observe that the pole structure of $G'_\smg$ is identical to that of $G_\ssb$ (in the diagonal component), both showcasing a gapped spectrum. However, they significantly differ in their zero structures, as seen by comparing \figref{fig: Gshallow}(b) and \figref{fig: Gssb}(b). Due to the restoration of symmetry, the low-energy zeros reemerge in the Green's function in the SMG phase. Additionally, its zero Fermi surface perfectly aligns with the original pole Fermi surface, fulfilling the Luttinger theorem's requirement for the Fermi volume.

Comparing the Green's function in the SMG phase derived from the strong coupling analysis \eqnref{eq: GSMG deep} and the weak coupling analysis \eqnref{eq: GSMG shallow} (see also \figref{fig: Gdeep} and \figref{fig: Gshallow}), we find that despite the apparent difference in high-energy spectral features, the zero Fermi surface defined by $G(0,\bk)=0$ remains a resilient low-energy feature. The persistent zero Fermi surface in the SMG phase is a consequence of the Luttinger theorem. 

Nonetheless, besides the low-energy zero structure, it is also intriguing to understand how the high-energy spectral feature deforms from the weak coupling case to the strong coupling case. However, this problem requires non-perturbative numerical simulations. Fortunately, the bilayer square lattice model \eqnref{eq: H} admits a sign-problem-free \cite{Troyer2005Ccond-mat/0408370} quantum Monte Carlo \cite{Fucito1981A, Scalapino1981M, Blankenbecler1981M, Hirsch1981E, Hirsch1985T} simulation. We will leave this interesting direction for future research.

\section{Probing Green's Function Zeros}\label{sec: probe}

While Green's function zeros are an important feature of the SMG insulator, they are not directly observable in experiments. Spectroscopy experiments, such as angle-resolved photoemission spectroscopy (ARPES), can directly probe the fermion's spectral function $A(\omega,\bk)=-2\Im G(\omega+\ii 0_+,\bk)$, which is the imaginary part of Green's function. By employing the Kramers-Kronig (KK) relation to recover the real part of Green's function from the spectral function, 
\eq{\Re G(\omega,\bk) = \frac{1}{2\pi}\scP\int d\omega' \frac{A(\omega',\bk)}{\omega'-\omega},}
we can indirectly study the zero structure of the Green's function.

\begin{figure}[htbp]
\begin{center}
\includegraphics[scale=0.65]{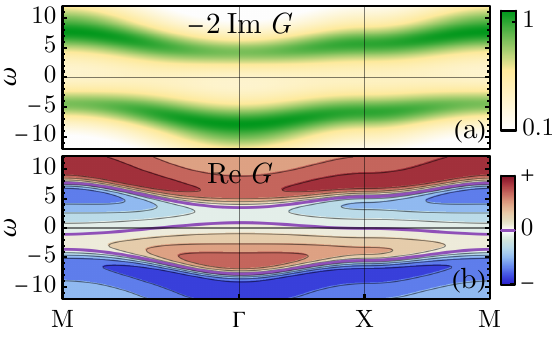}
\caption{(a) Broadened spectral function from the one in \figref{fig: Gdeep}. (b) Reconstructed Green's function real part by the KK relation, showing robust Green's function zeros (purple contour) crossing $\omega=0$.}
\label{fig: Gbroad}
\end{center}
\end{figure}

However, the spectral function might be broadened in experimental data due to noise or dissipation. We are interested in studying how sensitive the reconstructed Green's function zero is to these disturbances, in order to understand the stability of the method. Following \secref{sec: strong}, we start from the strong coupling limit and use the CPT approach to calculate Green's function. To account for the spectral broadening effect, we replace $\omega$ with $\omega +\ii \delta$, where $\delta$ is relatively large, say, about the order of the hopping $t$. Based on the broadened spectral function in \figref{fig: Gbroad}(a), we use the KK relation to reconstruct the real part, as shown in \figref{fig: Gbroad}(b). We find that the zero Fermi surface maintains the same shape, but the zero ``dispersion'' bandwidth gets larger.

The increase in bandwidth can be understood by taking the SMG Green's function $G_\text{SMG}(\omega,\bk)$ in \eqnref{eq: GSMG deep}, and solving for its zeros $\Re G(\omega+\ii\delta,\bk)=0$. To the leading order of $1/J$ and $\delta$, the solution is given by
\eq{\omega(\bk)=-\Big(1+\frac{\delta^2}{\alpha}\Big)\frac{\alpha\epsilon_\bk}{J^2}+\cdots,}
meaning that the bandwidth of Green's function zero dispersion will increase by $\delta^2/\alpha$, but the corresponding Luttinger surface remains unchanged. Therefore, the Green's function zero in the SMG phase is a robust feature that can be potentially identified from spectroscopy measurements, even in the presence of noises or dissipations.

\section{Summary and Discussions}\label{sec: summary}

In this paper, we investigated the Fermi surface SMG in a bilayer square lattice model. A crucial finding of this study lies in the robust Green's function zero in the SMG phase. Traditionally, a Fermi liquid state is characterized by poles in the Green's function along the Fermi surface. However, as the fermion system is driven into the SMG state by interaction effects, these poles are replaced by zeros. This is a robust phenomenon underlined by the constraints of the Luttinger theorem.

Our exploration is not limited to theoretical assertions. We also offer a tangible demonstration of this occurrence in the bilayer square lattice model. By applying both strong and weak coupling analyses, we provide a comprehensive portrayal of the fermion Green's function across different interaction regimes. We highlight that the emergence of the zero Fermi surface is not an ephemeral or fine-tuned phenomenon, but rather a robust and enduring feature of the SMG phase. We show that even when the system is subjected to spectral broadening, the zero Fermi surface persists, retaining the Fermi volume.

The results of this study confirm the robustness of the zero Fermi surface and underscore the possibility of observing it in experimental setups, such as through ARPES. Despite not being directly observable, the zero structure of the Green's function could be inferred indirectly via the KK relation.

The bilayer square lattice model may be relevant to the nickelate superconductor recently discovered in pressurized La$_3$Ni$_2$O$_7$ \cite{Sun2023S, Hou2023E2307.09865}, which is a layered two-dimensional material where each layer consists of nickel atoms arranged in a bilayer square lattice. The Fermi surface is dominated by $d_{z^2}$ and $d_{x^2-y^2}$ electrons of Ni. The $d_{z^2}$ electron has a relatively small intra-layer hopping $t$ due to the rather localized $d_{z^2}$ orbital wave function in the $xy$-plane but enjoys a large interlayer antiferromagnetic Heisenberg interaction $J$ due to the super-exchange mechanism mediated by the apical oxygen. This likely puts the $d_{z^2}$ electrons in an SMG insulator phase in the bilayer square lattice model and opens up opportunities to investigate the proposed Green's function zeros in real materials. The potential implication of SMG physics on the nickelate high-$T_c$ superconductor still requires further theoretical research in the future.


\begin{acknowledgments}
We acknowledge the helpful discussions with Liujun Zou, Zi-Xiang Li, Fan Yang, Yang Qi, Subir Sachdev, and Ya-Hui Zhang. All authors are supported by the NSF Grant DMR-2238360.

\end{acknowledgments}

\bibliography{ref}

\begin{thebibliography}{136}%
\makeatletter
\providecommand \@ifxundefined [1]{%
 \@ifx{#1\undefined}
}%
\providecommand \@ifnum [1]{%
 \ifnum #1\expandafter \@firstoftwo
 \else \expandafter \@secondoftwo
 \fi
}%
\providecommand \@ifx [1]{%
 \ifx #1\expandafter \@firstoftwo
 \else \expandafter \@secondoftwo
 \fi
}%
\providecommand \natexlab [1]{#1}%
\providecommand \enquote  [1]{``#1''}%
\providecommand \bibnamefont  [1]{#1}%
\providecommand \bibfnamefont [1]{#1}%
\providecommand \citenamefont [1]{#1}%
\providecommand \href@noop [0]{\@secondoftwo}%
\providecommand \href [0]{\begingroup \@sanitize@url \@href}%
\providecommand \@href[1]{\@@startlink{#1}\@@href}%
\providecommand \@@href[1]{\endgroup#1\@@endlink}%
\providecommand \@sanitize@url [0]{\catcode `\\12\catcode `\$12\catcode
  `\&12\catcode `\#12\catcode `\^12\catcode `\_12\catcode `\%12\relax}%
\providecommand \@@startlink[1]{}%
\providecommand \@@endlink[0]{}%
\providecommand \url  [0]{\begingroup\@sanitize@url \@url }%
\providecommand \@url [1]{\endgroup\@href {#1}{\urlprefix }}%
\providecommand \urlprefix  [0]{URL }%
\providecommand \Eprint [0]{\href }%
\providecommand \doibase [0]{https://doi.org/}%
\providecommand \selectlanguage [0]{\@gobble}%
\providecommand \bibinfo  [0]{\@secondoftwo}%
\providecommand \bibfield  [0]{\@secondoftwo}%
\providecommand \translation [1]{[#1]}%
\providecommand \BibitemOpen [0]{}%
\providecommand \bibitemStop [0]{}%
\providecommand \bibitemNoStop [0]{.\EOS\space}%
\providecommand \EOS [0]{\spacefactor3000\relax}%
\providecommand \BibitemShut  [1]{\csname bibitem#1\endcsname}%
\let\auto@bib@innerbib\@empty
\bibitem [{\citenamefont {{Fidkowski}}\ and\ \citenamefont
  {{Kitaev}}(2010)}]{Fid2010ISPT}%
  \BibitemOpen
  \bibfield  {author} {\bibinfo {author} {\bibfnamefont {L.}~\bibnamefont
  {{Fidkowski}}}\ and\ \bibinfo {author} {\bibfnamefont {A.}~\bibnamefont
  {{Kitaev}}},\ }\bibfield  {title} {\bibinfo {title} {{Effects of interactions
  on the topological classification of free fermion systems}},\ }\href
  {https://doi.org/10.1103/PhysRevB.81.134509} {\bibfield  {journal} {\bibinfo
  {journal} {\prb}\ }\textbf {\bibinfo {volume} {81}},\ \bibinfo {eid} {134509}
  (\bibinfo {year} {2010})},\ \Eprint {https://arxiv.org/abs/0904.2197}
  {arXiv:0904.2197 [cond-mat.str-el]} \BibitemShut {NoStop}%
\bibitem [{\citenamefont {{Fidkowski}}\ and\ \citenamefont
  {{Kitaev}}(2011)}]{Fid2011ISPT}%
  \BibitemOpen
  \bibfield  {author} {\bibinfo {author} {\bibfnamefont {L.}~\bibnamefont
  {{Fidkowski}}}\ and\ \bibinfo {author} {\bibfnamefont {A.}~\bibnamefont
  {{Kitaev}}},\ }\bibfield  {title} {\bibinfo {title} {{Topological phases of
  fermions in one dimension}},\ }\href
  {https://doi.org/10.1103/PhysRevB.83.075103} {\bibfield  {journal} {\bibinfo
  {journal} {\prb}\ }\textbf {\bibinfo {volume} {83}},\ \bibinfo {eid} {075103}
  (\bibinfo {year} {2011})},\ \Eprint {https://arxiv.org/abs/1008.4138}
  {arXiv:1008.4138 [cond-mat.str-el]} \BibitemShut {NoStop}%
\bibitem [{\citenamefont {{Wang}}\ and\ \citenamefont
  {{Wen}}(2023)}]{Wang1307.7480}%
  \BibitemOpen
  \bibfield  {author} {\bibinfo {author} {\bibfnamefont {J.}~\bibnamefont
  {{Wang}}}\ and\ \bibinfo {author} {\bibfnamefont {X.-G.}\ \bibnamefont
  {{Wen}}},\ }\bibfield  {title} {\bibinfo {title} {{Non-Perturbative
  Regularization of 1+1D Anomaly-Free Chiral Fermions and Bosons: On the
  equivalence of anomaly matching conditions and boundary gapping rules}},\
  }\href {https://doi.org/10.1103/PhysRevB.107.014311} {\bibfield  {journal}
  {\bibinfo  {journal} {Phys. Rev. B}\ }\textbf {\bibinfo {volume} {107}},\
  \bibinfo {eid} {014311} (\bibinfo {year} {2023})},\ \Eprint
  {https://arxiv.org/abs/1307.7480} {arXiv:1307.7480 [hep-lat]} \BibitemShut
  {NoStop}%
\bibitem [{\citenamefont {{Slagle}}\ \emph {et~al.}(2015)\citenamefont
  {{Slagle}}, \citenamefont {{You}},\ and\ \citenamefont
  {{Xu}}}]{Slagle1409.7401}%
  \BibitemOpen
  \bibfield  {author} {\bibinfo {author} {\bibfnamefont {K.}~\bibnamefont
  {{Slagle}}}, \bibinfo {author} {\bibfnamefont {Y.-Z.}\ \bibnamefont
  {{You}}},\ and\ \bibinfo {author} {\bibfnamefont {C.}~\bibnamefont {{Xu}}},\
  }\bibfield  {title} {\bibinfo {title} {{Exotic quantum phase transitions of
  strongly interacting topological insulators}},\ }\href
  {https://doi.org/10.1103/PhysRevB.91.115121} {\bibfield  {journal} {\bibinfo
  {journal} {\prb}\ }\textbf {\bibinfo {volume} {91}},\ \bibinfo {eid} {115121}
  (\bibinfo {year} {2015})},\ \Eprint {https://arxiv.org/abs/1409.7401}
  {arXiv:1409.7401 [cond-mat.str-el]} \BibitemShut {NoStop}%
\bibitem [{\citenamefont {{Ayyar}}\ and\ \citenamefont
  {{Chandrasekharan}}(2015)}]{Ayyar1410.6474}%
  \BibitemOpen
  \bibfield  {author} {\bibinfo {author} {\bibfnamefont {V.}~\bibnamefont
  {{Ayyar}}}\ and\ \bibinfo {author} {\bibfnamefont {S.}~\bibnamefont
  {{Chandrasekharan}}},\ }\bibfield  {title} {\bibinfo {title} {{Massive
  fermions without fermion bilinear condensates}},\ }\href
  {https://doi.org/10.1103/PhysRevD.91.065035} {\bibfield  {journal} {\bibinfo
  {journal} {\prd}\ }\textbf {\bibinfo {volume} {91}},\ \bibinfo {eid} {065035}
  (\bibinfo {year} {2015})},\ \Eprint {https://arxiv.org/abs/1410.6474}
  {arXiv:1410.6474 [hep-lat]} \BibitemShut {NoStop}%
\bibitem [{\citenamefont {{Catterall}}(2016)}]{Catterall1510.04153}%
  \BibitemOpen
  \bibfield  {author} {\bibinfo {author} {\bibfnamefont {S.}~\bibnamefont
  {{Catterall}}},\ }\bibfield  {title} {\bibinfo {title} {{Fermion mass without
  symmetry breaking}},\ }\href {https://doi.org/10.1007/JHEP01(2016)121}
  {\bibfield  {journal} {\bibinfo  {journal} {Journal of High Energy Physics}\
  }\textbf {\bibinfo {volume} {1}},\ \bibinfo {eid} {121} (\bibinfo {year}
  {2016})},\ \Eprint {https://arxiv.org/abs/1510.04153} {arXiv:1510.04153
  [hep-lat]} \BibitemShut {NoStop}%
\bibitem [{\citenamefont {{Tong}}(2022)}]{Tong2022SMG}%
  \BibitemOpen
  \bibfield  {author} {\bibinfo {author} {\bibfnamefont {D.}~\bibnamefont
  {{Tong}}},\ }\bibfield  {title} {\bibinfo {title} {{Comments on symmetric
  mass generation in 2d and 4d}},\ }\href
  {https://doi.org/10.1007/JHEP07(2022)001} {\bibfield  {journal} {\bibinfo
  {journal} {Journal of High Energy Physics}\ }\textbf {\bibinfo {volume}
  {2022}},\ \bibinfo {eid} {1} (\bibinfo {year} {2022})},\ \Eprint
  {https://arxiv.org/abs/2104.03997} {arXiv:2104.03997 [hep-th]} \BibitemShut
  {NoStop}%
\bibitem [{\citenamefont {{Wang}}\ and\ \citenamefont
  {{You}}(2022)}]{YZY2022SMGreview}%
  \BibitemOpen
  \bibfield  {author} {\bibinfo {author} {\bibfnamefont {J.}~\bibnamefont
  {{Wang}}}\ and\ \bibinfo {author} {\bibfnamefont {Y.-Z.}\ \bibnamefont
  {{You}}},\ }\bibfield  {title} {\bibinfo {title} {{Symmetric Mass
  Generation}},\ }\href {https://doi.org/10.3390/sym14071475} {\bibfield
  {journal} {\bibinfo  {journal} {Symmetry}\ }\textbf {\bibinfo {volume}
  {14}},\ \bibinfo {pages} {1475} (\bibinfo {year} {2022})},\ \Eprint
  {https://arxiv.org/abs/2204.14271} {arXiv:2204.14271 [cond-mat.str-el]}
  \BibitemShut {NoStop}%
\bibitem [{\citenamefont {{Nambu}}(1960)}]{Nambu1960higgs}%
  \BibitemOpen
  \bibfield  {author} {\bibinfo {author} {\bibfnamefont {Y.}~\bibnamefont
  {{Nambu}}},\ }\bibfield  {title} {\bibinfo {title} {{Quasi-Particles and
  Gauge Invariance in the Theory of Superconductivity}},\ }\href
  {https://doi.org/10.1103/PhysRev.117.648} {\bibfield  {journal} {\bibinfo
  {journal} {Physical Review}\ }\textbf {\bibinfo {volume} {117}},\ \bibinfo
  {pages} {648} (\bibinfo {year} {1960})}\BibitemShut {NoStop}%
\bibitem [{\citenamefont {{Nambu}}\ and\ \citenamefont
  {{Jona-Lasinio}}(1961)}]{Nambu1961higgs}%
  \BibitemOpen
  \bibfield  {author} {\bibinfo {author} {\bibfnamefont {Y.}~\bibnamefont
  {{Nambu}}}\ and\ \bibinfo {author} {\bibfnamefont {G.}~\bibnamefont
  {{Jona-Lasinio}}},\ }\bibfield  {title} {\bibinfo {title} {{Dynamical Model
  of Elementary Particles Based on an Analogy with Superconductivity. I}},\
  }\href {https://doi.org/10.1103/PhysRev.122.345} {\bibfield  {journal}
  {\bibinfo  {journal} {Physical Review}\ }\textbf {\bibinfo {volume} {122}},\
  \bibinfo {pages} {345} (\bibinfo {year} {1961})}\BibitemShut {NoStop}%
\bibitem [{\citenamefont {{Goldstone}}\ \emph {et~al.}(1962)\citenamefont
  {{Goldstone}}, \citenamefont {{Salam}},\ and\ \citenamefont
  {{Weinberg}}}]{Goldstone1962higgs}%
  \BibitemOpen
  \bibfield  {author} {\bibinfo {author} {\bibfnamefont {J.}~\bibnamefont
  {{Goldstone}}}, \bibinfo {author} {\bibfnamefont {A.}~\bibnamefont
  {{Salam}}},\ and\ \bibinfo {author} {\bibfnamefont {S.}~\bibnamefont
  {{Weinberg}}},\ }\bibfield  {title} {\bibinfo {title} {{Broken Symmetries}},\
  }\href {https://doi.org/10.1103/PhysRev.127.965} {\bibfield  {journal}
  {\bibinfo  {journal} {Physical Review}\ }\textbf {\bibinfo {volume} {127}},\
  \bibinfo {pages} {965} (\bibinfo {year} {1962})}\BibitemShut {NoStop}%
\bibitem [{\citenamefont {{Anderson}}(1963)}]{Anderson1963higgs}%
  \BibitemOpen
  \bibfield  {author} {\bibinfo {author} {\bibfnamefont {P.~W.}\ \bibnamefont
  {{Anderson}}},\ }\bibfield  {title} {\bibinfo {title} {{Plasmons, Gauge
  Invariance, and Mass}},\ }\href {https://doi.org/10.1103/PhysRev.130.439}
  {\bibfield  {journal} {\bibinfo  {journal} {Physical Review}\ }\textbf
  {\bibinfo {volume} {130}},\ \bibinfo {pages} {439} (\bibinfo {year}
  {1963})}\BibitemShut {NoStop}%
\bibitem [{\citenamefont {{Englert}}\ and\ \citenamefont
  {{Brout}}(1964)}]{Englert1964higgs}%
  \BibitemOpen
  \bibfield  {author} {\bibinfo {author} {\bibfnamefont {F.}~\bibnamefont
  {{Englert}}}\ and\ \bibinfo {author} {\bibfnamefont {R.}~\bibnamefont
  {{Brout}}},\ }\bibfield  {title} {\bibinfo {title} {{Broken Symmetry and the
  Mass of Gauge Vector Mesons}},\ }\href
  {https://doi.org/10.1103/PhysRevLett.13.321} {\bibfield  {journal} {\bibinfo
  {journal} {\prl}\ }\textbf {\bibinfo {volume} {13}},\ \bibinfo {pages} {321}
  (\bibinfo {year} {1964})}\BibitemShut {NoStop}%
\bibitem [{\citenamefont {{Higgs}}(1964)}]{Higgs1964higgs}%
  \BibitemOpen
  \bibfield  {author} {\bibinfo {author} {\bibfnamefont {P.~W.}\ \bibnamefont
  {{Higgs}}},\ }\bibfield  {title} {\bibinfo {title} {{Broken Symmetries and
  the Masses of Gauge Bosons}},\ }\href
  {https://doi.org/10.1103/PhysRevLett.13.508} {\bibfield  {journal} {\bibinfo
  {journal} {\prl}\ }\textbf {\bibinfo {volume} {13}},\ \bibinfo {pages} {508}
  (\bibinfo {year} {1964})}\BibitemShut {NoStop}%
\bibitem [{\citenamefont {Nielsen}\ and\ \citenamefont
  {Ninomiya}(1981)}]{Nielsen1981b}%
  \BibitemOpen
  \bibfield  {author} {\bibinfo {author} {\bibfnamefont {H.~B.}\ \bibnamefont
  {Nielsen}}\ and\ \bibinfo {author} {\bibfnamefont {M.}~\bibnamefont
  {Ninomiya}},\ }\bibfield  {title} {\bibinfo {title} {Absence of neutrinos on
  a lattice: (ii). intuitive topological proof},\ }\href
  {https://doi.org/https://doi.org/10.1016/0550-3213(81)90524-1} {\bibfield
  {journal} {\bibinfo  {journal} {Nuclear Physics B}\ }\textbf {\bibinfo
  {volume} {193}},\ \bibinfo {pages} {173} (\bibinfo {year}
  {1981})}\BibitemShut {NoStop}%
\bibitem [{\citenamefont {{Nielsen}}\ and\ \citenamefont
  {{Ninomiya}}(1981{\natexlab{a}})}]{Nielsen1981a}%
  \BibitemOpen
  \bibfield  {author} {\bibinfo {author} {\bibfnamefont {H.~B.}\ \bibnamefont
  {{Nielsen}}}\ and\ \bibinfo {author} {\bibfnamefont {M.}~\bibnamefont
  {{Ninomiya}}},\ }\bibfield  {title} {\bibinfo {title} {{Absence of neutrinos
  on a lattice (I). Proof by homotopy theory}},\ }\href
  {https://doi.org/10.1016/0550-3213(81)90361-8} {\bibfield  {journal}
  {\bibinfo  {journal} {Nuclear Physics B}\ }\textbf {\bibinfo {volume}
  {185}},\ \bibinfo {pages} {20} (\bibinfo {year}
  {1981}{\natexlab{a}})}\BibitemShut {NoStop}%
\bibitem [{\citenamefont {{Nielsen}}\ and\ \citenamefont
  {{Ninomiya}}(1981{\natexlab{b}})}]{Nielsen1981NoGo}%
  \BibitemOpen
  \bibfield  {author} {\bibinfo {author} {\bibfnamefont {H.~B.}\ \bibnamefont
  {{Nielsen}}}\ and\ \bibinfo {author} {\bibfnamefont {M.}~\bibnamefont
  {{Ninomiya}}},\ }\bibfield  {title} {\bibinfo {title} {{A no-go theorem for
  regularizing chiral fermions}},\ }\href
  {https://doi.org/10.1016/0370-2693(81)91026-1} {\bibfield  {journal}
  {\bibinfo  {journal} {Physics Letters B}\ }\textbf {\bibinfo {volume}
  {105}},\ \bibinfo {pages} {219} (\bibinfo {year}
  {1981}{\natexlab{b}})}\BibitemShut {NoStop}%
\bibitem [{\citenamefont {Swift}(1984)}]{Swift1984}%
  \BibitemOpen
  \bibfield  {author} {\bibinfo {author} {\bibfnamefont {P.}~\bibnamefont
  {Swift}},\ }\bibfield  {title} {\bibinfo {title} {The electroweak theory on
  the lattice},\ }\href
  {https://doi.org/https://doi.org/10.1016/0370-2693(84)90350-2} {\bibfield
  {journal} {\bibinfo  {journal} {Physics Letters B}\ }\textbf {\bibinfo
  {volume} {145}},\ \bibinfo {pages} {256} (\bibinfo {year}
  {1984})}\BibitemShut {NoStop}%
\bibitem [{\citenamefont {Eichten}\ and\ \citenamefont
  {Preskill}(1986)}]{Eichten1986Chiral}%
  \BibitemOpen
  \bibfield  {author} {\bibinfo {author} {\bibfnamefont {E.}~\bibnamefont
  {Eichten}}\ and\ \bibinfo {author} {\bibfnamefont {J.}~\bibnamefont
  {Preskill}},\ }\bibfield  {title} {\bibinfo {title} {{Chiral Gauge Theories
  on the Lattice}},\ }\href {https://doi.org/10.1016/0550-3213(86)90207-5}
  {\bibfield  {journal} {\bibinfo  {journal} {Nucl. Phys. B}\ }\textbf
  {\bibinfo {volume} {268}},\ \bibinfo {pages} {179} (\bibinfo {year}
  {1986})}\BibitemShut {NoStop}%
\bibitem [{\citenamefont {Smit}(1986)}]{Smit1986}%
  \BibitemOpen
  \bibfield  {author} {\bibinfo {author} {\bibfnamefont {J.}~\bibnamefont
  {Smit}},\ }\bibfield  {title} {\bibinfo {title} {{Fermions on a Lattice}},\
  }\href@noop {} {\bibfield  {journal} {\bibinfo  {journal} {Acta Phys. Polon.
  B}\ }\textbf {\bibinfo {volume} {17}},\ \bibinfo {pages} {531} (\bibinfo
  {year} {1986})}\BibitemShut {NoStop}%
\bibitem [{\citenamefont {Kaplan}(1992)}]{KaplanRLB1992}%
  \BibitemOpen
  \bibfield  {author} {\bibinfo {author} {\bibfnamefont {D.~B.}\ \bibnamefont
  {Kaplan}},\ }\bibfield  {title} {\bibinfo {title} {A method for simulating
  chiral fermions on the lattice},\ }\href
  {https://doi.org/https://doi.org/10.1016/0370-2693(92)91112-M} {\bibfield
  {journal} {\bibinfo  {journal} {Physics Letters B}\ }\textbf {\bibinfo
  {volume} {288}},\ \bibinfo {pages} {342} (\bibinfo {year}
  {1992})}\BibitemShut {NoStop}%
\bibitem [{\citenamefont {{Wen}}(2013)}]{Wen1305.1045}%
  \BibitemOpen
  \bibfield  {author} {\bibinfo {author} {\bibfnamefont {X.-G.}\ \bibnamefont
  {{Wen}}},\ }\bibfield  {title} {\bibinfo {title} {{A Lattice Non-Perturbative
  Definition of an SO(10) Chiral Gauge Theory and Its Induced Standard
  Model}},\ }\href {https://doi.org/10.1088/0256-307X/30/11/111101} {\bibfield
  {journal} {\bibinfo  {journal} {Chinese Physics Letters}\ }\textbf {\bibinfo
  {volume} {30}},\ \bibinfo {eid} {111101} (\bibinfo {year} {2013})},\ \Eprint
  {https://arxiv.org/abs/1305.1045} {arXiv:1305.1045 [hep-lat]} \BibitemShut
  {NoStop}%
\bibitem [{\citenamefont {{You}}\ \emph
  {et~al.}(2014{\natexlab{a}})\citenamefont {{You}}, \citenamefont {{BenTov}},\
  and\ \citenamefont {{Xu}}}]{You1402.4151}%
  \BibitemOpen
  \bibfield  {author} {\bibinfo {author} {\bibfnamefont {Y.-Z.}\ \bibnamefont
  {{You}}}, \bibinfo {author} {\bibfnamefont {Y.}~\bibnamefont {{BenTov}}},\
  and\ \bibinfo {author} {\bibfnamefont {C.}~\bibnamefont {{Xu}}},\ }\bibfield
  {title} {\bibinfo {title} {{Interacting Topological Superconductors and
  possible Origin of $16n$ Chiral Fermions in the Standard Model}},\
  }\href@noop {} {\bibfield  {journal} {\bibinfo  {journal} {ArXiv e-prints}\ }
  (\bibinfo {year} {2014}{\natexlab{a}})},\ \Eprint
  {https://arxiv.org/abs/1402.4151} {arXiv:1402.4151 [cond-mat.str-el]}
  \BibitemShut {NoStop}%
\bibitem [{\citenamefont {{You}}\ and\ \citenamefont
  {{Xu}}(2015)}]{You1412.4784}%
  \BibitemOpen
  \bibfield  {author} {\bibinfo {author} {\bibfnamefont {Y.-Z.}\ \bibnamefont
  {{You}}}\ and\ \bibinfo {author} {\bibfnamefont {C.}~\bibnamefont {{Xu}}},\
  }\bibfield  {title} {\bibinfo {title} {{Interacting topological insulator and
  emergent grand unified theory}},\ }\href
  {https://doi.org/10.1103/PhysRevB.91.125147} {\bibfield  {journal} {\bibinfo
  {journal} {\prb}\ }\textbf {\bibinfo {volume} {91}},\ \bibinfo {eid} {125147}
  (\bibinfo {year} {2015})},\ \Eprint {https://arxiv.org/abs/1412.4784}
  {arXiv:1412.4784 [cond-mat.str-el]} \BibitemShut {NoStop}%
\bibitem [{\citenamefont {{BenTov}}(2015)}]{BenTov1412.0154}%
  \BibitemOpen
  \bibfield  {author} {\bibinfo {author} {\bibfnamefont {Y.}~\bibnamefont
  {{BenTov}}},\ }\bibfield  {title} {\bibinfo {title} {{Fermion masses without
  symmetry breaking in two spacetime dimensions}},\ }\href
  {https://doi.org/10.1007/JHEP07(2015)034} {\bibfield  {journal} {\bibinfo
  {journal} {Journal of High Energy Physics}\ }\textbf {\bibinfo {volume}
  {7}},\ \bibinfo {eid} {34} (\bibinfo {year} {2015})},\ \Eprint
  {https://arxiv.org/abs/1412.0154} {arXiv:1412.0154 [cond-mat.str-el]}
  \BibitemShut {NoStop}%
\bibitem [{\citenamefont {{Ayyar}}\ and\ \citenamefont
  {{Chandrasekharan}}(2016{\natexlab{a}})}]{Ayyar1511.09071}%
  \BibitemOpen
  \bibfield  {author} {\bibinfo {author} {\bibfnamefont {V.}~\bibnamefont
  {{Ayyar}}}\ and\ \bibinfo {author} {\bibfnamefont {S.}~\bibnamefont
  {{Chandrasekharan}}},\ }\bibfield  {title} {\bibinfo {title} {{Origin of
  fermion masses without spontaneous symmetry breaking}},\ }\href
  {https://doi.org/10.1103/PhysRevD.93.081701} {\bibfield  {journal} {\bibinfo
  {journal} {\prd}\ }\textbf {\bibinfo {volume} {93}},\ \bibinfo {eid} {081701}
  (\bibinfo {year} {2016}{\natexlab{a}})},\ \Eprint
  {https://arxiv.org/abs/1511.09071} {arXiv:1511.09071 [hep-lat]} \BibitemShut
  {NoStop}%
\bibitem [{\citenamefont {{Ayyar}}\ and\ \citenamefont
  {{Chandrasekharan}}(2016{\natexlab{b}})}]{Ayyar1606.06312}%
  \BibitemOpen
  \bibfield  {author} {\bibinfo {author} {\bibfnamefont {V.}~\bibnamefont
  {{Ayyar}}}\ and\ \bibinfo {author} {\bibfnamefont {S.}~\bibnamefont
  {{Chandrasekharan}}},\ }\bibfield  {title} {\bibinfo {title} {{Fermion masses
  through four-fermion condensates}},\ }\href
  {https://doi.org/10.1007/JHEP10(2016)058} {\bibfield  {journal} {\bibinfo
  {journal} {Journal of High Energy Physics}\ }\textbf {\bibinfo {volume}
  {10}},\ \bibinfo {eid} {58} (\bibinfo {year} {2016}{\natexlab{b}})},\ \Eprint
  {https://arxiv.org/abs/1606.06312} {arXiv:1606.06312 [hep-lat]} \BibitemShut
  {NoStop}%
\bibitem [{\citenamefont {{Ayyar}}(2016)}]{Ayyar1611.00280}%
  \BibitemOpen
  \bibfield  {author} {\bibinfo {author} {\bibfnamefont {V.}~\bibnamefont
  {{Ayyar}}},\ }\bibfield  {title} {\bibinfo {title} {{Search for a continuum
  limit of the PMS phase}},\ }\href@noop {} {\bibfield  {journal} {\bibinfo
  {journal} {ArXiv e-prints}\ } (\bibinfo {year} {2016})},\ \Eprint
  {https://arxiv.org/abs/1611.00280} {arXiv:1611.00280 [hep-lat]} \BibitemShut
  {NoStop}%
\bibitem [{\citenamefont {{DeMarco}}\ and\ \citenamefont
  {{Wen}}(2017)}]{DeMarco1706.04648}%
  \BibitemOpen
  \bibfield  {author} {\bibinfo {author} {\bibfnamefont {M.}~\bibnamefont
  {{DeMarco}}}\ and\ \bibinfo {author} {\bibfnamefont {X.-G.}\ \bibnamefont
  {{Wen}}},\ }\bibfield  {title} {\bibinfo {title} {{A Novel Non-Perturbative
  Lattice Regularization of an Anomaly-Free $1 + 1d$ Chiral $SU(2)$ Gauge
  Theory}},\ }\href@noop {} {\bibfield  {journal} {\bibinfo  {journal} {arXiv
  e-prints}\ ,\ \bibinfo {eid} {arXiv:1706.04648}} (\bibinfo {year} {2017})},\
  \Eprint {https://arxiv.org/abs/1706.04648} {arXiv:1706.04648 [hep-lat]}
  \BibitemShut {NoStop}%
\bibitem [{\citenamefont {{Ayyar}}\ and\ \citenamefont
  {{Chandrasekharan}}(2017)}]{Ayyar1709.06048}%
  \BibitemOpen
  \bibfield  {author} {\bibinfo {author} {\bibfnamefont {V.}~\bibnamefont
  {{Ayyar}}}\ and\ \bibinfo {author} {\bibfnamefont {S.}~\bibnamefont
  {{Chandrasekharan}}},\ }\bibfield  {title} {\bibinfo {title} {{Generating a
  nonperturbative mass gap using Feynman diagrams in an asymptotically free
  theory}},\ }\href {https://doi.org/10.1103/PhysRevD.96.114506} {\bibfield
  {journal} {\bibinfo  {journal} {\prd}\ }\textbf {\bibinfo {volume} {96}},\
  \bibinfo {eid} {114506} (\bibinfo {year} {2017})},\ \Eprint
  {https://arxiv.org/abs/1709.06048} {arXiv:1709.06048 [hep-lat]} \BibitemShut
  {NoStop}%
\bibitem [{\citenamefont {{Schaich}}\ and\ \citenamefont
  {{Catterall}}(2018)}]{Schaich1710.08137}%
  \BibitemOpen
  \bibfield  {author} {\bibinfo {author} {\bibfnamefont {D.}~\bibnamefont
  {{Schaich}}}\ and\ \bibinfo {author} {\bibfnamefont {S.}~\bibnamefont
  {{Catterall}}},\ }\bibfield  {title} {\bibinfo {title} {{Phases of a strongly
  coupled four-fermion theory}},\ }in\ \href
  {https://doi.org/10.1051/epjconf/201817503004} {\emph {\bibinfo {booktitle}
  {European Physical Journal Web of Conferences}}},\ \bibinfo {series}
  {European Physical Journal Web of Conferences}, Vol.\ \bibinfo {volume}
  {175}\ (\bibinfo {year} {2018})\ p.\ \bibinfo {pages} {03004},\ \Eprint
  {https://arxiv.org/abs/1710.08137} {arXiv:1710.08137 [hep-lat]} \BibitemShut
  {NoStop}%
\bibitem [{\citenamefont {{Butt}}\ and\ \citenamefont
  {{Catterall}}(2018)}]{Butt1811.01015}%
  \BibitemOpen
  \bibfield  {author} {\bibinfo {author} {\bibfnamefont {N.}~\bibnamefont
  {{Butt}}}\ and\ \bibinfo {author} {\bibfnamefont {S.}~\bibnamefont
  {{Catterall}}},\ }\bibfield  {title} {\bibinfo {title} {{Four fermion
  condensates in SU(2) Yang-Mills-Higgs theory on a lattice}},\ }in\ \href@noop
  {} {\emph {\bibinfo {booktitle} {The 36th Annual International Symposium on
  Lattice Field Theory. 22-28 July}}}\ (\bibinfo {year} {2018})\ p.\ \bibinfo
  {pages} {294},\ \Eprint {https://arxiv.org/abs/1811.01015} {arXiv:1811.01015
  [hep-lat]} \BibitemShut {NoStop}%
\bibitem [{\citenamefont {{Butt}}\ \emph {et~al.}(2018)\citenamefont {{Butt}},
  \citenamefont {{Catterall}},\ and\ \citenamefont
  {{Schaich}}}]{Butt1810.06117}%
  \BibitemOpen
  \bibfield  {author} {\bibinfo {author} {\bibfnamefont {N.}~\bibnamefont
  {{Butt}}}, \bibinfo {author} {\bibfnamefont {S.}~\bibnamefont
  {{Catterall}}},\ and\ \bibinfo {author} {\bibfnamefont {D.}~\bibnamefont
  {{Schaich}}},\ }\bibfield  {title} {\bibinfo {title} {{SO(4) invariant
  Higgs-Yukawa model with reduced staggered fermions}},\ }\href
  {https://doi.org/10.1103/PhysRevD.98.114514} {\bibfield  {journal} {\bibinfo
  {journal} {\prd}\ }\textbf {\bibinfo {volume} {98}},\ \bibinfo {eid} {114514}
  (\bibinfo {year} {2018})},\ \Eprint {https://arxiv.org/abs/1810.06117}
  {arXiv:1810.06117 [hep-lat]} \BibitemShut {NoStop}%
\bibitem [{\citenamefont
  {{Kikukawa}}(2019{\natexlab{a}})}]{Kikukawa1710.11618}%
  \BibitemOpen
  \bibfield  {author} {\bibinfo {author} {\bibfnamefont {Y.}~\bibnamefont
  {{Kikukawa}}},\ }\bibfield  {title} {\bibinfo {title} {{On the
  gauge-invariant path-integral measure for the overlap Weyl fermions in 16 of
  SO(10)}},\ }\href {https://doi.org/10.1093/ptep/ptz115} {\bibfield  {journal}
  {\bibinfo  {journal} {Progress of Theoretical and Experimental Physics}\
  }\textbf {\bibinfo {volume} {2019}},\ \bibinfo {eid} {113B03} (\bibinfo
  {year} {2019}{\natexlab{a}})}\BibitemShut {NoStop}%
\bibitem [{\citenamefont
  {{Kikukawa}}(2019{\natexlab{b}})}]{Kikukawa1710.11101}%
  \BibitemOpen
  \bibfield  {author} {\bibinfo {author} {\bibfnamefont {Y.}~\bibnamefont
  {{Kikukawa}}},\ }\bibfield  {title} {\bibinfo {title} {{Why is the mission
  impossible? Decoupling the mirror Ginsparg-Wilson fermions in the lattice
  models for two-dimensional Abelian chiral gauge theories}},\ }\href
  {https://doi.org/10.1093/ptep/ptz055} {\bibfield  {journal} {\bibinfo
  {journal} {Progress of Theoretical and Experimental Physics}\ }\textbf
  {\bibinfo {volume} {2019}},\ \bibinfo {eid} {073B02} (\bibinfo {year}
  {2019}{\natexlab{b}})},\ \Eprint {https://arxiv.org/abs/1710.11101}
  {arXiv:1710.11101 [hep-lat]} \BibitemShut {NoStop}%
\bibitem [{\citenamefont {{Wang}}\ and\ \citenamefont
  {{Wen}}(2019)}]{Wang1807.05998}%
  \BibitemOpen
  \bibfield  {author} {\bibinfo {author} {\bibfnamefont {J.}~\bibnamefont
  {{Wang}}}\ and\ \bibinfo {author} {\bibfnamefont {X.-G.}\ \bibnamefont
  {{Wen}}},\ }\bibfield  {title} {\bibinfo {title} {{Solution to the 1 +1
  dimensional gauged chiral Fermion problem}},\ }\href
  {https://doi.org/10.1103/PhysRevD.99.111501} {\bibfield  {journal} {\bibinfo
  {journal} {\prd}\ }\textbf {\bibinfo {volume} {99}},\ \bibinfo {eid} {111501}
  (\bibinfo {year} {2019})},\ \Eprint {https://arxiv.org/abs/1807.05998}
  {arXiv:1807.05998 [hep-lat]} \BibitemShut {NoStop}%
\bibitem [{\citenamefont {{Wang}}\ and\ \citenamefont
  {{Wen}}(2020)}]{Wang1809.11171}%
  \BibitemOpen
  \bibfield  {author} {\bibinfo {author} {\bibfnamefont {J.}~\bibnamefont
  {{Wang}}}\ and\ \bibinfo {author} {\bibfnamefont {X.-G.}\ \bibnamefont
  {{Wen}}},\ }\bibfield  {title} {\bibinfo {title} {{Nonperturbative definition
  of the standard models}},\ }\href
  {https://doi.org/10.1103/PhysRevResearch.2.023356} {\bibfield  {journal}
  {\bibinfo  {journal} {Physical Review Research}\ }\textbf {\bibinfo {volume}
  {2}},\ \bibinfo {eid} {023356} (\bibinfo {year} {2020})},\ \Eprint
  {https://arxiv.org/abs/1809.11171} {arXiv:1809.11171 [hep-th]} \BibitemShut
  {NoStop}%
\bibitem [{\citenamefont {{Catterall}}\ \emph {et~al.}(2020)\citenamefont
  {{Catterall}}, \citenamefont {{Butt}},\ and\ \citenamefont
  {{Schaich}}}]{Catterall2002.00034}%
  \BibitemOpen
  \bibfield  {author} {\bibinfo {author} {\bibfnamefont {S.}~\bibnamefont
  {{Catterall}}}, \bibinfo {author} {\bibfnamefont {N.}~\bibnamefont
  {{Butt}}},\ and\ \bibinfo {author} {\bibfnamefont {D.}~\bibnamefont
  {{Schaich}}},\ }\bibfield  {title} {\bibinfo {title} {{Exotic Phases of a
  Higgs-Yukawa Model with Reduced Staggered Fermions}},\ }\href@noop {}
  {\bibfield  {journal} {\bibinfo  {journal} {arXiv e-prints}\ ,\ \bibinfo
  {eid} {arXiv:2002.00034}} (\bibinfo {year} {2020})},\ \Eprint
  {https://arxiv.org/abs/2002.00034} {arXiv:2002.00034 [hep-lat]} \BibitemShut
  {NoStop}%
\bibitem [{\citenamefont {{Razamat}}\ and\ \citenamefont
  {{Tong}}(2021)}]{Razamat2009.05037}%
  \BibitemOpen
  \bibfield  {author} {\bibinfo {author} {\bibfnamefont {S.~S.}\ \bibnamefont
  {{Razamat}}}\ and\ \bibinfo {author} {\bibfnamefont {D.}~\bibnamefont
  {{Tong}}},\ }\bibfield  {title} {\bibinfo {title} {{Gapped Chiral
  Fermions}},\ }\href {https://doi.org/10.1103/PhysRevX.11.011063} {\bibfield
  {journal} {\bibinfo  {journal} {Physical Review X}\ }\textbf {\bibinfo
  {volume} {11}},\ \bibinfo {eid} {011063} (\bibinfo {year} {2021})},\ \Eprint
  {https://arxiv.org/abs/2009.05037} {arXiv:2009.05037 [hep-th]} \BibitemShut
  {NoStop}%
\bibitem [{\citenamefont {{Catterall}}(2021)}]{Catterall2021SMG}%
  \BibitemOpen
  \bibfield  {author} {\bibinfo {author} {\bibfnamefont {S.}~\bibnamefont
  {{Catterall}}},\ }\bibfield  {title} {\bibinfo {title} {{Chiral lattice
  fermions from staggered fields}},\ }\href
  {https://doi.org/10.1103/PhysRevD.104.014503} {\bibfield  {journal} {\bibinfo
   {journal} {\prd}\ }\textbf {\bibinfo {volume} {104}},\ \bibinfo {eid}
  {014503} (\bibinfo {year} {2021})},\ \Eprint
  {https://arxiv.org/abs/2010.02290} {arXiv:2010.02290 [hep-lat]} \BibitemShut
  {NoStop}%
\bibitem [{\citenamefont {{Butt}}\ \emph
  {et~al.}(2021{\natexlab{a}})\citenamefont {{Butt}}, \citenamefont
  {{Catterall}}, \citenamefont {{Pradhan}},\ and\ \citenamefont
  {{Toga}}}]{Butt2101.01026}%
  \BibitemOpen
  \bibfield  {author} {\bibinfo {author} {\bibfnamefont {N.}~\bibnamefont
  {{Butt}}}, \bibinfo {author} {\bibfnamefont {S.}~\bibnamefont {{Catterall}}},
  \bibinfo {author} {\bibfnamefont {A.}~\bibnamefont {{Pradhan}}},\ and\
  \bibinfo {author} {\bibfnamefont {G.~C.}\ \bibnamefont {{Toga}}},\ }\bibfield
   {title} {\bibinfo {title} {{Anomalies and symmetric mass generation for
  K{\"a}hler-Dirac fermions}},\ }\href
  {https://doi.org/10.1103/PhysRevD.104.094504} {\bibfield  {journal} {\bibinfo
   {journal} {\prd}\ }\textbf {\bibinfo {volume} {104}},\ \bibinfo {eid}
  {094504} (\bibinfo {year} {2021}{\natexlab{a}})},\ \Eprint
  {https://arxiv.org/abs/2101.01026} {arXiv:2101.01026 [hep-th]} \BibitemShut
  {NoStop}%
\bibitem [{\citenamefont {{Butt}}\ \emph
  {et~al.}(2021{\natexlab{b}})\citenamefont {{Butt}}, \citenamefont
  {{Catterall}},\ and\ \citenamefont {{Toga}}}]{Butt2111.01001}%
  \BibitemOpen
  \bibfield  {author} {\bibinfo {author} {\bibfnamefont {N.}~\bibnamefont
  {{Butt}}}, \bibinfo {author} {\bibfnamefont {S.}~\bibnamefont
  {{Catterall}}},\ and\ \bibinfo {author} {\bibfnamefont {G.~C.}\ \bibnamefont
  {{Toga}}},\ }\bibfield  {title} {\bibinfo {title} {{Symmetric Mass Generation
  in Lattice Gauge Theory}},\ }\href@noop {} {\bibfield  {journal} {\bibinfo
  {journal} {arXiv e-prints}\ ,\ \bibinfo {eid} {arXiv:2111.01001}} (\bibinfo
  {year} {2021}{\natexlab{b}})},\ \Eprint {https://arxiv.org/abs/2111.01001}
  {arXiv:2111.01001 [hep-lat]} \BibitemShut {NoStop}%
\bibitem [{\citenamefont {{Zeng}}\ \emph {et~al.}(2022)\citenamefont {{Zeng}},
  \citenamefont {{Zhu}}, \citenamefont {{Wang}},\ and\ \citenamefont
  {{You}}}]{ZM2022SMG}%
  \BibitemOpen
  \bibfield  {author} {\bibinfo {author} {\bibfnamefont {M.}~\bibnamefont
  {{Zeng}}}, \bibinfo {author} {\bibfnamefont {Z.}~\bibnamefont {{Zhu}}},
  \bibinfo {author} {\bibfnamefont {J.}~\bibnamefont {{Wang}}},\ and\ \bibinfo
  {author} {\bibfnamefont {Y.-Z.}\ \bibnamefont {{You}}},\ }\bibfield  {title}
  {\bibinfo {title} {{Symmetric Mass Generation in the 1 +1 Dimensional Chiral
  Fermion 3-4-5-0 Model}},\ }\href
  {https://doi.org/10.1103/PhysRevLett.128.185301} {\bibfield  {journal}
  {\bibinfo  {journal} {\prl}\ }\textbf {\bibinfo {volume} {128}},\ \bibinfo
  {eid} {185301} (\bibinfo {year} {2022})},\ \Eprint
  {https://arxiv.org/abs/2202.12355} {arXiv:2202.12355 [cond-mat.str-el]}
  \BibitemShut {NoStop}%
\bibitem [{\citenamefont {{Catterall}}\ and\ \citenamefont
  {{Pradhan}}(2022)}]{Catterall2201.00750}%
  \BibitemOpen
  \bibfield  {author} {\bibinfo {author} {\bibfnamefont {S.}~\bibnamefont
  {{Catterall}}}\ and\ \bibinfo {author} {\bibfnamefont {A.}~\bibnamefont
  {{Pradhan}}},\ }\bibfield  {title} {\bibinfo {title} {{Induced topological
  gravity and anomaly inflow from K{\"a}hler-Dirac fermions in odd
  dimensions}},\ }\href {https://doi.org/10.1103/PhysRevD.106.014509}
  {\bibfield  {journal} {\bibinfo  {journal} {\prd}\ }\textbf {\bibinfo
  {volume} {106}},\ \bibinfo {eid} {014509} (\bibinfo {year} {2022})},\ \Eprint
  {https://arxiv.org/abs/2201.00750} {arXiv:2201.00750 [hep-th]} \BibitemShut
  {NoStop}%
\bibitem [{\citenamefont {{Catterall}}(2023)}]{Catterall2023KDfermion}%
  \BibitemOpen
  \bibfield  {author} {\bibinfo {author} {\bibfnamefont {S.}~\bibnamefont
  {{Catterall}}},\ }\bibfield  {title} {\bibinfo {title} {{'t Hooft anomalies
  for staggered fermions}},\ }\href
  {https://doi.org/10.1103/PhysRevD.107.014501} {\bibfield  {journal} {\bibinfo
   {journal} {\prd}\ }\textbf {\bibinfo {volume} {107}},\ \bibinfo {eid}
  {014501} (\bibinfo {year} {2023})},\ \Eprint
  {https://arxiv.org/abs/2209.03828} {arXiv:2209.03828 [hep-lat]} \BibitemShut
  {NoStop}%
\bibitem [{\citenamefont {{Guo}}\ and\ \citenamefont
  {{You}}(2023)}]{Guo2023S2306.17420}%
  \BibitemOpen
  \bibfield  {author} {\bibinfo {author} {\bibfnamefont {Y.}~\bibnamefont
  {{Guo}}}\ and\ \bibinfo {author} {\bibfnamefont {Y.-Z.}\ \bibnamefont
  {{You}}},\ }\bibfield  {title} {\bibinfo {title} {{Symmetric Mass Generation
  of K{\"a}hler-Dirac Fermions from the Perspective of Symmetry-Protected
  Topological Phases}},\ }\href {https://doi.org/10.48550/arXiv.2306.17420}
  {\bibfield  {journal} {\bibinfo  {journal} {arXiv e-prints}\ ,\ \bibinfo
  {eid} {arXiv:2306.17420}} (\bibinfo {year} {2023})},\ \Eprint
  {https://arxiv.org/abs/2306.17420} {arXiv:2306.17420 [cond-mat.str-el]}
  \BibitemShut {NoStop}%
\bibitem [{\citenamefont {{Turner}}\ \emph {et~al.}(2011)\citenamefont
  {{Turner}}, \citenamefont {{Pollmann}},\ and\ \citenamefont
  {{Berg}}}]{Turner1008.4346}%
  \BibitemOpen
  \bibfield  {author} {\bibinfo {author} {\bibfnamefont {A.~M.}\ \bibnamefont
  {{Turner}}}, \bibinfo {author} {\bibfnamefont {F.}~\bibnamefont
  {{Pollmann}}},\ and\ \bibinfo {author} {\bibfnamefont {E.}~\bibnamefont
  {{Berg}}},\ }\bibfield  {title} {\bibinfo {title} {{Topological phases of
  one-dimensional fermions: An entanglement point of view}},\ }\href
  {https://doi.org/10.1103/PhysRevB.83.075102} {\bibfield  {journal} {\bibinfo
  {journal} {\prb}\ }\textbf {\bibinfo {volume} {83}},\ \bibinfo {eid} {075102}
  (\bibinfo {year} {2011})},\ \Eprint {https://arxiv.org/abs/1008.4346}
  {arXiv:1008.4346 [cond-mat.str-el]} \BibitemShut {NoStop}%
\bibitem [{\citenamefont {{Ryu}}\ and\ \citenamefont
  {{Zhang}}(2012)}]{Ryu1202.4484}%
  \BibitemOpen
  \bibfield  {author} {\bibinfo {author} {\bibfnamefont {S.}~\bibnamefont
  {{Ryu}}}\ and\ \bibinfo {author} {\bibfnamefont {S.-C.}\ \bibnamefont
  {{Zhang}}},\ }\bibfield  {title} {\bibinfo {title} {{Interacting topological
  phases and modular invariance}},\ }\href
  {https://doi.org/10.1103/PhysRevB.85.245132} {\bibfield  {journal} {\bibinfo
  {journal} {\prb}\ }\textbf {\bibinfo {volume} {85}},\ \bibinfo {eid} {245132}
  (\bibinfo {year} {2012})},\ \Eprint {https://arxiv.org/abs/1202.4484}
  {arXiv:1202.4484 [cond-mat.str-el]} \BibitemShut {NoStop}%
\bibitem [{\citenamefont {{Qi}}(2013)}]{Qi1202.3983}%
  \BibitemOpen
  \bibfield  {author} {\bibinfo {author} {\bibfnamefont {X.-L.}\ \bibnamefont
  {{Qi}}},\ }\bibfield  {title} {\bibinfo {title} {{A new class of (2 +
  1)-dimensional topological superconductors with $\mathbb{Z}_8$ topological
  classification}},\ }\href {https://doi.org/10.1088/1367-2630/15/6/065002}
  {\bibfield  {journal} {\bibinfo  {journal} {New Journal of Physics}\ }\textbf
  {\bibinfo {volume} {15}},\ \bibinfo {eid} {065002} (\bibinfo {year}
  {2013})},\ \Eprint {https://arxiv.org/abs/1202.3983} {arXiv:1202.3983
  [cond-mat.str-el]} \BibitemShut {NoStop}%
\bibitem [{\citenamefont {{Yao}}\ and\ \citenamefont
  {{Ryu}}(2013)}]{Yao1202.5805}%
  \BibitemOpen
  \bibfield  {author} {\bibinfo {author} {\bibfnamefont {H.}~\bibnamefont
  {{Yao}}}\ and\ \bibinfo {author} {\bibfnamefont {S.}~\bibnamefont {{Ryu}}},\
  }\bibfield  {title} {\bibinfo {title} {{Interaction effect on topological
  classification of superconductors in two dimensions}},\ }\href
  {https://doi.org/10.1103/PhysRevB.88.064507} {\bibfield  {journal} {\bibinfo
  {journal} {\prb}\ }\textbf {\bibinfo {volume} {88}},\ \bibinfo {eid} {064507}
  (\bibinfo {year} {2013})},\ \Eprint {https://arxiv.org/abs/1202.5805}
  {arXiv:1202.5805 [cond-mat.str-el]} \BibitemShut {NoStop}%
\bibitem [{\citenamefont {{Gu}}\ and\ \citenamefont
  {{Levin}}(2014)}]{Gu1304.4569}%
  \BibitemOpen
  \bibfield  {author} {\bibinfo {author} {\bibfnamefont {Z.-C.}\ \bibnamefont
  {{Gu}}}\ and\ \bibinfo {author} {\bibfnamefont {M.}~\bibnamefont {{Levin}}},\
  }\bibfield  {title} {\bibinfo {title} {{Effect of interactions on
  two-dimensional fermionic symmetry-protected topological phases with Z$_{2}$
  symmetry}},\ }\href {https://doi.org/10.1103/PhysRevB.89.201113} {\bibfield
  {journal} {\bibinfo  {journal} {\prb}\ }\textbf {\bibinfo {volume} {89}},\
  \bibinfo {eid} {201113} (\bibinfo {year} {2014})},\ \Eprint
  {https://arxiv.org/abs/1304.4569} {arXiv:1304.4569 [cond-mat.str-el]}
  \BibitemShut {NoStop}%
\bibitem [{\citenamefont {{Wang}}\ and\ \citenamefont
  {{Senthil}}(2014)}]{Wang1401.1142}%
  \BibitemOpen
  \bibfield  {author} {\bibinfo {author} {\bibfnamefont {C.}~\bibnamefont
  {{Wang}}}\ and\ \bibinfo {author} {\bibfnamefont {T.}~\bibnamefont
  {{Senthil}}},\ }\bibfield  {title} {\bibinfo {title} {{Interacting fermionic
  topological insulators/superconductors in three dimensions}},\ }\href
  {https://doi.org/10.1103/PhysRevB.89.195124} {\bibfield  {journal} {\bibinfo
  {journal} {\prb}\ }\textbf {\bibinfo {volume} {89}},\ \bibinfo {eid} {195124}
  (\bibinfo {year} {2014})},\ \Eprint {https://arxiv.org/abs/1401.1142}
  {arXiv:1401.1142 [cond-mat.str-el]} \BibitemShut {NoStop}%
\bibitem [{\citenamefont {{Metlitski}}\ \emph {et~al.}(2014)\citenamefont
  {{Metlitski}}, \citenamefont {{Fidkowski}}, \citenamefont {{Chen}},\ and\
  \citenamefont {{Vishwanath}}}]{Metlitski1406.3032}%
  \BibitemOpen
  \bibfield  {author} {\bibinfo {author} {\bibfnamefont {M.~A.}\ \bibnamefont
  {{Metlitski}}}, \bibinfo {author} {\bibfnamefont {L.}~\bibnamefont
  {{Fidkowski}}}, \bibinfo {author} {\bibfnamefont {X.}~\bibnamefont
  {{Chen}}},\ and\ \bibinfo {author} {\bibfnamefont {A.}~\bibnamefont
  {{Vishwanath}}},\ }\bibfield  {title} {\bibinfo {title} {{Interaction effects
  on 3D topological superconductors: surface topological order from vortex
  condensation, the 16 fold way and fermionic Kramers doublets}},\ }\href@noop
  {} {\bibfield  {journal} {\bibinfo  {journal} {arXiv e-prints}\ ,\ \bibinfo
  {eid} {arXiv:1406.3032}} (\bibinfo {year} {2014})},\ \Eprint
  {https://arxiv.org/abs/1406.3032} {arXiv:1406.3032 [cond-mat.str-el]}
  \BibitemShut {NoStop}%
\bibitem [{\citenamefont {{Kapustin}}\ \emph {et~al.}(2015)\citenamefont
  {{Kapustin}}, \citenamefont {{Thorngren}}, \citenamefont {{Turzillo}},\ and\
  \citenamefont {{Wang}}}]{Kapustin2015IFSPT}%
  \BibitemOpen
  \bibfield  {author} {\bibinfo {author} {\bibfnamefont {A.}~\bibnamefont
  {{Kapustin}}}, \bibinfo {author} {\bibfnamefont {R.}~\bibnamefont
  {{Thorngren}}}, \bibinfo {author} {\bibfnamefont {A.}~\bibnamefont
  {{Turzillo}}},\ and\ \bibinfo {author} {\bibfnamefont {Z.}~\bibnamefont
  {{Wang}}},\ }\bibfield  {title} {\bibinfo {title} {{Fermionic symmetry
  protected topological phases and cobordisms}},\ }\href
  {https://doi.org/10.1007/JHEP12(2015)052} {\bibfield  {journal} {\bibinfo
  {journal} {Journal of High Energy Physics}\ }\textbf {\bibinfo {volume}
  {2015}},\ \bibinfo {eid} {52} (\bibinfo {year} {2015})},\ \Eprint
  {https://arxiv.org/abs/1406.7329} {arXiv:1406.7329 [cond-mat.str-el]}
  \BibitemShut {NoStop}%
\bibitem [{\citenamefont {{You}}\ and\ \citenamefont {{Xu}}(2014)}]{YZY2014}%
  \BibitemOpen
  \bibfield  {author} {\bibinfo {author} {\bibfnamefont {Y.-Z.}\ \bibnamefont
  {{You}}}\ and\ \bibinfo {author} {\bibfnamefont {C.}~\bibnamefont {{Xu}}},\
  }\bibfield  {title} {\bibinfo {title} {{Symmetry-protected topological states
  of interacting fermions and bosons}},\ }\href
  {https://doi.org/10.1103/PhysRevB.90.245120} {\bibfield  {journal} {\bibinfo
  {journal} {\prb}\ }\textbf {\bibinfo {volume} {90}},\ \bibinfo {eid} {245120}
  (\bibinfo {year} {2014})},\ \Eprint {https://arxiv.org/abs/1409.0168}
  {arXiv:1409.0168 [cond-mat.str-el]} \BibitemShut {NoStop}%
\bibitem [{\citenamefont {{Cheng}}\ \emph {et~al.}(2015)\citenamefont
  {{Cheng}}, \citenamefont {{Bi}}, \citenamefont {{You}},\ and\ \citenamefont
  {{Gu}}}]{Cheng1501.01313}%
  \BibitemOpen
  \bibfield  {author} {\bibinfo {author} {\bibfnamefont {M.}~\bibnamefont
  {{Cheng}}}, \bibinfo {author} {\bibfnamefont {Z.}~\bibnamefont {{Bi}}},
  \bibinfo {author} {\bibfnamefont {Y.-Z.}\ \bibnamefont {{You}}},\ and\
  \bibinfo {author} {\bibfnamefont {Z.-C.}\ \bibnamefont {{Gu}}},\ }\bibfield
  {title} {\bibinfo {title} {{Classification of Symmetry-Protected Phases for
  Interacting Fermions in Two Dimensions}},\ }\href@noop {} {\bibfield
  {journal} {\bibinfo  {journal} {arXiv e-prints}\ ,\ \bibinfo {eid}
  {arXiv:1501.01313}} (\bibinfo {year} {2015})},\ \Eprint
  {https://arxiv.org/abs/1501.01313} {arXiv:1501.01313 [cond-mat.str-el]}
  \BibitemShut {NoStop}%
\bibitem [{\citenamefont {{Yoshida}}\ and\ \citenamefont
  {{Furusaki}}(2015)}]{Yoshida1505.06598}%
  \BibitemOpen
  \bibfield  {author} {\bibinfo {author} {\bibfnamefont {T.}~\bibnamefont
  {{Yoshida}}}\ and\ \bibinfo {author} {\bibfnamefont {A.}~\bibnamefont
  {{Furusaki}}},\ }\bibfield  {title} {\bibinfo {title} {{Correlation effects
  on topological crystalline insulators}},\ }\href
  {https://doi.org/10.1103/PhysRevB.92.085114} {\bibfield  {journal} {\bibinfo
  {journal} {\prb}\ }\textbf {\bibinfo {volume} {92}},\ \bibinfo {eid} {085114}
  (\bibinfo {year} {2015})},\ \Eprint {https://arxiv.org/abs/1505.06598}
  {arXiv:1505.06598 [cond-mat.str-el]} \BibitemShut {NoStop}%
\bibitem [{\citenamefont {{Gu}}\ and\ \citenamefont
  {{Qi}}(2015)}]{Gu1512.04919}%
  \BibitemOpen
  \bibfield  {author} {\bibinfo {author} {\bibfnamefont {Y.}~\bibnamefont
  {{Gu}}}\ and\ \bibinfo {author} {\bibfnamefont {X.-L.}\ \bibnamefont
  {{Qi}}},\ }\bibfield  {title} {\bibinfo {title} {{Axion field theory approach
  and the classification of interacting topological superconductors}},\
  }\href@noop {} {\bibfield  {journal} {\bibinfo  {journal} {ArXiv e-prints}\ }
  (\bibinfo {year} {2015})},\ \Eprint {https://arxiv.org/abs/1512.04919}
  {arXiv:1512.04919 [cond-mat.supr-con]} \BibitemShut {NoStop}%
\bibitem [{\citenamefont {{Song}}\ and\ \citenamefont
  {{Schnyder}}(2016)}]{Song1609.07469}%
  \BibitemOpen
  \bibfield  {author} {\bibinfo {author} {\bibfnamefont {X.-Y.}\ \bibnamefont
  {{Song}}}\ and\ \bibinfo {author} {\bibfnamefont {A.~P.}\ \bibnamefont
  {{Schnyder}}},\ }\bibfield  {title} {\bibinfo {title} {{Interaction effects
  on the classification of crystalline topological insulators and
  superconductors}},\ }\href@noop {} {\bibfield  {journal} {\bibinfo  {journal}
  {ArXiv e-prints}\ } (\bibinfo {year} {2016})},\ \Eprint
  {https://arxiv.org/abs/1609.07469} {arXiv:1609.07469 [cond-mat.str-el]}
  \BibitemShut {NoStop}%
\bibitem [{\citenamefont {{Queiroz}}\ \emph {et~al.}(2016)\citenamefont
  {{Queiroz}}, \citenamefont {{Khalaf}},\ and\ \citenamefont
  {{Stern}}}]{Queiroz1601.01596}%
  \BibitemOpen
  \bibfield  {author} {\bibinfo {author} {\bibfnamefont {R.}~\bibnamefont
  {{Queiroz}}}, \bibinfo {author} {\bibfnamefont {E.}~\bibnamefont
  {{Khalaf}}},\ and\ \bibinfo {author} {\bibfnamefont {A.}~\bibnamefont
  {{Stern}}},\ }\bibfield  {title} {\bibinfo {title} {{Dimensional Hierarchy of
  Fermionic Interacting Topological Phases}},\ }\href
  {https://doi.org/10.1103/PhysRevLett.117.206405} {\bibfield  {journal}
  {\bibinfo  {journal} {Physical Review Letters}\ }\textbf {\bibinfo {volume}
  {117}},\ \bibinfo {eid} {206405} (\bibinfo {year} {2016})},\ \Eprint
  {https://arxiv.org/abs/1601.01596} {arXiv:1601.01596 [cond-mat.str-el]}
  \BibitemShut {NoStop}%
\bibitem [{\citenamefont {{Witten}}(2016)}]{Witten1605.02391}%
  \BibitemOpen
  \bibfield  {author} {\bibinfo {author} {\bibfnamefont {E.}~\bibnamefont
  {{Witten}}},\ }\bibfield  {title} {\bibinfo {title} {{The ``parity'' anomaly
  on an unorientable manifold}},\ }\href
  {https://doi.org/10.1103/PhysRevB.94.195150} {\bibfield  {journal} {\bibinfo
  {journal} {\prb}\ }\textbf {\bibinfo {volume} {94}},\ \bibinfo {eid} {195150}
  (\bibinfo {year} {2016})},\ \Eprint {https://arxiv.org/abs/1605.02391}
  {arXiv:1605.02391 [hep-th]} \BibitemShut {NoStop}%
\bibitem [{\citenamefont {{Wang}}\ and\ \citenamefont
  {{Gu}}(2017)}]{Wang1703.10937}%
  \BibitemOpen
  \bibfield  {author} {\bibinfo {author} {\bibfnamefont {Q.-R.}\ \bibnamefont
  {{Wang}}}\ and\ \bibinfo {author} {\bibfnamefont {Z.-C.}\ \bibnamefont
  {{Gu}}},\ }\bibfield  {title} {\bibinfo {title} {{Towards a complete
  classification of fermionic symmetry protected topological phases in 3D and a
  general group supercohomology theory}},\ }\href@noop {} {\bibfield  {journal}
  {\bibinfo  {journal} {ArXiv e-prints}\ } (\bibinfo {year} {2017})},\ \Eprint
  {https://arxiv.org/abs/1703.10937} {arXiv:1703.10937 [cond-mat.str-el]}
  \BibitemShut {NoStop}%
\bibitem [{\citenamefont {{Kapustin}}\ and\ \citenamefont
  {{Thorngren}}(2017)}]{Kapustin1701.08264}%
  \BibitemOpen
  \bibfield  {author} {\bibinfo {author} {\bibfnamefont {A.}~\bibnamefont
  {{Kapustin}}}\ and\ \bibinfo {author} {\bibfnamefont {R.}~\bibnamefont
  {{Thorngren}}},\ }\bibfield  {title} {\bibinfo {title} {{Fermionic SPT phases
  in higher dimensions and bosonization}},\ }\href
  {https://doi.org/10.1007/JHEP10(2017)080} {\bibfield  {journal} {\bibinfo
  {journal} {Journal of High Energy Physics}\ }\textbf {\bibinfo {volume}
  {2017}},\ \bibinfo {eid} {80} (\bibinfo {year} {2017})},\ \Eprint
  {https://arxiv.org/abs/1701.08264} {arXiv:1701.08264 [cond-mat.str-el]}
  \BibitemShut {NoStop}%
\bibitem [{\citenamefont {{Wang}}\ \emph {et~al.}(2018)\citenamefont {{Wang}},
  \citenamefont {{Ohmori}}, \citenamefont {{Putrov}}, \citenamefont {{Zheng}},
  \citenamefont {{Wan}}, \citenamefont {{Guo}}, \citenamefont {{Lin}},
  \citenamefont {{Gao}},\ and\ \citenamefont
  {{Yau}}}]{Wang2018Tunneling1801.05416}%
  \BibitemOpen
  \bibfield  {author} {\bibinfo {author} {\bibfnamefont {J.}~\bibnamefont
  {{Wang}}}, \bibinfo {author} {\bibfnamefont {K.}~\bibnamefont {{Ohmori}}},
  \bibinfo {author} {\bibfnamefont {P.}~\bibnamefont {{Putrov}}}, \bibinfo
  {author} {\bibfnamefont {Y.}~\bibnamefont {{Zheng}}}, \bibinfo {author}
  {\bibfnamefont {Z.}~\bibnamefont {{Wan}}}, \bibinfo {author} {\bibfnamefont
  {M.}~\bibnamefont {{Guo}}}, \bibinfo {author} {\bibfnamefont
  {H.}~\bibnamefont {{Lin}}}, \bibinfo {author} {\bibfnamefont
  {P.}~\bibnamefont {{Gao}}},\ and\ \bibinfo {author} {\bibfnamefont {S.-T.}\
  \bibnamefont {{Yau}}},\ }\bibfield  {title} {\bibinfo {title} {{Tunneling
  topological vacua via extended operators: (Spin-)TQFT spectra and boundary
  deconfinement in various dimensions}},\ }\href
  {https://doi.org/10.1093/ptep/pty051} {\bibfield  {journal} {\bibinfo
  {journal} {Progress of Theoretical and Experimental Physics}\ }\textbf
  {\bibinfo {volume} {2018}},\ \bibinfo {eid} {053A01} (\bibinfo {year}
  {2018})},\ \Eprint {https://arxiv.org/abs/1801.05416} {arXiv:1801.05416
  [cond-mat.str-el]} \BibitemShut {NoStop}%
\bibitem [{\citenamefont {{Wang}}\ and\ \citenamefont
  {{Gu}}(2018)}]{Wang1811.00536}%
  \BibitemOpen
  \bibfield  {author} {\bibinfo {author} {\bibfnamefont {Q.-R.}\ \bibnamefont
  {{Wang}}}\ and\ \bibinfo {author} {\bibfnamefont {Z.-C.}\ \bibnamefont
  {{Gu}}},\ }\bibfield  {title} {\bibinfo {title} {{Construction and
  classification of symmetry protected topological phases in interacting
  fermion systems}},\ }\href {https://doi.org/10.48550/arXiv.1811.00536}
  {\bibfield  {journal} {\bibinfo  {journal} {arXiv e-prints}\ ,\ \bibinfo
  {eid} {arXiv:1811.00536}} (\bibinfo {year} {2018})},\ \Eprint
  {https://arxiv.org/abs/1811.00536} {arXiv:1811.00536 [cond-mat.str-el]}
  \BibitemShut {NoStop}%
\bibitem [{\citenamefont {{Guo}}\ \emph {et~al.}(2020)\citenamefont {{Guo}},
  \citenamefont {{Ohmori}}, \citenamefont {{Putrov}}, \citenamefont {{Wan}},\
  and\ \citenamefont {{Wang}}}]{Guo1812.11959}%
  \BibitemOpen
  \bibfield  {author} {\bibinfo {author} {\bibfnamefont {M.}~\bibnamefont
  {{Guo}}}, \bibinfo {author} {\bibfnamefont {K.}~\bibnamefont {{Ohmori}}},
  \bibinfo {author} {\bibfnamefont {P.}~\bibnamefont {{Putrov}}}, \bibinfo
  {author} {\bibfnamefont {Z.}~\bibnamefont {{Wan}}},\ and\ \bibinfo {author}
  {\bibfnamefont {J.}~\bibnamefont {{Wang}}},\ }\bibfield  {title} {\bibinfo
  {title} {{Fermionic Finite-Group Gauge Theories and Interacting
  Symmetric/Crystalline Orders via Cobordisms}},\ }\href
  {https://doi.org/10.1007/s00220-019-03671-6} {\bibfield  {journal} {\bibinfo
  {journal} {Communications in Mathematical Physics}\ }\textbf {\bibinfo
  {volume} {376}},\ \bibinfo {pages} {1073} (\bibinfo {year} {2020})},\ \Eprint
  {https://arxiv.org/abs/1812.11959} {arXiv:1812.11959 [hep-th]} \BibitemShut
  {NoStop}%
\bibitem [{\citenamefont {{Aasen}}\ \emph {et~al.}(2021)\citenamefont
  {{Aasen}}, \citenamefont {{Bonderson}},\ and\ \citenamefont
  {{Knapp}}}]{Aasen2109.10911}%
  \BibitemOpen
  \bibfield  {author} {\bibinfo {author} {\bibfnamefont {D.}~\bibnamefont
  {{Aasen}}}, \bibinfo {author} {\bibfnamefont {P.}~\bibnamefont
  {{Bonderson}}},\ and\ \bibinfo {author} {\bibfnamefont {C.}~\bibnamefont
  {{Knapp}}},\ }\bibfield  {title} {\bibinfo {title} {{Characterization and
  Classification of Fermionic Symmetry Enriched Topological Phases}},\ }\href
  {https://doi.org/10.48550/arXiv.2109.10911} {\bibfield  {journal} {\bibinfo
  {journal} {arXiv e-prints}\ ,\ \bibinfo {eid} {arXiv:2109.10911}} (\bibinfo
  {year} {2021})},\ \Eprint {https://arxiv.org/abs/2109.10911}
  {arXiv:2109.10911 [cond-mat.str-el]} \BibitemShut {NoStop}%
\bibitem [{\citenamefont {{Barkeshli}}\ \emph {et~al.}(2022)\citenamefont
  {{Barkeshli}}, \citenamefont {{Chen}}, \citenamefont {{Hsin}},\ and\
  \citenamefont {{Manjunath}}}]{Barkeshli2109.11039}%
  \BibitemOpen
  \bibfield  {author} {\bibinfo {author} {\bibfnamefont {M.}~\bibnamefont
  {{Barkeshli}}}, \bibinfo {author} {\bibfnamefont {Y.-A.}\ \bibnamefont
  {{Chen}}}, \bibinfo {author} {\bibfnamefont {P.-S.}\ \bibnamefont {{Hsin}}},\
  and\ \bibinfo {author} {\bibfnamefont {N.}~\bibnamefont {{Manjunath}}},\
  }\bibfield  {title} {\bibinfo {title} {{Classification of (2 +1 )D invertible
  fermionic topological phases with symmetry}},\ }\href
  {https://doi.org/10.1103/PhysRevB.105.235143} {\bibfield  {journal} {\bibinfo
   {journal} {\prb}\ }\textbf {\bibinfo {volume} {105}},\ \bibinfo {eid}
  {235143} (\bibinfo {year} {2022})},\ \Eprint
  {https://arxiv.org/abs/2109.11039} {arXiv:2109.11039 [cond-mat.str-el]}
  \BibitemShut {NoStop}%
\bibitem [{\citenamefont {{Zou}}\ and\ \citenamefont
  {{Chowdhury}}(2020{\natexlab{a}})}]{Zou2004.14391}%
  \BibitemOpen
  \bibfield  {author} {\bibinfo {author} {\bibfnamefont {L.}~\bibnamefont
  {{Zou}}}\ and\ \bibinfo {author} {\bibfnamefont {D.}~\bibnamefont
  {{Chowdhury}}},\ }\bibfield  {title} {\bibinfo {title} {{Deconfined
  Metal-Insulator Transitions in Quantum Hall Bilayers}},\ }\href
  {https://doi.org/10.48550/arXiv.2004.14391} {\bibfield  {journal} {\bibinfo
  {journal} {arXiv e-prints}\ ,\ \bibinfo {eid} {arXiv:2004.14391}} (\bibinfo
  {year} {2020}{\natexlab{a}})},\ \Eprint {https://arxiv.org/abs/2004.14391}
  {arXiv:2004.14391 [cond-mat.str-el]} \BibitemShut {NoStop}%
\bibitem [{\citenamefont {{Zhang}}\ and\ \citenamefont
  {{Sachdev}}(2020{\natexlab{a}})}]{yahui2020-1}%
  \BibitemOpen
  \bibfield  {author} {\bibinfo {author} {\bibfnamefont {Y.-H.}\ \bibnamefont
  {{Zhang}}}\ and\ \bibinfo {author} {\bibfnamefont {S.}~\bibnamefont
  {{Sachdev}}},\ }\bibfield  {title} {\bibinfo {title} {{From the pseudogap
  metal to the Fermi liquid using ancilla qubits}},\ }\href
  {https://doi.org/10.1103/PhysRevResearch.2.023172} {\bibfield  {journal}
  {\bibinfo  {journal} {Physical Review Research}\ }\textbf {\bibinfo {volume}
  {2}},\ \bibinfo {eid} {023172} (\bibinfo {year} {2020}{\natexlab{a}})},\
  \Eprint {https://arxiv.org/abs/2001.09159} {arXiv:2001.09159
  [cond-mat.str-el]} \BibitemShut {NoStop}%
\bibitem [{\citenamefont {{Zou}}\ and\ \citenamefont
  {{Chowdhury}}(2020{\natexlab{b}})}]{Zou2002.02972}%
  \BibitemOpen
  \bibfield  {author} {\bibinfo {author} {\bibfnamefont {L.}~\bibnamefont
  {{Zou}}}\ and\ \bibinfo {author} {\bibfnamefont {D.}~\bibnamefont
  {{Chowdhury}}},\ }\bibfield  {title} {\bibinfo {title} {{Deconfined metallic
  quantum criticality: A U(2) gauge-theoretic approach}},\ }\href
  {https://doi.org/10.1103/PhysRevResearch.2.023344} {\bibfield  {journal}
  {\bibinfo  {journal} {Physical Review Research}\ }\textbf {\bibinfo {volume}
  {2}},\ \bibinfo {eid} {023344} (\bibinfo {year} {2020}{\natexlab{b}})},\
  \Eprint {https://arxiv.org/abs/2002.02972} {arXiv:2002.02972
  [cond-mat.str-el]} \BibitemShut {NoStop}%
\bibitem [{\citenamefont {{Zhang}}\ and\ \citenamefont
  {{Sachdev}}(2020{\natexlab{b}})}]{yahui2020-2}%
  \BibitemOpen
  \bibfield  {author} {\bibinfo {author} {\bibfnamefont {Y.-H.}\ \bibnamefont
  {{Zhang}}}\ and\ \bibinfo {author} {\bibfnamefont {S.}~\bibnamefont
  {{Sachdev}}},\ }\bibfield  {title} {\bibinfo {title} {{Deconfined criticality
  and ghost Fermi surfaces at the onset of antiferromagnetism in a metal}},\
  }\href {https://doi.org/10.1103/PhysRevB.102.155124} {\bibfield  {journal}
  {\bibinfo  {journal} {\prb}\ }\textbf {\bibinfo {volume} {102}},\ \bibinfo
  {eid} {155124} (\bibinfo {year} {2020}{\natexlab{b}})},\ \Eprint
  {https://arxiv.org/abs/2006.01140} {arXiv:2006.01140 [cond-mat.str-el]}
  \BibitemShut {NoStop}%
\bibitem [{\citenamefont {{Nikolaenko}}\ \emph {et~al.}(2021)\citenamefont
  {{Nikolaenko}}, \citenamefont {{Tikhanovskaya}}, \citenamefont {{Sachdev}},\
  and\ \citenamefont {{Zhang}}}]{yahui2021}%
  \BibitemOpen
  \bibfield  {author} {\bibinfo {author} {\bibfnamefont {A.}~\bibnamefont
  {{Nikolaenko}}}, \bibinfo {author} {\bibfnamefont {M.}~\bibnamefont
  {{Tikhanovskaya}}}, \bibinfo {author} {\bibfnamefont {S.}~\bibnamefont
  {{Sachdev}}},\ and\ \bibinfo {author} {\bibfnamefont {Y.-H.}\ \bibnamefont
  {{Zhang}}},\ }\bibfield  {title} {\bibinfo {title} {{Small to large Fermi
  surface transition in a single-band model using randomly coupled ancillas}},\
  }\href {https://doi.org/10.1103/PhysRevB.103.235138} {\bibfield  {journal}
  {\bibinfo  {journal} {\prb}\ }\textbf {\bibinfo {volume} {103}},\ \bibinfo
  {eid} {235138} (\bibinfo {year} {2021})},\ \Eprint
  {https://arxiv.org/abs/2103.05009} {arXiv:2103.05009 [cond-mat.str-el]}
  \BibitemShut {NoStop}%
\bibitem [{\citenamefont {{Lu}}\ \emph {et~al.}(2022)\citenamefont {{Lu}},
  \citenamefont {{Zeng}}, \citenamefont {{Wang}},\ and\ \citenamefont
  {{You}}}]{YZY2022FSSMG}%
  \BibitemOpen
  \bibfield  {author} {\bibinfo {author} {\bibfnamefont {D.-C.}\ \bibnamefont
  {{Lu}}}, \bibinfo {author} {\bibfnamefont {M.}~\bibnamefont {{Zeng}}},
  \bibinfo {author} {\bibfnamefont {J.}~\bibnamefont {{Wang}}},\ and\ \bibinfo
  {author} {\bibfnamefont {Y.-Z.}\ \bibnamefont {{You}}},\ }\bibfield  {title}
  {\bibinfo {title} {{Fermi Surface Symmetric Mass Generation}},\ }\href
  {https://doi.org/10.48550/arXiv.2210.16304} {\bibfield  {journal} {\bibinfo
  {journal} {arXiv e-prints}\ ,\ \bibinfo {eid} {arXiv:2210.16304}} (\bibinfo
  {year} {2022})},\ \Eprint {https://arxiv.org/abs/2210.16304}
  {arXiv:2210.16304 [cond-mat.str-el]} \BibitemShut {NoStop}%
\bibitem [{\citenamefont {{Bulmash}}\ \emph {et~al.}(2015)\citenamefont
  {{Bulmash}}, \citenamefont {{Hosur}}, \citenamefont {{Zhang}},\ and\
  \citenamefont {{Qi}}}]{Bulmash1410.4202}%
  \BibitemOpen
  \bibfield  {author} {\bibinfo {author} {\bibfnamefont {D.}~\bibnamefont
  {{Bulmash}}}, \bibinfo {author} {\bibfnamefont {P.}~\bibnamefont {{Hosur}}},
  \bibinfo {author} {\bibfnamefont {S.-C.}\ \bibnamefont {{Zhang}}},\ and\
  \bibinfo {author} {\bibfnamefont {X.-L.}\ \bibnamefont {{Qi}}},\ }\bibfield
  {title} {\bibinfo {title} {{Unified Topological Response Theory For Gapped
  and Gapless Free Fermions}},\ }\href
  {https://doi.org/10.1103/PhysRevX.5.02101810.48550/arXiv.1410.4202}
  {\bibfield  {journal} {\bibinfo  {journal} {Physical Review X}\ }\textbf
  {\bibinfo {volume} {5}},\ \bibinfo {eid} {021018} (\bibinfo {year} {2015})},\
  \Eprint {https://arxiv.org/abs/1410.4202} {arXiv:1410.4202
  [cond-mat.mes-hall]} \BibitemShut {NoStop}%
\bibitem [{\citenamefont {{Lu}}\ \emph {et~al.}(2023)\citenamefont {{Lu}},
  \citenamefont {{Wang}},\ and\ \citenamefont {{You}}}]{yzy2023FSanomaly}%
  \BibitemOpen
  \bibfield  {author} {\bibinfo {author} {\bibfnamefont {D.-C.}\ \bibnamefont
  {{Lu}}}, \bibinfo {author} {\bibfnamefont {J.}~\bibnamefont {{Wang}}},\ and\
  \bibinfo {author} {\bibfnamefont {Y.-Z.}\ \bibnamefont {{You}}},\ }\bibfield
  {title} {\bibinfo {title} {{Definition and Classification of Fermi Surface
  Anomalies}},\ }\href {https://doi.org/10.48550/arXiv.2302.12731} {\bibfield
  {journal} {\bibinfo  {journal} {arXiv e-prints}\ ,\ \bibinfo {eid}
  {arXiv:2302.12731}} (\bibinfo {year} {2023})},\ \Eprint
  {https://arxiv.org/abs/2302.12731} {arXiv:2302.12731 [cond-mat.str-el]}
  \BibitemShut {NoStop}%
\bibitem [{\citenamefont {{Gurarie}}(2011)}]{Gurarie1011.2273}%
  \BibitemOpen
  \bibfield  {author} {\bibinfo {author} {\bibfnamefont {V.}~\bibnamefont
  {{Gurarie}}},\ }\bibfield  {title} {\bibinfo {title} {{Single-particle
  Green's functions and interacting topological insulators}},\ }\href
  {https://doi.org/10.1103/PhysRevB.83.085426} {\bibfield  {journal} {\bibinfo
  {journal} {\prb}\ }\textbf {\bibinfo {volume} {83}},\ \bibinfo {eid} {085426}
  (\bibinfo {year} {2011})},\ \Eprint {https://arxiv.org/abs/1011.2273}
  {arXiv:1011.2273 [cond-mat.mes-hall]} \BibitemShut {NoStop}%
\bibitem [{\citenamefont {{You}}\ \emph
  {et~al.}(2014{\natexlab{b}})\citenamefont {{You}}, \citenamefont {{Wang}},
  \citenamefont {{Oon}},\ and\ \citenamefont
  {{Xu}}}]{yzy2014greenfunctionzero}%
  \BibitemOpen
  \bibfield  {author} {\bibinfo {author} {\bibfnamefont {Y.-Z.}\ \bibnamefont
  {{You}}}, \bibinfo {author} {\bibfnamefont {Z.}~\bibnamefont {{Wang}}},
  \bibinfo {author} {\bibfnamefont {J.}~\bibnamefont {{Oon}}},\ and\ \bibinfo
  {author} {\bibfnamefont {C.}~\bibnamefont {{Xu}}},\ }\bibfield  {title}
  {\bibinfo {title} {{Topological number and fermion Green's function for
  strongly interacting topological superconductors}},\ }\href
  {https://doi.org/10.1103/PhysRevB.90.060502} {\bibfield  {journal} {\bibinfo
  {journal} {\prb}\ }\textbf {\bibinfo {volume} {90}},\ \bibinfo {eid} {060502}
  (\bibinfo {year} {2014}{\natexlab{b}})},\ \Eprint
  {https://arxiv.org/abs/1403.4938} {arXiv:1403.4938 [cond-mat.str-el]}
  \BibitemShut {NoStop}%
\bibitem [{\citenamefont {{Catterall}}\ and\ \citenamefont
  {{Schaich}}(2016)}]{Catterall1609.08541}%
  \BibitemOpen
  \bibfield  {author} {\bibinfo {author} {\bibfnamefont {S.}~\bibnamefont
  {{Catterall}}}\ and\ \bibinfo {author} {\bibfnamefont {D.}~\bibnamefont
  {{Schaich}}},\ }\bibfield  {title} {\bibinfo {title} {{Novel phases in
  strongly coupled four-fermion theories}},\ }\href@noop {} {\bibfield
  {journal} {\bibinfo  {journal} {ArXiv e-prints}\ } (\bibinfo {year}
  {2016})},\ \Eprint {https://arxiv.org/abs/1609.08541} {arXiv:1609.08541
  [hep-lat]} \BibitemShut {NoStop}%
\bibitem [{\citenamefont {{You}}\ \emph
  {et~al.}(2018{\natexlab{a}})\citenamefont {{You}}, \citenamefont {{He}},
  \citenamefont {{Xu}},\ and\ \citenamefont {{Vishwanath}}}]{YZY2018SMGDQCP-1}%
  \BibitemOpen
  \bibfield  {author} {\bibinfo {author} {\bibfnamefont {Y.-Z.}\ \bibnamefont
  {{You}}}, \bibinfo {author} {\bibfnamefont {Y.-C.}\ \bibnamefont {{He}}},
  \bibinfo {author} {\bibfnamefont {C.}~\bibnamefont {{Xu}}},\ and\ \bibinfo
  {author} {\bibfnamefont {A.}~\bibnamefont {{Vishwanath}}},\ }\bibfield
  {title} {\bibinfo {title} {{Symmetric Fermion Mass Generation as Deconfined
  Quantum Criticality}},\ }\href {https://doi.org/10.1103/PhysRevX.8.011026}
  {\bibfield  {journal} {\bibinfo  {journal} {Physical Review X}\ }\textbf
  {\bibinfo {volume} {8}},\ \bibinfo {eid} {011026} (\bibinfo {year}
  {2018}{\natexlab{a}})},\ \Eprint {https://arxiv.org/abs/1705.09313}
  {arXiv:1705.09313 [cond-mat.str-el]} \BibitemShut {NoStop}%
\bibitem [{\citenamefont {{Catterall}}\ and\ \citenamefont
  {{Butt}}(2018)}]{Catterall1708.06715}%
  \BibitemOpen
  \bibfield  {author} {\bibinfo {author} {\bibfnamefont {S.}~\bibnamefont
  {{Catterall}}}\ and\ \bibinfo {author} {\bibfnamefont {N.}~\bibnamefont
  {{Butt}}},\ }\bibfield  {title} {\bibinfo {title} {{Topology and strong four
  fermion interactions in four dimensions}},\ }\href
  {https://doi.org/10.1103/PhysRevD.97.094502} {\bibfield  {journal} {\bibinfo
  {journal} {\prd}\ }\textbf {\bibinfo {volume} {97}},\ \bibinfo {eid} {094502}
  (\bibinfo {year} {2018})},\ \Eprint {https://arxiv.org/abs/1708.06715}
  {arXiv:1708.06715 [hep-lat]} \BibitemShut {NoStop}%
\bibitem [{\citenamefont {{Xu}}\ and\ \citenamefont
  {{Xu}}(2021)}]{Xu2103.15865}%
  \BibitemOpen
  \bibfield  {author} {\bibinfo {author} {\bibfnamefont {Y.}~\bibnamefont
  {{Xu}}}\ and\ \bibinfo {author} {\bibfnamefont {C.}~\bibnamefont {{Xu}}},\
  }\bibfield  {title} {\bibinfo {title} {{Green's function Zero and Symmetric
  Mass Generation}},\ }\href@noop {} {\bibfield  {journal} {\bibinfo  {journal}
  {arXiv e-prints}\ ,\ \bibinfo {eid} {arXiv:2103.15865}} (\bibinfo {year}
  {2021})},\ \Eprint {https://arxiv.org/abs/2103.15865} {arXiv:2103.15865
  [cond-mat.str-el]} \BibitemShut {NoStop}%
\bibitem [{\citenamefont {{You}}\ \emph
  {et~al.}(2018{\natexlab{b}})\citenamefont {{You}}, \citenamefont {{He}},
  \citenamefont {{Vishwanath}},\ and\ \citenamefont {{Xu}}}]{YZY2018SMGDQCP-2}%
  \BibitemOpen
  \bibfield  {author} {\bibinfo {author} {\bibfnamefont {Y.-Z.}\ \bibnamefont
  {{You}}}, \bibinfo {author} {\bibfnamefont {Y.-C.}\ \bibnamefont {{He}}},
  \bibinfo {author} {\bibfnamefont {A.}~\bibnamefont {{Vishwanath}}},\ and\
  \bibinfo {author} {\bibfnamefont {C.}~\bibnamefont {{Xu}}},\ }\bibfield
  {title} {\bibinfo {title} {{From bosonic topological transition to symmetric
  fermion mass generation}},\ }\href
  {https://doi.org/10.1103/PhysRevB.97.125112} {\bibfield  {journal} {\bibinfo
  {journal} {\prb}\ }\textbf {\bibinfo {volume} {97}},\ \bibinfo {eid} {125112}
  (\bibinfo {year} {2018}{\natexlab{b}})},\ \Eprint
  {https://arxiv.org/abs/1711.00863} {arXiv:1711.00863 [cond-mat.str-el]}
  \BibitemShut {NoStop}%
\bibitem [{\citenamefont {{Wang}}\ \emph {et~al.}(2021)\citenamefont {{Wang}},
  \citenamefont {{Hickey}}, \citenamefont {{Ying}},\ and\ \citenamefont
  {{Burkov}}}]{Chong2021FSanomaly}%
  \BibitemOpen
  \bibfield  {author} {\bibinfo {author} {\bibfnamefont {C.}~\bibnamefont
  {{Wang}}}, \bibinfo {author} {\bibfnamefont {A.}~\bibnamefont {{Hickey}}},
  \bibinfo {author} {\bibfnamefont {X.}~\bibnamefont {{Ying}}},\ and\ \bibinfo
  {author} {\bibfnamefont {A.~A.}\ \bibnamefont {{Burkov}}},\ }\bibfield
  {title} {\bibinfo {title} {{Emergent anomalies and generalized Luttinger
  theorems in metals and semimetals}},\ }\href
  {https://doi.org/10.1103/PhysRevB.104.235113} {\bibfield  {journal} {\bibinfo
   {journal} {\prb}\ }\textbf {\bibinfo {volume} {104}},\ \bibinfo {eid}
  {235113} (\bibinfo {year} {2021})},\ \Eprint
  {https://arxiv.org/abs/2110.10692} {arXiv:2110.10692 [cond-mat.str-el]}
  \BibitemShut {NoStop}%
\bibitem [{\citenamefont {{Wen}}(2021)}]{XiaoGang2021FSanomaly}%
  \BibitemOpen
  \bibfield  {author} {\bibinfo {author} {\bibfnamefont {X.-G.}\ \bibnamefont
  {{Wen}}},\ }\bibfield  {title} {\bibinfo {title} {{Low-energy effective field
  theories of fermion liquids and the mixed U (1 ) {\texttimes}R$^{d}$
  anomaly}},\ }\href {https://doi.org/10.1103/PhysRevB.103.165126} {\bibfield
  {journal} {\bibinfo  {journal} {\prb}\ }\textbf {\bibinfo {volume} {103}},\
  \bibinfo {eid} {165126} (\bibinfo {year} {2021})},\ \Eprint
  {https://arxiv.org/abs/2101.08772} {arXiv:2101.08772 [cond-mat.str-el]}
  \BibitemShut {NoStop}%
\bibitem [{\citenamefont {{Cheng}}\ \emph {et~al.}(2016)\citenamefont
  {{Cheng}}, \citenamefont {{Zaletel}}, \citenamefont {{Barkeshli}},
  \citenamefont {{Vishwanath}},\ and\ \citenamefont
  {{Bonderson}}}]{Meng2016Topo}%
  \BibitemOpen
  \bibfield  {author} {\bibinfo {author} {\bibfnamefont {M.}~\bibnamefont
  {{Cheng}}}, \bibinfo {author} {\bibfnamefont {M.}~\bibnamefont {{Zaletel}}},
  \bibinfo {author} {\bibfnamefont {M.}~\bibnamefont {{Barkeshli}}}, \bibinfo
  {author} {\bibfnamefont {A.}~\bibnamefont {{Vishwanath}}},\ and\ \bibinfo
  {author} {\bibfnamefont {P.}~\bibnamefont {{Bonderson}}},\ }\bibfield
  {title} {\bibinfo {title} {{Translational Symmetry and Microscopic
  Constraints on Symmetry-Enriched Topological Phases: A View from the
  Surface}},\ }\href {https://doi.org/10.1103/PhysRevX.6.041068} {\bibfield
  {journal} {\bibinfo  {journal} {Physical Review X}\ }\textbf {\bibinfo
  {volume} {6}},\ \bibinfo {eid} {041068} (\bibinfo {year} {2016})},\ \Eprint
  {https://arxiv.org/abs/1511.02263} {arXiv:1511.02263 [cond-mat.str-el]}
  \BibitemShut {NoStop}%
\bibitem [{\citenamefont {{Cho}}\ \emph {et~al.}(2017)\citenamefont {{Cho}},
  \citenamefont {{Hsieh}},\ and\ \citenamefont {{Ryu}}}]{Shinsei2017FSLSM}%
  \BibitemOpen
  \bibfield  {author} {\bibinfo {author} {\bibfnamefont {G.~Y.}\ \bibnamefont
  {{Cho}}}, \bibinfo {author} {\bibfnamefont {C.-T.}\ \bibnamefont {{Hsieh}}},\
  and\ \bibinfo {author} {\bibfnamefont {S.}~\bibnamefont {{Ryu}}},\ }\bibfield
   {title} {\bibinfo {title} {{Anomaly manifestation of Lieb-Schultz-Mattis
  theorem and topological phases}},\ }\href
  {https://doi.org/10.1103/PhysRevB.96.195105} {\bibfield  {journal} {\bibinfo
  {journal} {\prb}\ }\textbf {\bibinfo {volume} {96}},\ \bibinfo {eid} {195105}
  (\bibinfo {year} {2017})},\ \Eprint {https://arxiv.org/abs/1705.03892}
  {arXiv:1705.03892 [cond-mat.str-el]} \BibitemShut {NoStop}%
\bibitem [{\citenamefont {{Bultinck}}\ and\ \citenamefont
  {{Cheng}}(2018)}]{Bultinck1808.00324}%
  \BibitemOpen
  \bibfield  {author} {\bibinfo {author} {\bibfnamefont {N.}~\bibnamefont
  {{Bultinck}}}\ and\ \bibinfo {author} {\bibfnamefont {M.}~\bibnamefont
  {{Cheng}}},\ }\bibfield  {title} {\bibinfo {title} {{Filling constraints on
  fermionic topological order in zero magnetic field}},\ }\href
  {https://doi.org/10.1103/PhysRevB.98.161119} {\bibfield  {journal} {\bibinfo
  {journal} {\prb}\ }\textbf {\bibinfo {volume} {98}},\ \bibinfo {eid} {161119}
  (\bibinfo {year} {2018})},\ \Eprint {https://arxiv.org/abs/1808.00324}
  {arXiv:1808.00324 [cond-mat.str-el]} \BibitemShut {NoStop}%
\bibitem [{\citenamefont {{Thorngren}}\ and\ \citenamefont
  {{Else}}(2018)}]{Else2018FSLSMcrystalline}%
  \BibitemOpen
  \bibfield  {author} {\bibinfo {author} {\bibfnamefont {R.}~\bibnamefont
  {{Thorngren}}}\ and\ \bibinfo {author} {\bibfnamefont {D.~V.}\ \bibnamefont
  {{Else}}},\ }\bibfield  {title} {\bibinfo {title} {{Gauging Spatial
  Symmetries and the Classification of Topological Crystalline Phases}},\
  }\href {https://doi.org/10.1103/PhysRevX.8.011040} {\bibfield  {journal}
  {\bibinfo  {journal} {Physical Review X}\ }\textbf {\bibinfo {volume} {8}},\
  \bibinfo {eid} {011040} (\bibinfo {year} {2018})}\BibitemShut {NoStop}%
\bibitem [{\citenamefont {{Cheng}}\ and\ \citenamefont
  {{Seiberg}}(2022)}]{Seiberg2022emanant}%
  \BibitemOpen
  \bibfield  {author} {\bibinfo {author} {\bibfnamefont {M.}~\bibnamefont
  {{Cheng}}}\ and\ \bibinfo {author} {\bibfnamefont {N.}~\bibnamefont
  {{Seiberg}}},\ }\bibfield  {title} {\bibinfo {title} {{Lieb-Schultz-Mattis,
  Luttinger, and 't Hooft -- anomaly matching in lattice systems}},\ }\href
  {https://doi.org/10.48550/arXiv.2211.12543} {\bibfield  {journal} {\bibinfo
  {journal} {arXiv e-prints}\ ,\ \bibinfo {eid} {arXiv:2211.12543}} (\bibinfo
  {year} {2022})},\ \Eprint {https://arxiv.org/abs/2211.12543}
  {arXiv:2211.12543 [cond-mat.str-el]} \BibitemShut {NoStop}%
\bibitem [{\citenamefont {{Else}}\ \emph {et~al.}(2021)\citenamefont {{Else}},
  \citenamefont {{Thorngren}},\ and\ \citenamefont
  {{Senthil}}}]{Senthil2007.07896}%
  \BibitemOpen
  \bibfield  {author} {\bibinfo {author} {\bibfnamefont {D.~V.}\ \bibnamefont
  {{Else}}}, \bibinfo {author} {\bibfnamefont {R.}~\bibnamefont
  {{Thorngren}}},\ and\ \bibinfo {author} {\bibfnamefont {T.}~\bibnamefont
  {{Senthil}}},\ }\bibfield  {title} {\bibinfo {title} {{Non-Fermi Liquids as
  Ersatz Fermi Liquids: General Constraints on Compressible Metals}},\ }\href
  {https://doi.org/10.1103/PhysRevX.11.021005} {\bibfield  {journal} {\bibinfo
  {journal} {Physical Review X}\ }\textbf {\bibinfo {volume} {11}},\ \bibinfo
  {eid} {021005} (\bibinfo {year} {2021})},\ \Eprint
  {https://arxiv.org/abs/2007.07896} {arXiv:2007.07896 [cond-mat.str-el]}
  \BibitemShut {NoStop}%
\bibitem [{\citenamefont {{Else}}\ and\ \citenamefont
  {{Senthil}}(2021)}]{Else2010.10523}%
  \BibitemOpen
  \bibfield  {author} {\bibinfo {author} {\bibfnamefont {D.~V.}\ \bibnamefont
  {{Else}}}\ and\ \bibinfo {author} {\bibfnamefont {T.}~\bibnamefont
  {{Senthil}}},\ }\bibfield  {title} {\bibinfo {title} {{Strange Metals as
  Ersatz Fermi Liquids}},\ }\href
  {https://doi.org/10.1103/PhysRevLett.127.086601} {\bibfield  {journal}
  {\bibinfo  {journal} {\prl}\ }\textbf {\bibinfo {volume} {127}},\ \bibinfo
  {eid} {086601} (\bibinfo {year} {2021})},\ \Eprint
  {https://arxiv.org/abs/2010.10523} {arXiv:2010.10523 [cond-mat.str-el]}
  \BibitemShut {NoStop}%
\bibitem [{\citenamefont {{Darius Shi}}\ \emph {et~al.}(2022)\citenamefont
  {{Darius Shi}}, \citenamefont {{Goldman}}, \citenamefont {{Else}},\ and\
  \citenamefont {{Senthil}}}]{Darius-Shi2204.07585}%
  \BibitemOpen
  \bibfield  {author} {\bibinfo {author} {\bibfnamefont {Z.}~\bibnamefont
  {{Darius Shi}}}, \bibinfo {author} {\bibfnamefont {H.}~\bibnamefont
  {{Goldman}}}, \bibinfo {author} {\bibfnamefont {D.~V.}\ \bibnamefont
  {{Else}}},\ and\ \bibinfo {author} {\bibfnamefont {T.}~\bibnamefont
  {{Senthil}}},\ }\bibfield  {title} {\bibinfo {title} {{Gifts from anomalies:
  Exact results for Landau phase transitions in metals}},\ }\href
  {https://doi.org/10.21468/SciPostPhys.13.5.102} {\bibfield  {journal}
  {\bibinfo  {journal} {SciPost Phys.}\ }\textbf {\bibinfo {volume} {13}},\
  \bibinfo {eid} {102} (\bibinfo {year} {2022})},\ \Eprint
  {https://arxiv.org/abs/2204.07585} {arXiv:2204.07585 [cond-mat.str-el]}
  \BibitemShut {NoStop}%
\bibitem [{\citenamefont {{S{\'e}n{\'e}chal}}\ \emph
  {et~al.}(2000)\citenamefont {{S{\'e}n{\'e}chal}}, \citenamefont {{Perez}},\
  and\ \citenamefont {{Pioro-Ladri{\`e}re}}}]{CPT2000}%
  \BibitemOpen
  \bibfield  {author} {\bibinfo {author} {\bibfnamefont {D.}~\bibnamefont
  {{S{\'e}n{\'e}chal}}}, \bibinfo {author} {\bibfnamefont {D.}~\bibnamefont
  {{Perez}}},\ and\ \bibinfo {author} {\bibfnamefont {M.}~\bibnamefont
  {{Pioro-Ladri{\`e}re}}},\ }\bibfield  {title} {\bibinfo {title} {{Spectral
  Weight of the Hubbard Model through Cluster Perturbation Theory}},\ }\href
  {https://doi.org/10.1103/PhysRevLett.84.522} {\bibfield  {journal} {\bibinfo
  {journal} {\prl}\ }\textbf {\bibinfo {volume} {84}},\ \bibinfo {pages} {522}
  (\bibinfo {year} {2000})},\ \Eprint {https://arxiv.org/abs/cond-mat/9908045}
  {arXiv:cond-mat/9908045 [cond-mat.str-el]} \BibitemShut {NoStop}%
\bibitem [{\citenamefont {{Wang}}\ and\ \citenamefont
  {{Qi}}(2023)}]{Wang2023P2212.05737}%
  \BibitemOpen
  \bibfield  {author} {\bibinfo {author} {\bibfnamefont {X.-C.}\ \bibnamefont
  {{Wang}}}\ and\ \bibinfo {author} {\bibfnamefont {Y.}~\bibnamefont {{Qi}}},\
  }\bibfield  {title} {\bibinfo {title} {{Phase fluctuations in two-dimensional
  superconductors and pseudogap phenomenon}},\ }\href
  {https://doi.org/10.1103/PhysRevB.107.224502} {\bibfield  {journal} {\bibinfo
   {journal} {\prb}\ }\textbf {\bibinfo {volume} {107}},\ \bibinfo {eid}
  {224502} (\bibinfo {year} {2023})},\ \Eprint
  {https://arxiv.org/abs/2212.05737} {arXiv:2212.05737 [cond-mat.supr-con]}
  \BibitemShut {NoStop}%
\bibitem [{\citenamefont {{Altshuler}}\ \emph {et~al.}(1998)\citenamefont
  {{Altshuler}}, \citenamefont {{Chubukov}}, \citenamefont {{Dashevskii}},
  \citenamefont {{Finkel'stein}},\ and\ \citenamefont
  {{Morr}}}]{Altshuler1998Lcond-mat/9703120}%
  \BibitemOpen
  \bibfield  {author} {\bibinfo {author} {\bibfnamefont {B.~L.}\ \bibnamefont
  {{Altshuler}}}, \bibinfo {author} {\bibfnamefont {A.~V.}\ \bibnamefont
  {{Chubukov}}}, \bibinfo {author} {\bibfnamefont {A.}~\bibnamefont
  {{Dashevskii}}}, \bibinfo {author} {\bibfnamefont {A.~M.}\ \bibnamefont
  {{Finkel'stein}}},\ and\ \bibinfo {author} {\bibfnamefont {D.~K.}\
  \bibnamefont {{Morr}}},\ }\bibfield  {title} {\bibinfo {title} {{Luttinger
  theorem for a spin-density-wave state}},\ }\href
  {https://doi.org/10.1209/epl/i1998-00164-y} {\bibfield  {journal} {\bibinfo
  {journal} {EPL (Europhysics Letters)}\ }\textbf {\bibinfo {volume} {41}},\
  \bibinfo {pages} {401} (\bibinfo {year} {1998})},\ \Eprint
  {https://arxiv.org/abs/cond-mat/9703120} {arXiv:cond-mat/9703120
  [cond-mat.str-el]} \BibitemShut {NoStop}%
\bibitem [{\citenamefont {{Oshikawa}}(2000)}]{Oshikawacond-mat/0002392}%
  \BibitemOpen
  \bibfield  {author} {\bibinfo {author} {\bibfnamefont {M.}~\bibnamefont
  {{Oshikawa}}},\ }\bibfield  {title} {\bibinfo {title} {{Topological Approach
  to Luttinger's Theorem and the Fermi Surface of a Kondo Lattice}},\ }\href
  {https://doi.org/10.1103/PhysRevLett.84.3370} {\bibfield  {journal} {\bibinfo
   {journal} {\prl}\ }\textbf {\bibinfo {volume} {84}},\ \bibinfo {pages}
  {3370} (\bibinfo {year} {2000})},\ \Eprint
  {https://arxiv.org/abs/cond-mat/0002392} {arXiv:cond-mat/0002392
  [cond-mat.str-el]} \BibitemShut {NoStop}%
\bibitem [{\citenamefont {{Dzyaloshinskii}}(2003)}]{2003LuttingerSurface}%
  \BibitemOpen
  \bibfield  {author} {\bibinfo {author} {\bibfnamefont {I.}~\bibnamefont
  {{Dzyaloshinskii}}},\ }\bibfield  {title} {\bibinfo {title} {{Some
  consequences of the Luttinger theorem: The Luttinger surfaces in non-Fermi
  liquids and Mott insulators}},\ }\href
  {https://doi.org/10.1103/PhysRevB.68.085113} {\bibfield  {journal} {\bibinfo
  {journal} {\prb}\ }\textbf {\bibinfo {volume} {68}},\ \bibinfo {eid} {085113}
  (\bibinfo {year} {2003})}\BibitemShut {NoStop}%
\bibitem [{\citenamefont {{Gnezdilov}}\ and\ \citenamefont
  {{Wang}}(2022)}]{Gnezdilov2022S2111.09906}%
  \BibitemOpen
  \bibfield  {author} {\bibinfo {author} {\bibfnamefont {N.~V.}\ \bibnamefont
  {{Gnezdilov}}}\ and\ \bibinfo {author} {\bibfnamefont {Y.}~\bibnamefont
  {{Wang}}},\ }\bibfield  {title} {\bibinfo {title} {{Solvable model for a
  charge-4 e superconductor}},\ }\href
  {https://doi.org/10.1103/PhysRevB.106.094508} {\bibfield  {journal} {\bibinfo
   {journal} {\prb}\ }\textbf {\bibinfo {volume} {106}},\ \bibinfo {eid}
  {094508} (\bibinfo {year} {2022})},\ \Eprint
  {https://arxiv.org/abs/2111.09906} {arXiv:2111.09906 [cond-mat.str-el]}
  \BibitemShut {NoStop}%
\bibitem [{\citenamefont {Luttinger}(1960)}]{LuttingerRP1960}%
  \BibitemOpen
  \bibfield  {author} {\bibinfo {author} {\bibfnamefont {J.~M.}\ \bibnamefont
  {Luttinger}},\ }\bibfield  {title} {\bibinfo {title} {Fermi surface and some
  simple equilibrium properties of a system of interacting fermions},\ }\href
  {https://doi.org/10.1103/PhysRev.119.1153} {\bibfield  {journal} {\bibinfo
  {journal} {Phys. Rev.}\ }\textbf {\bibinfo {volume} {119}},\ \bibinfo {pages}
  {1153} (\bibinfo {year} {1960})}\BibitemShut {NoStop}%
\bibitem [{\citenamefont {{Stanescu}}\ \emph {et~al.}(2007)\citenamefont
  {{Stanescu}}, \citenamefont {{Phillips}},\ and\ \citenamefont
  {{Choy}}}]{Stanescu2007Tcond-mat/0602280}%
  \BibitemOpen
  \bibfield  {author} {\bibinfo {author} {\bibfnamefont {T.~D.}\ \bibnamefont
  {{Stanescu}}}, \bibinfo {author} {\bibfnamefont {P.}~\bibnamefont
  {{Phillips}}},\ and\ \bibinfo {author} {\bibfnamefont {T.-P.}\ \bibnamefont
  {{Choy}}},\ }\bibfield  {title} {\bibinfo {title} {{Theory of the Luttinger
  surface in doped Mott insulators}},\ }\href
  {https://doi.org/10.1103/PhysRevB.75.104503} {\bibfield  {journal} {\bibinfo
  {journal} {\prb}\ }\textbf {\bibinfo {volume} {75}},\ \bibinfo {eid} {104503}
  (\bibinfo {year} {2007})},\ \Eprint {https://arxiv.org/abs/cond-mat/0602280}
  {arXiv:cond-mat/0602280 [cond-mat.str-el]} \BibitemShut {NoStop}%
\bibitem [{\citenamefont {{Scheurer}}\ \emph {et~al.}(2018)\citenamefont
  {{Scheurer}}, \citenamefont {{Chatterjee}}, \citenamefont {{Wu}},
  \citenamefont {{Ferrero}}, \citenamefont {{Georges}},\ and\ \citenamefont
  {{Sachdev}}}]{Sachdev1711.09925}%
  \BibitemOpen
  \bibfield  {author} {\bibinfo {author} {\bibfnamefont {M.~S.}\ \bibnamefont
  {{Scheurer}}}, \bibinfo {author} {\bibfnamefont {S.}~\bibnamefont
  {{Chatterjee}}}, \bibinfo {author} {\bibfnamefont {W.}~\bibnamefont {{Wu}}},
  \bibinfo {author} {\bibfnamefont {M.}~\bibnamefont {{Ferrero}}}, \bibinfo
  {author} {\bibfnamefont {A.}~\bibnamefont {{Georges}}},\ and\ \bibinfo
  {author} {\bibfnamefont {S.}~\bibnamefont {{Sachdev}}},\ }\bibfield  {title}
  {\bibinfo {title} {{Topological order in the pseudogap metal}},\ }\href
  {https://doi.org/10.1073/pnas.1720580115} {\bibfield  {journal} {\bibinfo
  {journal} {Proceedings of the National Academy of Science}\ }\textbf
  {\bibinfo {volume} {115}},\ \bibinfo {pages} {E3665} (\bibinfo {year}
  {2018})},\ \Eprint {https://arxiv.org/abs/1711.09925} {arXiv:1711.09925
  [cond-mat.str-el]} \BibitemShut {NoStop}%
\bibitem [{\citenamefont {Xiang}\ and\ \citenamefont {Wu}(2022)}]{xiang2022d}%
  \BibitemOpen
  \bibfield  {author} {\bibinfo {author} {\bibfnamefont {T.}~\bibnamefont
  {Xiang}}\ and\ \bibinfo {author} {\bibfnamefont {C.}~\bibnamefont {Wu}},\
  }\href {https://doi.org/10.1017/9781009218566} {\emph {\bibinfo {title}
  {D-wave Superconductivity}}}\ (\bibinfo  {publisher} {Cambridge University
  Press},\ \bibinfo {year} {2022})\ Chap.~\bibinfo {chapter} {4}\BibitemShut
  {NoStop}%
\bibitem [{\citenamefont {{Zhai}}\ \emph {et~al.}(2009)\citenamefont {{Zhai}},
  \citenamefont {{Wang}},\ and\ \citenamefont {{Lee}}}]{Zhai2009A0905.1711}%
  \BibitemOpen
  \bibfield  {author} {\bibinfo {author} {\bibfnamefont {H.}~\bibnamefont
  {{Zhai}}}, \bibinfo {author} {\bibfnamefont {F.}~\bibnamefont {{Wang}}},\
  and\ \bibinfo {author} {\bibfnamefont {D.-H.}\ \bibnamefont {{Lee}}},\
  }\bibfield  {title} {\bibinfo {title} {{Antiferromagnetically driven
  electronic correlations in iron pnictides and cuprates}},\ }\href
  {https://doi.org/10.1103/PhysRevB.80.064517} {\bibfield  {journal} {\bibinfo
  {journal} {\prb}\ }\textbf {\bibinfo {volume} {80}},\ \bibinfo {eid} {064517}
  (\bibinfo {year} {2009})},\ \Eprint {https://arxiv.org/abs/0905.1711}
  {arXiv:0905.1711 [cond-mat.supr-con]} \BibitemShut {NoStop}%
\bibitem [{\citenamefont {{R{\"u}ger}}\ \emph {et~al.}(2014)\citenamefont
  {{R{\"u}ger}}, \citenamefont {{Tocchio}}, \citenamefont {{Valent{\'\i}}},\
  and\ \citenamefont {{Gros}}}]{Ruger2014T1311.6504}%
  \BibitemOpen
  \bibfield  {author} {\bibinfo {author} {\bibfnamefont {R.}~\bibnamefont
  {{R{\"u}ger}}}, \bibinfo {author} {\bibfnamefont {L.~F.}\ \bibnamefont
  {{Tocchio}}}, \bibinfo {author} {\bibfnamefont {R.}~\bibnamefont
  {{Valent{\'\i}}}},\ and\ \bibinfo {author} {\bibfnamefont {C.}~\bibnamefont
  {{Gros}}},\ }\bibfield  {title} {\bibinfo {title} {{The phase diagram of the
  square lattice bilayer Hubbard model: a variational Monte Carlo study}},\
  }\href {https://doi.org/10.1088/1367-2630/16/3/033010} {\bibfield  {journal}
  {\bibinfo  {journal} {New Journal of Physics}\ }\textbf {\bibinfo {volume}
  {16}},\ \bibinfo {eid} {033010} (\bibinfo {year} {2014})},\ \Eprint
  {https://arxiv.org/abs/1311.6504} {arXiv:1311.6504 [cond-mat.str-el]}
  \BibitemShut {NoStop}%
\bibitem [{\citenamefont {{Rhim}}\ and\ \citenamefont
  {{Yang}}(2019)}]{Rhim2019C1808.05926}%
  \BibitemOpen
  \bibfield  {author} {\bibinfo {author} {\bibfnamefont {J.-W.}\ \bibnamefont
  {{Rhim}}}\ and\ \bibinfo {author} {\bibfnamefont {B.-J.}\ \bibnamefont
  {{Yang}}},\ }\bibfield  {title} {\bibinfo {title} {{Classification of flat
  bands according to the band-crossing singularity of Bloch wave functions}},\
  }\href {https://doi.org/10.1103/PhysRevB.99.045107} {\bibfield  {journal}
  {\bibinfo  {journal} {\prb}\ }\textbf {\bibinfo {volume} {99}},\ \bibinfo
  {eid} {045107} (\bibinfo {year} {2019})},\ \Eprint
  {https://arxiv.org/abs/1808.05926} {arXiv:1808.05926 [cond-mat.str-el]}
  \BibitemShut {NoStop}%
\bibitem [{\citenamefont {Chao}\ \emph {et~al.}(1978)\citenamefont {Chao},
  \citenamefont {Spa\l{}ek},\ and\ \citenamefont {Ole\ifmmode~\acute{s}\else
  \'{s}\fi{}}}]{Chao1978tJ}%
  \BibitemOpen
  \bibfield  {author} {\bibinfo {author} {\bibfnamefont {K.~A.}\ \bibnamefont
  {Chao}}, \bibinfo {author} {\bibfnamefont {J.}~\bibnamefont {Spa\l{}ek}},\
  and\ \bibinfo {author} {\bibfnamefont {A.~M.}\ \bibnamefont
  {Ole\ifmmode~\acute{s}\else \'{s}\fi{}}},\ }\bibfield  {title} {\bibinfo
  {title} {Canonical perturbation expansion of the hubbard model},\ }\href
  {https://doi.org/10.1103/PhysRevB.18.3453} {\bibfield  {journal} {\bibinfo
  {journal} {Phys. Rev. B}\ }\textbf {\bibinfo {volume} {18}},\ \bibinfo
  {pages} {3453} (\bibinfo {year} {1978})}\BibitemShut {NoStop}%
\bibitem [{\citenamefont {Gutzwiller}(1964)}]{Gutzwiller1964E}%
  \BibitemOpen
  \bibfield  {author} {\bibinfo {author} {\bibfnamefont {M.~C.}\ \bibnamefont
  {Gutzwiller}},\ }\bibfield  {title} {\bibinfo {title} {Effect of correlation
  on the ferromagnetism of transition metals},\ }\href
  {https://doi.org/10.1103/PhysRev.134.A923} {\bibfield  {journal} {\bibinfo
  {journal} {Phys. Rev.}\ }\textbf {\bibinfo {volume} {134}},\ \bibinfo {pages}
  {A923} (\bibinfo {year} {1964})}\BibitemShut {NoStop}%
\bibitem [{\citenamefont {{Hou}}\ and\ \citenamefont
  {{You}}(2022)}]{YZY2022nuSMG}%
  \BibitemOpen
  \bibfield  {author} {\bibinfo {author} {\bibfnamefont {W.}~\bibnamefont
  {{Hou}}}\ and\ \bibinfo {author} {\bibfnamefont {Y.-Z.}\ \bibnamefont
  {{You}}},\ }\bibfield  {title} {\bibinfo {title} {{Variational Monte Carlo
  Study of Symmetric Mass Generation in a Bilayer Honeycomb Lattice Model}},\
  }\href {https://doi.org/10.48550/arXiv.2212.13364} {\bibfield  {journal}
  {\bibinfo  {journal} {arXiv e-prints}\ ,\ \bibinfo {eid} {arXiv:2212.13364}}
  (\bibinfo {year} {2022})},\ \Eprint {https://arxiv.org/abs/2212.13364}
  {arXiv:2212.13364 [cond-mat.str-el]} \BibitemShut {NoStop}%
\bibitem [{\citenamefont {Luttinger}\ and\ \citenamefont
  {Ward}(1960)}]{Luttinger1960G}%
  \BibitemOpen
  \bibfield  {author} {\bibinfo {author} {\bibfnamefont {J.~M.}\ \bibnamefont
  {Luttinger}}\ and\ \bibinfo {author} {\bibfnamefont {J.~C.}\ \bibnamefont
  {Ward}},\ }\bibfield  {title} {\bibinfo {title} {Ground-state energy of a
  many-fermion system. ii},\ }\href {https://doi.org/10.1103/PhysRev.118.1417}
  {\bibfield  {journal} {\bibinfo  {journal} {Phys. Rev.}\ }\textbf {\bibinfo
  {volume} {118}},\ \bibinfo {pages} {1417} (\bibinfo {year}
  {1960})}\BibitemShut {NoStop}%
\bibitem [{\citenamefont {{Senthil}}\ \emph {et~al.}(2003)\citenamefont
  {{Senthil}}, \citenamefont {{Sachdev}},\ and\ \citenamefont
  {{Vojta}}}]{Senthil2003Topo}%
  \BibitemOpen
  \bibfield  {author} {\bibinfo {author} {\bibfnamefont {T.}~\bibnamefont
  {{Senthil}}}, \bibinfo {author} {\bibfnamefont {S.}~\bibnamefont
  {{Sachdev}}},\ and\ \bibinfo {author} {\bibfnamefont {M.}~\bibnamefont
  {{Vojta}}},\ }\bibfield  {title} {\bibinfo {title} {{Fractionalized Fermi
  Liquids}},\ }\href {https://doi.org/10.1103/PhysRevLett.90.216403} {\bibfield
   {journal} {\bibinfo  {journal} {\prl}\ }\textbf {\bibinfo {volume} {90}},\
  \bibinfo {eid} {216403} (\bibinfo {year} {2003})},\ \Eprint
  {https://arxiv.org/abs/cond-mat/0209144} {arXiv:cond-mat/0209144
  [cond-mat.str-el]} \BibitemShut {NoStop}%
\bibitem [{\citenamefont {{Senthil}}\ \emph {et~al.}(2004)\citenamefont
  {{Senthil}}, \citenamefont {{Vojta}},\ and\ \citenamefont
  {{Sachdev}}}]{Senthilcond-mat/0305193}%
  \BibitemOpen
  \bibfield  {author} {\bibinfo {author} {\bibfnamefont {T.}~\bibnamefont
  {{Senthil}}}, \bibinfo {author} {\bibfnamefont {M.}~\bibnamefont {{Vojta}}},\
  and\ \bibinfo {author} {\bibfnamefont {S.}~\bibnamefont {{Sachdev}}},\
  }\bibfield  {title} {\bibinfo {title} {{Weak magnetism and non-Fermi liquids
  near heavy-fermion critical points}},\ }\href
  {https://doi.org/10.1103/PhysRevB.69.035111} {\bibfield  {journal} {\bibinfo
  {journal} {\prb}\ }\textbf {\bibinfo {volume} {69}},\ \bibinfo {eid} {035111}
  (\bibinfo {year} {2004})},\ \Eprint {https://arxiv.org/abs/cond-mat/0305193}
  {arXiv:cond-mat/0305193 [cond-mat.str-el]} \BibitemShut {NoStop}%
\bibitem [{\citenamefont {{Paramekanti}}\ and\ \citenamefont
  {{Vishwanath}}(2004)}]{Paramekanticond-mat/0406619}%
  \BibitemOpen
  \bibfield  {author} {\bibinfo {author} {\bibfnamefont {A.}~\bibnamefont
  {{Paramekanti}}}\ and\ \bibinfo {author} {\bibfnamefont {A.}~\bibnamefont
  {{Vishwanath}}},\ }\bibfield  {title} {\bibinfo {title} {{Extending
  Luttinger's theorem to $\mathbb{Z}_{2}$ fractionalized phases of matter}},\
  }\href {https://doi.org/10.1103/PhysRevB.70.245118} {\bibfield  {journal}
  {\bibinfo  {journal} {\prb}\ }\textbf {\bibinfo {volume} {70}},\ \bibinfo
  {eid} {245118} (\bibinfo {year} {2004})},\ \Eprint
  {https://arxiv.org/abs/cond-mat/0406619} {arXiv:cond-mat/0406619
  [cond-mat.str-el]} \BibitemShut {NoStop}%
\bibitem [{\citenamefont {{Oshikawa}}\ and\ \citenamefont
  {{Senthil}}(2006)}]{Senthil2006Topo}%
  \BibitemOpen
  \bibfield  {author} {\bibinfo {author} {\bibfnamefont {M.}~\bibnamefont
  {{Oshikawa}}}\ and\ \bibinfo {author} {\bibfnamefont {T.}~\bibnamefont
  {{Senthil}}},\ }\bibfield  {title} {\bibinfo {title} {{Fractionalization,
  Topological Order, and Quasiparticle Statistics}},\ }\href
  {https://doi.org/10.1103/PhysRevLett.96.060601} {\bibfield  {journal}
  {\bibinfo  {journal} {\prl}\ }\textbf {\bibinfo {volume} {96}},\ \bibinfo
  {eid} {060601} (\bibinfo {year} {2006})},\ \Eprint
  {https://arxiv.org/abs/cond-mat/0506008} {arXiv:cond-mat/0506008
  [cond-mat.str-el]} \BibitemShut {NoStop}%
\bibitem [{\citenamefont {{Nandkishore}}\ \emph {et~al.}(2012)\citenamefont
  {{Nandkishore}}, \citenamefont {{Metlitski}},\ and\ \citenamefont
  {{Senthil}}}]{Senthil2012Topo}%
  \BibitemOpen
  \bibfield  {author} {\bibinfo {author} {\bibfnamefont {R.}~\bibnamefont
  {{Nandkishore}}}, \bibinfo {author} {\bibfnamefont {M.~A.}\ \bibnamefont
  {{Metlitski}}},\ and\ \bibinfo {author} {\bibfnamefont {T.}~\bibnamefont
  {{Senthil}}},\ }\bibfield  {title} {\bibinfo {title} {{Orthogonal metals: The
  simplest non-Fermi liquids}},\ }\href
  {https://doi.org/10.1103/PhysRevB.86.045128} {\bibfield  {journal} {\bibinfo
  {journal} {\prb}\ }\textbf {\bibinfo {volume} {86}},\ \bibinfo {eid} {045128}
  (\bibinfo {year} {2012})},\ \Eprint {https://arxiv.org/abs/1201.5998}
  {arXiv:1201.5998 [cond-mat.str-el]} \BibitemShut {NoStop}%
\bibitem [{\citenamefont {{Sachdev}}(2019)}]{Sachdev2019T1801.01125}%
  \BibitemOpen
  \bibfield  {author} {\bibinfo {author} {\bibfnamefont {S.}~\bibnamefont
  {{Sachdev}}},\ }\bibfield  {title} {\bibinfo {title} {{Topological order,
  emergent gauge fields, and Fermi surface reconstruction}},\ }\href
  {https://doi.org/10.1088/1361-6633/aae110} {\bibfield  {journal} {\bibinfo
  {journal} {Reports on Progress in Physics}\ }\textbf {\bibinfo {volume}
  {82}},\ \bibinfo {eid} {014001} (\bibinfo {year} {2019})},\ \Eprint
  {https://arxiv.org/abs/1801.01125} {arXiv:1801.01125 [cond-mat.str-el]}
  \BibitemShut {NoStop}%
\bibitem [{\citenamefont {{Skolimowski}}\ and\ \citenamefont
  {{Fabrizio}}(2022)}]{Skolimowski2022L2202.00426}%
  \BibitemOpen
  \bibfield  {author} {\bibinfo {author} {\bibfnamefont {J.}~\bibnamefont
  {{Skolimowski}}}\ and\ \bibinfo {author} {\bibfnamefont {M.}~\bibnamefont
  {{Fabrizio}}},\ }\bibfield  {title} {\bibinfo {title} {{Luttinger's theorem
  in the presence of Luttinger surfaces}},\ }\href
  {https://doi.org/10.1103/PhysRevB.106.045109} {\bibfield  {journal} {\bibinfo
   {journal} {\prb}\ }\textbf {\bibinfo {volume} {106}},\ \bibinfo {eid}
  {045109} (\bibinfo {year} {2022})},\ \Eprint
  {https://arxiv.org/abs/2202.00426} {arXiv:2202.00426 [cond-mat.str-el]}
  \BibitemShut {NoStop}%
\bibitem [{\citenamefont {{S{\'e}n{\'e}chal}}\ \emph
  {et~al.}(2002)\citenamefont {{S{\'e}n{\'e}chal}}, \citenamefont {{Perez}},\
  and\ \citenamefont {{Plouffe}}}]{senechal2002cluster}%
  \BibitemOpen
  \bibfield  {author} {\bibinfo {author} {\bibfnamefont {D.}~\bibnamefont
  {{S{\'e}n{\'e}chal}}}, \bibinfo {author} {\bibfnamefont {D.}~\bibnamefont
  {{Perez}}},\ and\ \bibinfo {author} {\bibfnamefont {D.}~\bibnamefont
  {{Plouffe}}},\ }\bibfield  {title} {\bibinfo {title} {{Cluster perturbation
  theory for Hubbard models}},\ }\href
  {https://doi.org/10.1103/PhysRevB.66.075129} {\bibfield  {journal} {\bibinfo
  {journal} {\prb}\ }\textbf {\bibinfo {volume} {66}},\ \bibinfo {eid} {075129}
  (\bibinfo {year} {2002})},\ \Eprint {https://arxiv.org/abs/cond-mat/0205044}
  {arXiv:cond-mat/0205044 [cond-mat.str-el]} \BibitemShut {NoStop}%
\bibitem [{\citenamefont {{Lee}}\ \emph {et~al.}(2006)\citenamefont {{Lee}},
  \citenamefont {{Nagaosa}},\ and\ \citenamefont
  {{Wen}}}]{Lee2004Dcond-mat/0410445}%
  \BibitemOpen
  \bibfield  {author} {\bibinfo {author} {\bibfnamefont {P.~A.}\ \bibnamefont
  {{Lee}}}, \bibinfo {author} {\bibfnamefont {N.}~\bibnamefont {{Nagaosa}}},\
  and\ \bibinfo {author} {\bibfnamefont {X.-G.}\ \bibnamefont {{Wen}}},\
  }\bibfield  {title} {\bibinfo {title} {{Doping a Mott Insulator: Physics of
  High Temperature Superconductivity}},\ }\href
  {https://doi.org/10.1103/RevModPhys.78.17} {\bibfield  {journal} {\bibinfo
  {journal} {Rev. Mod. Phys.}\ }\textbf {\bibinfo {volume} {78}},\ \bibinfo
  {pages} {17} (\bibinfo {year} {2006})},\ \Eprint
  {https://arxiv.org/abs/cond-mat/0410445} {arXiv:cond-mat/0410445
  [cond-mat.str-el]} \BibitemShut {NoStop}%
\bibitem [{\citenamefont {{Keimer}}\ \emph {et~al.}(2015)\citenamefont
  {{Keimer}}, \citenamefont {{Kivelson}}, \citenamefont {{Norman}},
  \citenamefont {{Uchida}},\ and\ \citenamefont
  {{Zaanen}}}]{Keimer2014H1409.4673}%
  \BibitemOpen
  \bibfield  {author} {\bibinfo {author} {\bibfnamefont {B.}~\bibnamefont
  {{Keimer}}}, \bibinfo {author} {\bibfnamefont {S.~A.}\ \bibnamefont
  {{Kivelson}}}, \bibinfo {author} {\bibfnamefont {M.~R.}\ \bibnamefont
  {{Norman}}}, \bibinfo {author} {\bibfnamefont {S.}~\bibnamefont {{Uchida}}},\
  and\ \bibinfo {author} {\bibfnamefont {J.}~\bibnamefont {{Zaanen}}},\
  }\bibfield  {title} {\bibinfo {title} {{From quantum matter to
  high-temperature superconductivity in copper oxides}},\ }\href
  {https://doi.org/10.1038/nature14165} {\bibfield  {journal} {\bibinfo
  {journal} {Nature}\ }\textbf {\bibinfo {volume} {518}},\ \bibinfo {eid} {179}
  (\bibinfo {year} {2015})},\ \Eprint {https://arxiv.org/abs/1409.4673}
  {arXiv:1409.4673 [cond-mat.supr-con]} \BibitemShut {NoStop}%
\bibitem [{\citenamefont {{Franz}}\ and\ \citenamefont
  {{Millis}}(1998)}]{Franz1998Pcond-mat/9805401}%
  \BibitemOpen
  \bibfield  {author} {\bibinfo {author} {\bibfnamefont {M.}~\bibnamefont
  {{Franz}}}\ and\ \bibinfo {author} {\bibfnamefont {A.~J.}\ \bibnamefont
  {{Millis}}},\ }\bibfield  {title} {\bibinfo {title} {{Phase fluctuations and
  spectral properties of underdoped cuprates}},\ }\href
  {https://doi.org/10.1103/PhysRevB.58.14572} {\bibfield  {journal} {\bibinfo
  {journal} {\prb}\ }\textbf {\bibinfo {volume} {58}},\ \bibinfo {pages}
  {14572} (\bibinfo {year} {1998})},\ \Eprint
  {https://arxiv.org/abs/cond-mat/9805401} {arXiv:cond-mat/9805401
  [cond-mat.supr-con]} \BibitemShut {NoStop}%
\bibitem [{\citenamefont {{Kwon}}\ and\ \citenamefont
  {{Dorsey}}(1999)}]{Kwon1999Econd-mat/9809225}%
  \BibitemOpen
  \bibfield  {author} {\bibinfo {author} {\bibfnamefont {H.-J.}\ \bibnamefont
  {{Kwon}}}\ and\ \bibinfo {author} {\bibfnamefont {A.~T.}\ \bibnamefont
  {{Dorsey}}},\ }\bibfield  {title} {\bibinfo {title} {{Effect of phase
  fluctuations on the single-particle properties of underdoped cuprates}},\
  }\href {https://doi.org/10.1103/PhysRevB.59.6438} {\bibfield  {journal}
  {\bibinfo  {journal} {\prb}\ }\textbf {\bibinfo {volume} {59}},\ \bibinfo
  {pages} {6438} (\bibinfo {year} {1999})},\ \Eprint
  {https://arxiv.org/abs/cond-mat/9809225} {arXiv:cond-mat/9809225
  [cond-mat.supr-con]} \BibitemShut {NoStop}%
\bibitem [{\citenamefont {{Kwon}}\ \emph {et~al.}(2001)\citenamefont {{Kwon}},
  \citenamefont {{Dorsey}},\ and\ \citenamefont
  {{Hirschfeld}}}]{Kwon2001Ocond-mat/0006290}%
  \BibitemOpen
  \bibfield  {author} {\bibinfo {author} {\bibfnamefont {H.-J.}\ \bibnamefont
  {{Kwon}}}, \bibinfo {author} {\bibfnamefont {A.~T.}\ \bibnamefont
  {{Dorsey}}},\ and\ \bibinfo {author} {\bibfnamefont {P.~J.}\ \bibnamefont
  {{Hirschfeld}}},\ }\bibfield  {title} {\bibinfo {title} {{Observability of
  Quantum Phase Fluctuations in Cuprate Superconductors}},\ }\href
  {https://doi.org/10.1103/PhysRevLett.86.3875} {\bibfield  {journal} {\bibinfo
   {journal} {\prl}\ }\textbf {\bibinfo {volume} {86}},\ \bibinfo {pages}
  {3875} (\bibinfo {year} {2001})},\ \Eprint
  {https://arxiv.org/abs/cond-mat/0006290} {arXiv:cond-mat/0006290
  [cond-mat.supr-con]} \BibitemShut {NoStop}%
\bibitem [{\citenamefont {{Franz}}\ and\ \citenamefont
  {{Te{\v{s}}anovi{\'c}}}(2001)}]{Franz2001Acond-mat/0012445}%
  \BibitemOpen
  \bibfield  {author} {\bibinfo {author} {\bibfnamefont {M.}~\bibnamefont
  {{Franz}}}\ and\ \bibinfo {author} {\bibfnamefont {Z.}~\bibnamefont
  {{Te{\v{s}}anovi{\'c}}}},\ }\bibfield  {title} {\bibinfo {title} {{Algebraic
  Fermi Liquid from Phase Fluctuations: ``Topological`` Fermions, Vortex
  ``Berryons,'' and QED$_{3}$ Theory of Cuprate Superconductors}},\ }\href
  {https://doi.org/10.1103/PhysRevLett.87.257003} {\bibfield  {journal}
  {\bibinfo  {journal} {\prl}\ }\textbf {\bibinfo {volume} {87}},\ \bibinfo
  {eid} {257003} (\bibinfo {year} {2001})},\ \Eprint
  {https://arxiv.org/abs/cond-mat/0012445} {arXiv:cond-mat/0012445
  [cond-mat.supr-con]} \BibitemShut {NoStop}%
\bibitem [{\citenamefont {{Curty}}\ and\ \citenamefont
  {{Beck}}(2003)}]{Curty2003Tcond-mat/0401124}%
  \BibitemOpen
  \bibfield  {author} {\bibinfo {author} {\bibfnamefont {P.}~\bibnamefont
  {{Curty}}}\ and\ \bibinfo {author} {\bibfnamefont {H.}~\bibnamefont
  {{Beck}}},\ }\bibfield  {title} {\bibinfo {title} {{Thermodynamics and Phase
  Diagram of High Temperature Superconductors}},\ }\href
  {https://doi.org/10.1103/PhysRevLett.91.257002} {\bibfield  {journal}
  {\bibinfo  {journal} {\prl}\ }\textbf {\bibinfo {volume} {91}},\ \bibinfo
  {eid} {257002} (\bibinfo {year} {2003})},\ \Eprint
  {https://arxiv.org/abs/cond-mat/0401124} {arXiv:cond-mat/0401124
  [cond-mat.supr-con]} \BibitemShut {NoStop}%
\bibitem [{\citenamefont {{Li}}\ and\ \citenamefont
  {{Yao}}(2018{\natexlab{a}})}]{Li2018P1805.05530}%
  \BibitemOpen
  \bibfield  {author} {\bibinfo {author} {\bibfnamefont {T.}~\bibnamefont
  {{Li}}}\ and\ \bibinfo {author} {\bibfnamefont {D.-W.}\ \bibnamefont
  {{Yao}}},\ }\bibfield  {title} {\bibinfo {title} {{Pairing origin of the
  pseudogap as observed in ARPES measurement in the underdoped cuprates}},\
  }\href {https://doi.org/10.48550/arXiv.1805.05530} {\bibfield  {journal}
  {\bibinfo  {journal} {arXiv e-prints}\ ,\ \bibinfo {eid} {arXiv:1805.05530}}
  (\bibinfo {year} {2018}{\natexlab{a}})},\ \Eprint
  {https://arxiv.org/abs/1805.05530} {arXiv:1805.05530 [cond-mat.supr-con]}
  \BibitemShut {NoStop}%
\bibitem [{\citenamefont {{Li}}\ and\ \citenamefont
  {{Yao}}(2018{\natexlab{b}})}]{Li2018W1803.08226}%
  \BibitemOpen
  \bibfield  {author} {\bibinfo {author} {\bibfnamefont {T.}~\bibnamefont
  {{Li}}}\ and\ \bibinfo {author} {\bibfnamefont {D.-W.}\ \bibnamefont
  {{Yao}}},\ }\bibfield  {title} {\bibinfo {title} {{Why is the antinodal
  quasiparticle in the electron-doped cuprate
  Pr$_{1.3-x}$La$_{0.7}$Ce$_{x}$CuO$_{4}$ immune to the antiferromagnetic
  band-folding effect?}},\ }\href {https://doi.org/10.1209/0295-5075/124/47001}
  {\bibfield  {journal} {\bibinfo  {journal} {EPL (Europhysics Letters)}\
  }\textbf {\bibinfo {volume} {124}},\ \bibinfo {pages} {47001} (\bibinfo
  {year} {2018}{\natexlab{b}})},\ \Eprint {https://arxiv.org/abs/1803.08226}
  {arXiv:1803.08226 [cond-mat.str-el]} \BibitemShut {NoStop}%
\bibitem [{\citenamefont {{Ye}}\ and\ \citenamefont
  {{Chubukov}}(2019)}]{Ye2019H1905.11412}%
  \BibitemOpen
  \bibfield  {author} {\bibinfo {author} {\bibfnamefont {M.}~\bibnamefont
  {{Ye}}}\ and\ \bibinfo {author} {\bibfnamefont {A.~V.}\ \bibnamefont
  {{Chubukov}}},\ }\bibfield  {title} {\bibinfo {title} {{Hubbard model on a
  triangular lattice: Pseudogap due to spin density wave fluctuations}},\
  }\href {https://doi.org/10.1103/PhysRevB.100.035135} {\bibfield  {journal}
  {\bibinfo  {journal} {\prb}\ }\textbf {\bibinfo {volume} {100}},\ \bibinfo
  {eid} {035135} (\bibinfo {year} {2019})},\ \Eprint
  {https://arxiv.org/abs/1905.11412} {arXiv:1905.11412 [cond-mat.str-el]}
  \BibitemShut {NoStop}%
\bibitem [{\citenamefont {{Troyer}}\ and\ \citenamefont
  {{Wiese}}(2005)}]{Troyer2005Ccond-mat/0408370}%
  \BibitemOpen
  \bibfield  {author} {\bibinfo {author} {\bibfnamefont {M.}~\bibnamefont
  {{Troyer}}}\ and\ \bibinfo {author} {\bibfnamefont {U.-J.}\ \bibnamefont
  {{Wiese}}},\ }\bibfield  {title} {\bibinfo {title} {{Computational Complexity
  and Fundamental Limitations to Fermionic Quantum Monte Carlo Simulations}},\
  }\href {https://doi.org/10.1103/PhysRevLett.94.170201} {\bibfield  {journal}
  {\bibinfo  {journal} {\prl}\ }\textbf {\bibinfo {volume} {94}},\ \bibinfo
  {eid} {170201} (\bibinfo {year} {2005})},\ \Eprint
  {https://arxiv.org/abs/cond-mat/0408370} {arXiv:cond-mat/0408370
  [cond-mat.stat-mech]} \BibitemShut {NoStop}%
\bibitem [{\citenamefont {Fucito}\ \emph {et~al.}(1981)\citenamefont {Fucito},
  \citenamefont {Marinari}, \citenamefont {Parisi},\ and\ \citenamefont
  {Rebbi}}]{Fucito1981A}%
  \BibitemOpen
  \bibfield  {author} {\bibinfo {author} {\bibfnamefont {F.}~\bibnamefont
  {Fucito}}, \bibinfo {author} {\bibfnamefont {E.}~\bibnamefont {Marinari}},
  \bibinfo {author} {\bibfnamefont {G.}~\bibnamefont {Parisi}},\ and\ \bibinfo
  {author} {\bibfnamefont {C.}~\bibnamefont {Rebbi}},\ }\bibfield  {title}
  {\bibinfo {title} {A proposal for monte carlo simulations of fermionic
  systems},\ }\href
  {https://doi.org/https://doi.org/10.1016/0550-3213(81)90055-9} {\bibfield
  {journal} {\bibinfo  {journal} {Nuclear Physics B}\ }\textbf {\bibinfo
  {volume} {180}},\ \bibinfo {pages} {369} (\bibinfo {year}
  {1981})}\BibitemShut {NoStop}%
\bibitem [{\citenamefont {Scalapino}\ and\ \citenamefont
  {Sugar}(1981)}]{Scalapino1981M}%
  \BibitemOpen
  \bibfield  {author} {\bibinfo {author} {\bibfnamefont {D.~J.}\ \bibnamefont
  {Scalapino}}\ and\ \bibinfo {author} {\bibfnamefont {R.~L.}\ \bibnamefont
  {Sugar}},\ }\bibfield  {title} {\bibinfo {title} {Method for performing monte
  carlo calculations for systems with fermions},\ }\href
  {https://doi.org/10.1103/PhysRevLett.46.519} {\bibfield  {journal} {\bibinfo
  {journal} {Phys. Rev. Lett.}\ }\textbf {\bibinfo {volume} {46}},\ \bibinfo
  {pages} {519} (\bibinfo {year} {1981})}\BibitemShut {NoStop}%
\bibitem [{\citenamefont {Blankenbecler}\ \emph {et~al.}(1981)\citenamefont
  {Blankenbecler}, \citenamefont {Scalapino},\ and\ \citenamefont
  {Sugar}}]{Blankenbecler1981M}%
  \BibitemOpen
  \bibfield  {author} {\bibinfo {author} {\bibfnamefont {R.}~\bibnamefont
  {Blankenbecler}}, \bibinfo {author} {\bibfnamefont {D.~J.}\ \bibnamefont
  {Scalapino}},\ and\ \bibinfo {author} {\bibfnamefont {R.~L.}\ \bibnamefont
  {Sugar}},\ }\bibfield  {title} {\bibinfo {title} {Monte carlo calculations of
  coupled boson-fermion systems. i},\ }\href
  {https://doi.org/10.1103/PhysRevD.24.2278} {\bibfield  {journal} {\bibinfo
  {journal} {Phys. Rev. D}\ }\textbf {\bibinfo {volume} {24}},\ \bibinfo
  {pages} {2278} (\bibinfo {year} {1981})}\BibitemShut {NoStop}%
\bibitem [{\citenamefont {Hirsch}\ \emph {et~al.}(1981)\citenamefont {Hirsch},
  \citenamefont {Scalapino}, \citenamefont {Sugar},\ and\ \citenamefont
  {Blankenbecler}}]{Hirsch1981E}%
  \BibitemOpen
  \bibfield  {author} {\bibinfo {author} {\bibfnamefont {J.~E.}\ \bibnamefont
  {Hirsch}}, \bibinfo {author} {\bibfnamefont {D.~J.}\ \bibnamefont
  {Scalapino}}, \bibinfo {author} {\bibfnamefont {R.~L.}\ \bibnamefont
  {Sugar}},\ and\ \bibinfo {author} {\bibfnamefont {R.}~\bibnamefont
  {Blankenbecler}},\ }\bibfield  {title} {\bibinfo {title} {Efficient monte
  carlo procedure for systems with fermions},\ }\href
  {https://doi.org/10.1103/PhysRevLett.47.1628} {\bibfield  {journal} {\bibinfo
   {journal} {Phys. Rev. Lett.}\ }\textbf {\bibinfo {volume} {47}},\ \bibinfo
  {pages} {1628} (\bibinfo {year} {1981})}\BibitemShut {NoStop}%
\bibitem [{\citenamefont {Hirsch}(1985)}]{Hirsch1985T}%
  \BibitemOpen
  \bibfield  {author} {\bibinfo {author} {\bibfnamefont {J.~E.}\ \bibnamefont
  {Hirsch}},\ }\bibfield  {title} {\bibinfo {title} {Two-dimensional hubbard
  model: Numerical simulation study},\ }\href
  {https://doi.org/10.1103/PhysRevB.31.4403} {\bibfield  {journal} {\bibinfo
  {journal} {Phys. Rev. B}\ }\textbf {\bibinfo {volume} {31}},\ \bibinfo
  {pages} {4403} (\bibinfo {year} {1985})}\BibitemShut {NoStop}%
\bibitem [{\citenamefont {Sun}\ \emph {et~al.}(2023)\citenamefont {Sun},
  \citenamefont {Huo}, \citenamefont {Hu}, \citenamefont {Li}, \citenamefont
  {Liu}, \citenamefont {Han}, \citenamefont {Tang}, \citenamefont {Mao},
  \citenamefont {Yang}, \citenamefont {Wang}, \citenamefont {Cheng},
  \citenamefont {Yao}, \citenamefont {Zhang},\ and\ \citenamefont
  {Wang}}]{Sun2023S}%
  \BibitemOpen
  \bibfield  {author} {\bibinfo {author} {\bibfnamefont {H.}~\bibnamefont
  {Sun}}, \bibinfo {author} {\bibfnamefont {M.}~\bibnamefont {Huo}}, \bibinfo
  {author} {\bibfnamefont {X.}~\bibnamefont {Hu}}, \bibinfo {author}
  {\bibfnamefont {J.}~\bibnamefont {Li}}, \bibinfo {author} {\bibfnamefont
  {Z.}~\bibnamefont {Liu}}, \bibinfo {author} {\bibfnamefont {Y.}~\bibnamefont
  {Han}}, \bibinfo {author} {\bibfnamefont {L.}~\bibnamefont {Tang}}, \bibinfo
  {author} {\bibfnamefont {Z.}~\bibnamefont {Mao}}, \bibinfo {author}
  {\bibfnamefont {P.}~\bibnamefont {Yang}}, \bibinfo {author} {\bibfnamefont
  {B.}~\bibnamefont {Wang}}, \bibinfo {author} {\bibfnamefont {J.}~\bibnamefont
  {Cheng}}, \bibinfo {author} {\bibfnamefont {D.-X.}\ \bibnamefont {Yao}},
  \bibinfo {author} {\bibfnamefont {G.-M.}\ \bibnamefont {Zhang}},\ and\
  \bibinfo {author} {\bibfnamefont {M.}~\bibnamefont {Wang}},\ }\bibfield
  {title} {\bibinfo {title} {Signatures of superconductivity near 80 k in a
  nickelate under high pressure},\ }\bibfield  {journal} {\bibinfo  {journal}
  {Nature}\ }\href {https://doi.org/10.1038/s41586-023-06408-7}
  {10.1038/s41586-023-06408-7} (\bibinfo {year} {2023})\BibitemShut {NoStop}%
\bibitem [{\citenamefont {{Hou}}\ \emph {et~al.}(2023)\citenamefont {{Hou}},
  \citenamefont {{Yang}}, \citenamefont {{Liu}}, \citenamefont {{Li}},
  \citenamefont {{Shan}}, \citenamefont {{Ma}}, \citenamefont {{Wang}},
  \citenamefont {{Wang}}, \citenamefont {{Guo}}, \citenamefont {{Sun}},
  \citenamefont {{Uwatoko}}, \citenamefont {{Wang}}, \citenamefont {{Zhang}},
  \citenamefont {{Wang}},\ and\ \citenamefont {{Cheng}}}]{Hou2023E2307.09865}%
  \BibitemOpen
  \bibfield  {author} {\bibinfo {author} {\bibfnamefont {J.}~\bibnamefont
  {{Hou}}}, \bibinfo {author} {\bibfnamefont {P.~T.}\ \bibnamefont {{Yang}}},
  \bibinfo {author} {\bibfnamefont {Z.~Y.}\ \bibnamefont {{Liu}}}, \bibinfo
  {author} {\bibfnamefont {J.~Y.}\ \bibnamefont {{Li}}}, \bibinfo {author}
  {\bibfnamefont {P.~F.}\ \bibnamefont {{Shan}}}, \bibinfo {author}
  {\bibfnamefont {L.}~\bibnamefont {{Ma}}}, \bibinfo {author} {\bibfnamefont
  {G.}~\bibnamefont {{Wang}}}, \bibinfo {author} {\bibfnamefont {N.~N.}\
  \bibnamefont {{Wang}}}, \bibinfo {author} {\bibfnamefont {H.~Z.}\
  \bibnamefont {{Guo}}}, \bibinfo {author} {\bibfnamefont {J.~P.}\ \bibnamefont
  {{Sun}}}, \bibinfo {author} {\bibfnamefont {Y.}~\bibnamefont {{Uwatoko}}},
  \bibinfo {author} {\bibfnamefont {M.}~\bibnamefont {{Wang}}}, \bibinfo
  {author} {\bibfnamefont {G.~M.}\ \bibnamefont {{Zhang}}}, \bibinfo {author}
  {\bibfnamefont {B.~S.}\ \bibnamefont {{Wang}}},\ and\ \bibinfo {author}
  {\bibfnamefont {J.~G.}\ \bibnamefont {{Cheng}}},\ }\bibfield  {title}
  {\bibinfo {title} {{Emergence of high-temperature superconducting phase in
  the pressurized La3Ni2O7 crystals}},\ }\href
  {https://doi.org/10.48550/arXiv.2307.09865} {\bibfield  {journal} {\bibinfo
  {journal} {arXiv e-prints}\ ,\ \bibinfo {eid} {arXiv:2307.09865}} (\bibinfo
  {year} {2023})},\ \Eprint {https://arxiv.org/abs/2307.09865}
  {arXiv:2307.09865 [cond-mat.supr-con]} \BibitemShut {NoStop}%
\end{thebibliography}%

\clearpage
\appendix

\section{Cluster Perturbation Theory}
\label{append:cpt}

\begin{figure}[h]
    \centering
    \includegraphics[width=0.3\textwidth]{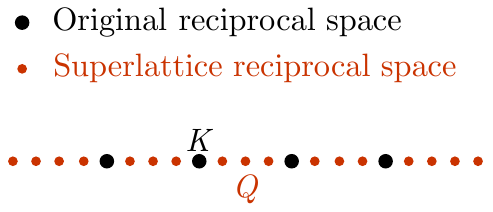}
    \caption{Reciprocal lattice in 1d for a 4-site cluster. $K$ labels the reciprocal lattice vector for the original lattice and $Q$ labels the reciprocal lattice vector for the superlattice. More precisely, $K_s=\frac{2\pi}{a}s$ and $Q_s=\frac{2\pi}{La}s$, where $a$ is the lattice constant of the original lattice, $L=4$ here and $s\in\dsZ$ .}
    \label{fig:reciprocal-lattice}
\end{figure}

Here we review the details of cluster perturbation theory (CPT) originally developed in \cite{CPT2000}. 
Denote the superlattice lattice points by $\bR$, then the position of any original lattice point would be given by $\bR+\br$, where $\br$ is the relative position of the lattice point to the location $\bR$ of the cluster containing that particular lattice point. For clusters of size $L$, the generic Green's function in real space can be denoted by $G_{i,j}^{\bR,\bR'}$, with $i,j=1,...,L$, where the time-dependence is implicitly assumed and same goes for the frequency-dependence in Fourier space. Due to the translation invariance of the clusters on the \textit{superlattice}, the real space Green's function can be firstly partially Fourier-transformed to give,
\begin{equation}
   G_{i,j}^{\bR,\bR'}=\frac{1}{N}\sum_{\bq} G(\bq)_{ij}\e^{i\bq\cdot (\bR-\bR')},
   \label{eq-G-real-space}
\end{equation}
where the $\bq$-summation is over the Brillouin zone (BZ) of the superlattice and $N$ is the number of clusters on the superlattice, which goes to infinity in the thermodynamic limit. In contrast to the translation invariance of the $(\bR,\bR')$-part of $G_{i,j}^{\bR,\bR'}$, or equivalently it only depends on the difference $\bR-\bR'$ as can be seen in Eq.~(\ref{eq-G-real-space}),  the $(i,j)$-part of the Green's function loses translation invariance due to the introduction of clusters. This is so because correlation between two points within the same cluster is not manifestly the same with the correlation between another pair of equally separated points \textit{across} clusters. Therefore, it takes two lattice momenta to fully characterize $G_{i,j}^{\bR,\bR'}$ in Fourier space. More precisely, we have, 
\begin{equation}
    G(\bk,\bk')=\frac{1}{NL}\sum_{\bR,\bR'}\sum_{i,j}G_{i,j}^{\bR,\bR'}\e^{i\bk\cdot(\bR+\br_i)-i\bk'\cdot(\bR'+\br_j)}.
    \label{eq-G-k-space}
\end{equation}
Then we can plug Eq.~(\ref{eq-G-real-space}) into Eq.~(\ref{eq-G-k-space}) and integrate out the superlattice lattice vectors $\bR,\bR'$ to obtain the following,
\begin{equation}
\label{eq-G-k-space-2}
    G(\bk,\bk')=\frac{1}{L}\sum_{i,j}\sum_{\bq}G(\bq)_{ij}\tilde{\delta}_{\bk,\bq}\tilde{\delta}_{\bk',\bq}\e^{i(\bk\cdot\br_i-\bk'\cdot\br_j)},
\end{equation}
where the $\tilde{\delta}$-function denotes the fact that the two wavevectors are equivalent only up to a superlattice reciprocal lattice vector $\bQ$ because $\bQ\cdot\bR=2\pi\dsZ$ in the phase factor. More precisely, we have
\begin{equation}
    \tilde{\delta}_{\bk,\bq}=\sum_{s=1}^{L}\delta_{\bk,\bq+\bQ_s},
\end{equation}
where $\bQ_s$ with $s=1,...,L$ are the $L$ inequivalent wave vectors in the reciprocal lattice of the original lattice (see the 1d case shown in Fig.~\ref{fig:reciprocal-lattice}). Then we can perform the $\bq$-summation in Eq.~(\ref{eq-G-k-space-2}) to have, 
\begin{equation}
    \begin{split}
     G(\bk,\bk')&=\frac{1}{L}\sum_{i,j}\sum_{s,s'}G(\bk-\bQ_s)_{ij}\delta_{\bk'-\bk,\bQ_s-\bQ_{s'}}\e^{i(\bk\cdot\br_i-\bk'\cdot\br_j)}\\
        &=\sum_{i,j}\sum_{\Delta\bQ}G(\bk)_{ij}\delta_{\bk'-\bk,\Delta\bQ}\e^{i(\bk\cdot\br_i-\bk'\cdot\br_j)},
    \end{split}
\end{equation}
where we have used the fact that $G(\bq)_{ij}$ is invariant under the shift by a superlattice reciprocal lattice vector $\bQ_s$.  

The translation invariant approximation for the Green's function on the original lattice is obtained when $\Delta\bQ=0$, i.e. $\bk=\bk'$.  Therefore, the Green's function becomes,
\begin{equation} 
\label{eq-Green-final}
     G(\bk) =\sum_{i,j}G(\bk)_{ij}\e^{i\bk\cdot(\br_i-\br_j)}.
\end{equation}

\begin{figure}
    \centering
    \includegraphics[width=0.3\textwidth]{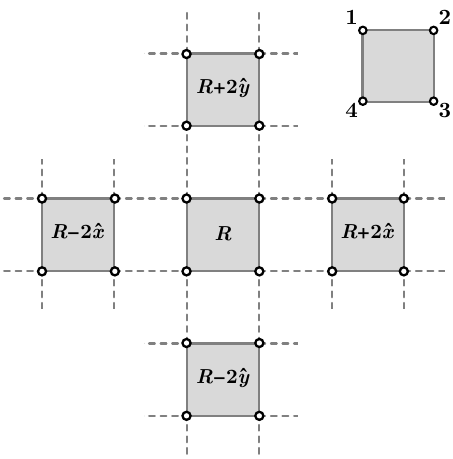}
    \caption{Cluster diagram showing the hopping between neighboring clusters (dashed line). The four sites inside each cluster are numbered as shown.}
    \label{fig:cluster}
\end{figure}

Now we just need to calculate $G_{i,j}(\bk)$ using cluster perturbation. The idea is to treat hopping between clusters as perturbation when consider strong on-site interactions. In particular,
\begin{equation}
    \hat{H}=\hat{H}_0+\hat{V},
\end{equation}
where $\hat{H}_0$ contains intra-cluster terms and $\hat{V}$ contains inter-cluster hopping. Considering nearest-neighbor hopping between the square clusters used in the main text. The cluster construction is reproduced in Fig.~\ref{fig:cluster} with the 4 sites in each cluster labeled by $1,2,3,4$. The hopping matrix is given by (setting lattice constant $a=1$)
\begin{equation}
\begin{split}
V_{i,j}^{\bR,\bR'}=&-t\delta_{\bR,\bR'-2\hat{x}}(\delta_{i,2}\delta_{j,1}+\delta_{i,3}\delta_{j,4})\\
&-t\delta_{\bR,\bR'+2\hat{x}}(\delta_{i,1}\delta_{j,2}+\delta_{i,4}\delta_{j,3})\\
&-t\delta_{\bR,\bR'-2\hat{y}}(\delta_{i,1}\delta_{j,4}+\delta_{i,2}\delta_{j,3})\\
&-t\delta_{\bR,\bR'+2\hat{y}}(\delta_{i,3}\delta_{j,2}+\delta_{i,4}\delta_{j,1})
\end{split}
\end{equation}
Fourier transforming $V_{i,j}^{\bR,\bR'}$ into the superlattice reciprocal space, we have 
\begin{equation}
\begin{split}
V_{i,j}(\bq)&=-t\e^{i2q_x}(\delta_{i,2}\delta_{j,1}+\delta_{i,3}\delta_{j,4})\\
&\quad -t\e^{-i2q_x}(\delta_{i,1}\delta_{j,2}+\delta_{i,4}\delta_{j,3})\\
&\quad -t\e^{i2q_y}(\delta_{i,1}\delta_{j,4}+\delta_{i,2}\delta_{j,3})\\
&\quad -t\e^{-i2q_y}(\delta_{i,3}\delta_{j,2}+\delta_{i,4}\delta_{j,1})\\
&=-t\begin{pmatrix}
0 &  e^{-i2\bq_x} & 0 & e^{i2\bq_y}\\
e^{i2\bq_x} &0 & e^{i2\bq_y} & 0\\
0 &  e^{-i2\bq_y} & 0&  e^{i2\bq_x}\\
 e^{-i2\bq_y} & 0 &  e^{-i2\bq_x} & 0
\end{pmatrix}_{i,j},
\end{split}
\end{equation}
which is the form presented in Eq.~(\ref{eq-V}) in the main text. Then the interacting Green's function is given by
\begin{equation}
\begin{split}
    \hat{G}(\bq)&=\frac{1}{\omega-\hat{H}}=\frac{1}{\omega-\hat{H}_0-\hat{V}(\bq)}\\
    &=\frac{\hat{G}_0}{1-\hat{V}(\bq)\hat{G}_0},
    \end{split}
\end{equation}
where $\hat{G}_0\equiv (\omega-\hat{H}_0)^{-1}$ is the intra-cluster Green's function that can be easily obtained by exact diagonalization as long as the cluster size is not too big. The obtained $G(\bq)_{ij}$ can now be plugged into Eq.~(\ref{eq-Green-final}) to calculate the CPT Green's function for the interacting system. 

\end{document}